\documentclass[fleqn,usenatbib]{mnras}
\usepackage{natbib}
\usepackage{verbatim}
\usepackage{ulem}

\usepackage[T1]{fontenc}
\usepackage{ae,aecompl}
\usepackage{comment}

\usepackage{graphicx}	
\usepackage{amsmath}	
\usepackage{amssymb}	

\usepackage{float}
\usepackage{scrextend}
\usepackage{xcolor}
\usepackage{subcaption}
\usepackage[switch]{lineno}
\usepackage{booktabs}
\usepackage{longtable}
\usepackage[para,flushleft]{threeparttable}
\usepackage{multirow}
\usepackage{newtxtext,newtxmath}

\title[MWL study of OT~081]{Multi-wavelength study of OT~081: broadband modelling of a transitional blazar}
\author[H.~Abe~et.~al.]{\parbox{\textwidth}{\large{
MAGIC Collaboration: H.~Abe$^{1}$,
S.~Abe$^{1}$,
V.~A.~Acciari$^{2}$,
I.~Agudo$^{3}$,
T.~Aniello$^{4}$,
S.~Ansoldi$^{5,6}$,
L.~A.~Antonelli$^{4}$,
A.~Arbet Engels$^{7}$,
C.~Arcaro$^{8}$,
M.~Artero$^{9}$
% ...
}
\newline
\emph{\normalsize Affiliations are listed at the end of the paper}
}
}

\date{Accepted XXX. Received YYY; in original form ZZZ}
\pubyear{2024}
\begin{document}
\label{firstpage}
\pagerange{\pageref{firstpage}--\pageref{lastpage}}
\maketitle
\clearpage
\begin{abstract}
OT~081 is a well-known, luminous blazar that is remarkably variable in many energy bands. We present the first broadband study of the source which includes very-high-energy (VHE, $E>$100\,GeV) $\gamma$-ray data taken by the MAGIC and H.E.S.S. imaging Cherenkov telescopes.
The discovery of VHE $\gamma$-ray emission happened during a high state of $\gamma$-ray activity in July 2016, observed by many instruments from radio to VHE $\gamma$-rays. We identify four states of activity of the source, one of which includes VHE $\gamma$-ray emission.
Variability in the VHE domain is found on daily timescales. The intrinsic VHE spectrum can be described by a power-law
with index $3.27\pm0.44_{\rm stat}\pm0.15_{\rm sys}$ (MAGIC) and $3.39\pm0.58_{\rm stat}\pm0.64_{\rm sys}$ (H.E.S.S.) in the energy range of 55--300\,GeV and 120--500\,GeV, respectively. The broadband emission cannot be sucessfully reproduced by a simple one-zone synchrotron self-Compton model. Instead, an additional external Compton component is required.
We test a lepto-hadronic model that reproduces the dataset well and a proton-synchrotron dominated model that requires an extreme proton luminosity. Emission models that are able to successfully represent the data place the emitting region well outside of the Broad Line Region (BLR) to a location at which the radiative environment is dominated by the infrared thermal radiation field of the dusty torus. In the scenario described by this flaring activity, the source appears to be an FSRQ, in contrast with past categorizations. This suggests that the source can be considered to be a transitional blazar, intermediate between BL~Lac and FSRQ objects.
\end{abstract}




\section{Introduction}
\label{sec:introduction}
Blazars are radio-loud active galactic nuclei (AGNs) whose relativistic jets are aligned with the Earth's line of sight. Within the wide blazar category, BL Lacs are commonly identified as those sources that show no or very weak emission lines in their optical spectra~\citep{1991ApJ...374..431S,1991ApJS...76..813S,2014A&ARv..22...73F}. This is attributed to a weak or absent broad-line region (BLR) with the relativistic jet emission dominating over the accretion disc spectrum~\citep{1995PASP..107..803U}. BL Lac objects represent the majority of the 84 blazars detected at very-high energies (VHE, $E >$ 100\,GeV)\footnote{\url{http://tevcat.uchicago.edu}/,\cite{2008ICRC....3.1341W}}.

The broadband spectral energy distribution (SED) that is characteristic of blazars~\citep{2017MNRAS.469..255G} has a doubled-bumped structure. Regarding the underlying radiative processes, the low-energy bump is universally attributed to synchrotron radiation, while the origin of the high-energy bump is under debate with two possible broad emission scenarios including also a combination of them. Within the leptonic framework, the high-energy emission can be explained as inverse Compton scattering of low-energy target photons. On the other hand, in hadronic models, hadrons can also be accelerated within the blazar jet, and synchrotron radiation by protons and secondary particles created in proton-photon interactions can also be responsible for the $\gamma$-ray emission. The broadband SED is used to identify three different types of BL Lac objects: low-, intermediate- and high-frequency-peaked BL Lacs~\citep[see][and references therein]{1995ApJ...444..567P,2007Ap&SS.309...95B}. For low-frequency-peaked BL Lacs (LBLs), the synchrotron emission peaks in the submillimeter to infrared wavebands, whereas it peaks in the ultraviolet (UV) to X-ray bands for high-frequency-peaked BL Lacs (HBLs). The peaks in intermediate BL Lacs (IBLs) occur somewhere in between. 

OT~081, also known as PKS~1749+096 and 4C~09.57, is a blazar located at a redshift of $z$=0.320 $\pm~ 0.005$~\citep{1988A&A...191L..16S}. Initially classified as an HBL blazar by \cite{1999A&AS..139..545K} and \cite{2000A&A...363..887D} because of its radio spectrum peaking above 10~GHz, a few years later it was instead suggested to be a flat spectrum radio quasar (FSRQ), presenting an inverted radio spectrum during flares~\citep{2005A&A...435..839T,2007ApJS..171...61H}.
In contrast with BL Lac objects, FSRQs exhibit strong and broad emission lines (with typical equivalent widths of $|EW_{\rm{rest}}|>5$\,\AA\,in the rest frame) in their optical spectra, indicating the presence of BLRs.
In the case of OT~081, measurements performed during low-flux periods clearly identified the emission lines H$\beta$ and [O III] 5007$~$\AA$~$ with equivalent widths of $EW\sim10~\textrm{\AA}$ ($|EW_{\rm{rest}}|\sim8~\textrm{\AA}$) \citep{1988A&A...191L..16S}.
Emission lines are not always present in the available spectra, and for this reason the source was categorized as a BL Lac in several cases \citep{1981AJ.....86..325A,1984AJ.....89..323G,1985ApJ...298..630L,1987AJ.....93....1B,1988AJ.....96.1215P,1991ApJ...374..431S,1997A&A...325..109S,1999ApJ...515..140S,2002A&A...381..810P,2010ApJ...716...30A}.
Recently, \cite{2021MNRAS.506.1540L} claimed to have detected quasi-periodic oscillations with periods of $\sim$~850\,d in the long-term radio (at frequencies of 4.8, 8, and 14.5~,GHz) light curves of OT~081. These oscillations are compatible with the presence of a massive black hole (BH) binary system or with blobs moving through helical structures inside the jet.
The variability of the light curve in the radio band is remarkable (i.e. flux density changes by a factor of 8) as documented by UMRAO observations~\citep{2000PASJ...52.1037I}\footnote{\url{https://dept.astro.lsa.umich.edu/obs/radiotel/gif/1749_096.gif}} and others~\citep{1997A&AS..122..271R,1998A&AS..132..305T}.
OT~081 has also shown strong variability in the optical~\citep{1995AJ....110..529C} and X-ray~\citep{1996ApJ...463..424U} bands.

Blazars often show a high polarization degree~\citep{1978ApJ...219L..85M}, and OT~081 has exhibited optical polarization up to 32\%~\citep{1986MNRAS.221..739B}.
Emission in the high-energy $\gamma$-ray range (HE, $E > 100~\mathrm{MeV}$) was reported by~\cite{2009ApJ...700..597A}. 
The source is listed as 3FHL~J1751.5+0938 in the third catalogue of hard Fermi-LAT sources~\citep[3FHL,][]{2017ApJS..232...18A}.
In the fourth catalogue of AGN detected by \textit{Fermi}-LAT \citep[4LAC,][]{2020ApJ...892..105A}, OT~081 is classified as a low-synchrotron-peaked blazar (LSP).\par
Emission in the VHE $\gamma$-ray band was detected for the first time by MAGIC (Major Atmospheric Gamma-ray Imaging Cherenkov telescopes) and H.E.S.S. (High Energy Stereoscopic System) in July 2016~\citep{2016ATel.9267....1M,2017ICRC...35..652S}  during a bright flare that was also detected by \textit{Fermi}-LAT \citep{2016ATel.9231....1B,2016ATel.9260....1C,2018MNRAS.480.2324K} and at other wavelengths \citep[e.g.][]{2016ATel.9259....1B}. The results of the dedicated multi-wavelength (MWL) campaign during that flare are reported in this work.\par
Prior to the flaring activity reported in this paper, the MWL studies of the source were conducted only up to the HE $\gamma$-ray band using \textit{Fermi}-LAT data. The high-energy bump of the SED was therefore only constrained by \textit{Fermi}-LAT observations.
As pointed out in \cite{2013MNRAS.436..304P}, the broadband SED of OT~081 is not well described by a synchrotron self-Compton model \citep[SSC,][]{1981ApJ...243..700K,1985ApJ...298..128B,1985ApJ...298..114M}, in contrast with high-frequency-peaked BL Lac objects. Instead, external photon fields, more typically associated with FSRQs, are required to explain the SED using a leptonic model.
From the works mentioned so far in this introduction, it emerges that the source showed a behaviour of a BL Lac in many occasions, while more recent studies instead noted more FSRQ characteristics. This transitional nature is already visible in the above mentioned presence of emission lines in more recent observations, together with the need of external photon fields in the modelling for the broadband SED. From the study performed in this paper, the inclusion of VHE $\gamma$-ray data adds a piece to the puzzle in the understanding of this source and supports more the FSRQ behaviour as it will be seen in Secs.~\ref{sec:SED} and \ref{sec:conclusions}.\par
The paper is structured as follows. In Sec.~\ref{sec:observations}, we report on the details of the observations performed by the different instruments involved and provide descriptions of the dedicated analyses. In Sec.~\ref{sec:LC}, the MWL light curves are discussed. 
In Sec.~\ref{sec:SED}, the broadband SEDs are discussed together with the theoretical interpretation within the framework of different state-of-the-art emission models. In Sec.~\ref{sec:location}, we use the $\gamma$-ray data to derive constraints on the location of the emission region. Conclusions are drawn in Sec.~\ref{sec:conclusions}. Details on the proton-synchrotron dominated model are given in Appendix~\ref{sec:appendix}. The cosmological parameters used in the present work are $H_{0}$ = 70 km/s/Mpc, $\Omega_{m}$=0.3, and $\Omega_{\Lambda}$=0.7.

\section{Observations and analysis}
\label{sec:observations}
\subsection{VHE $\gamma$-rays}

\begin{table*}           
\caption{Parameters of the power-law fit to the VHE $\gamma$-ray spectra observed by MAGIC and H.E.S.S. (this work). For both experiments, the observation time $T_{\mathrm{obs}}$, the effective observing time $t_{\mathrm{eff}}$, the minimum energy for the extraction of spectral parameters $E_{\mathrm{min}}$, the decorrelation energy $E_{\mathrm{dec}}$, the flux $f_0$ at decorrelation energy, the photon index $\Gamma$, and the flux $f_{200\mathrm{GeV}}$ at 200\,GeV are provided. The parameters are obtained from the observations during the high-flux state in VHE $\gamma$-rays, later defined as P3.}
\label{table:VHEspectrum}      
\centering
\resizebox{\textwidth}{!}{   
\begin{tabular}{ l c c c c c c c c c c}  
\hline\hline 
Exp. & $T_{\mathrm{obs}}$ & $t_{\mathrm{eff}}$ &  & $E_{\mathrm{min}}$ & $E_{\rm dec} $ & $f_{0}$ & $\Gamma$  &  $f_{ 200 {\rm GeV}}$ \\  
&  [MJD] & [h]& & [$\mathrm{GeV}$] & [$\mathrm{GeV}$] & $[\mathrm{cm}^{-2} \mathrm{s}^{-1} \mathrm{TeV}^{-1}]$ & & $[\mathrm{cm}^{-2} \mathrm{s}^{-1} \mathrm{TeV}^{-1}]$\\
\hline\hline   
\multirow{2}{*}{MAGIC} & \multirow{2}{*}{57593.9} & \multirow{2}{*}{1.6} & \multicolumn{1}{l}{obs.} & \multicolumn{1}{l}{57} & \multicolumn{1}{l}{125}&  \multicolumn{1}{l}{$(1.17\pm0.15_{\rm stat}\pm0.12_{\rm sys}) \times 10^{-9}$} & \multicolumn{1}{l}{$3.67\pm0.38_{\rm stat}\pm0.15_{\rm sys}$} &  \multicolumn{1}{l}{$(2.08 \pm0.45_{\rm stat}\pm0.22_{\rm sys})\times 10^{-10}$} \\\cline{4-9} 
\multirow{2}{*}{} & \multirow{2}{*}{} & \multirow{2}{*}{} & \multicolumn{1}{l}{intr.} & \multicolumn{1}{l}{57} & \multicolumn{1}{l}{125}& \multicolumn{1}{l}{$(1.46\pm0.20_{\rm stat}\pm0.16_{\rm sys}) \times 10^{-9}$} & \multicolumn{1}{l}{$3.27\pm0.44_{\rm stat}\pm0.15_{\rm sys}$} & \multicolumn{1}{l}{$(3.13 \pm0.65_{\rm stat}\pm0.34_{\rm sys}) \times 10^{-10}$} \\\cline{1-9} 
\multirow{2}{*}{H.E.S.S.} & \multirow{2}{*}{57591.8 -- 57593.9} & \multirow{2}{*}{3.1}& \multicolumn{1}{l}{obs.} & \multicolumn{1}{l}{119} & \multicolumn{1}{l}{173}& \multicolumn{1}{l}{$(3.93\pm 0.41_{\rm stat}\pm 0.78_{\rm sys}) \times 10^{-10}$ } & \multicolumn{1}{l}{$4.42\pm0.38_{\rm stat}\pm0.64_{\rm sys}$}& \multicolumn{1}{l}{$(2.07 \pm  0.25_{\rm stat}\pm 0.41_{\rm sys})\times 10^{-10}$} \\\cline{4-9} 
\multirow{2}{*}{} & \multirow{2}{*}{} & \multirow{2}{*}{} & \multicolumn{1}{l}{intr.} &  \multicolumn{1}{l}{120} & \multicolumn{1}{l}{183} & \multicolumn{1}{l}{$(4.50\pm 0.58_{\rm stat}\pm 0.90_{\rm sys}) \times 10^{-10}$ }  & \multicolumn{1}{l}{$3.39\pm0.58_{\rm stat}\pm0.64_{\rm sys}$} & \multicolumn{1}{l}{$(3.33 \pm0.46_{\rm stat}\pm 0.67_{\rm sys})\times 10^{-10}$} \\\cline{1-9} 
\end{tabular}
}

\end{table*}

\subsubsection{MAGIC}
MAGIC is a stereoscopic system consisting of two 17\,m diameter Imaging Atmospheric Cherenkov Telescopes (IACTs) located at the Observatorio del Roque de los Muchachos on the Canary Island of La Palma. The current sensitivity for low-zenith-angle observations ($15^\circ<zd<30^\circ$) above 218\,GeV is $0.83\pm0.03\,\%$ of the Crab Nebula's flux in 50\,h~\citep{2016APh....72...76A}.

The observations reported in the present work were performed in wobble mode, with the telescopes pointing $0.4^\circ$ away from the source and moving in 4 symmetrical positions w.r.t. the pointing position, in order to simultaneously collect the signal and background~\citep{1994APh.....2..137F}. A total of 2.03\,h of data were collected in the zenith angle range from 15$^\circ$ to 30$^\circ$, and the analysis was performed using the standard MAGIC analysis framework MARS~\citep{zanin2013, 2016APh....72...76A}. 
After applying quality cuts based on weather conditions, the remaining observing time amounts to 1.79\,h. 
The energy range of the MAGIC analysis performed in this work extends from 55 to 300\,GeV. 
A full description of the MAGIC systematic uncertainties can be found in~\cite{2016APh....72...76A} and references therein.

\begin{figure}
\centering
\includegraphics[width=8.5 cm]{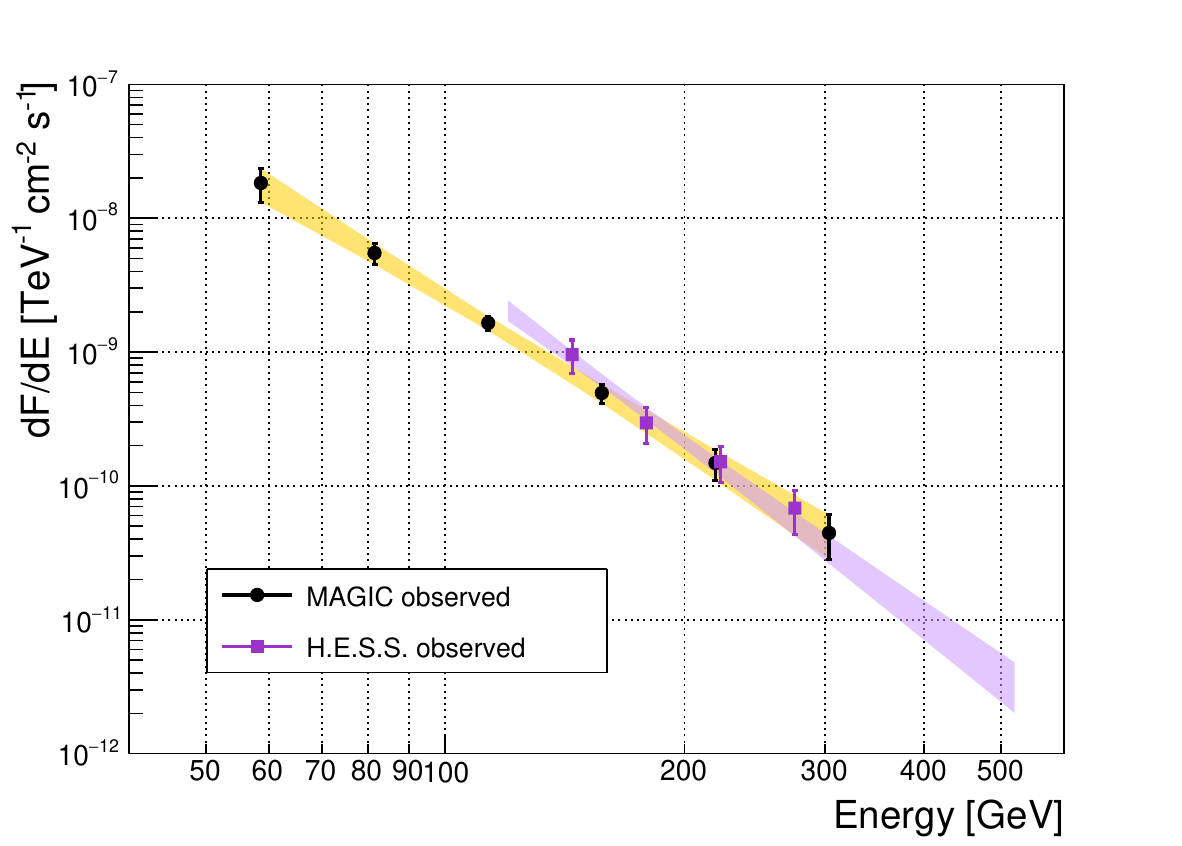}
 \caption{Observed differential energy spectra of the VHE $\gamma$-ray emission:  MAGIC (orange band, black points corresponding to MJD~57593.9) and H.E.S.S. (violet band from the forward folding technique, dark violet squares for the unfolded spectrum corresponding to MJD~57591 -- 57593). Parameters are described in the text and listed in Table~\ref{table:VHEspectrum}.}
 \label{fig:VHE_spectrum} 
\end{figure}

MAGIC started to observe OT~081 on MJD~57591 (22 July 2016), triggered by an optical flare and reports of HE photons detected by the \textit{Fermi}-LAT~\citep{2016ATel.9231....1B,2016ATel.9259....1B,2016ATel.9260....1C}.
However, due to the strong moonlight and bad weather conditions, the first observation surviving the quality cuts took place on MJD~57593.9 (24 July 2016). These data were taken under dark conditions, resulting in the detection of the source with a significance of 9.6 $\sigma$ in 1.64\,h of observation.
The flux on that night was $(9.8 \pm 1.5) \times 10^{-11}
\textrm{ph}\,\textrm{cm}^{-2}\,\textrm{s}^{-1}$ above 100\,GeV (20\% of the Crab Nebula flux). The observed spectrum is well described by a power-law function, parameterised as $dF/dE=f_0 \left(E/E_{\rm{dec}}\right)^{-\Gamma}$. The normalisation constant $f_0$, the spectral index $\Gamma$, and the decorrelation energy $E_{\rm{dec}}$ (the energy corresponding to the smallest error on the flux) are reported in Table~\ref{table:VHEspectrum}. The spectrum is shown in Fig.~\ref{fig:VHE_spectrum}. The intrinsic spectrum after correcting for the absorption due to interaction with the extragalactic background light (EBL) is estimated using the model from \cite{2011MNRAS.410.2556D}. The resulting intrinsic VHE spectrum is also compatible with a power-law fit.

MAGIC followed the source for two more nights, MJD~57594 and MJD~57596 (25 and 27 July 2016). Data from MJD~57594 were discarded because of adverse weather conditions. The data taken on MJD~57596 during 9 minutes of integration, assuming the same spectral shape as measured by MAGIC (see Table~\ref{table:VHEspectrum}), yield a 95\% confidence level (C.L.) flux upper limit (UL) of $15.7\times 10^{-11} \rm{ph}\,\textrm{cm}^{-2}\,\textrm{s}^{-1}$ above 100\,GeV including systematics.

\subsubsection{H.E.S.S.}
\label{sec:HESS}

H.E.S.S. is a stereoscopic system of five IACTs, located at approximately 1800 meters above sea level in the Khomas Highland plateau of Namibia (23$^\circ$16$'$18$''$ South, 16$^\circ$30$'$00$''$ East).

In its current phase, H.E.S.S. includes four 12-meter telescopes, as well as a 28-meter telescope in its centre that lowers the energy threshold. As a whole, the system is sensitive to $\gamma$-ray energies from $\sim$ 30~GeV to $\sim$ 30~TeV. The sensitivity to low-energy events provided by the larger dish is particularly beneficial for the study of extragalactic sources like AGN, which show characteristic soft VHE $\gamma$-ray spectra. The analysis presented here only includes monoscopic data from the 28-meter telescope.

Data were collected in the context of a Target of Opportunity (ToO) program, developed in the H.E.S.S. collaboration, that searches for intrinsic variability in blazars \citep{2017ICRC...35..652S}. This observation strategy allows for rapid reaction to various triggers from MWL observations shared by other experiments. 

Following the \textit{Fermi}-LAT trigger on a flaring state of OT~081~\citep{2016ATel.9231....1B} and the subsequent X-ray and UV observations that showed correlated $\gamma$-ray/X-ray/UV/optical activity of the source, H.E.S.S started observations of the source on MJD~57591 (22 July 2016). Observations continued for 6 consecutive nights, ending on MJD~57596 (27 July 2016). A total of 26 observation runs of 28 minutes each were obtained in the zenith angle range from 33$^\circ$ to 47$^\circ$ with a mean zenith angle of 38$^\circ$. All data pass the standard data-quality selection criteria~\citep{2017A&A...600A..89H}, which translates to a total of 11.7\,h of observations (10.1\,h after acceptance correction due to the wobble offsets around the nominal source position) available for analysis.
Data presented here are analysed using the \textit{Model} Analysis adapted for H.E.S.S. phase II \citep{2009APh....32..231D,2015arXiv150902902H}. 
\textit{Standard cuts} from \textit{Model} Analysis are applied to the data. A standard Reflected-Region Background method~\citep{2007A&A...466.1219B} is used to estimate the background in the region of interest (ROI) centred on the source. Using the \citet{li1983analysis} formalism, the source is detected at a significance of 6.5 $\sigma$ considering the full dataset. 
Results have been cross-checked with an independent analysis calibration chain and procedure, using the \textit{ImPACT} reconstruction method~\citep{2014APh....56...26P}. The consistency of the independent results proves the robustness of the conclusions.

A Bayesian Block analysis~\citep{2013ApJ...764..167S} was applied to the H.E.S.S. dataset to assess time variability of the $\gamma$-ray emission. This analysis reveals the identification of two flux states of the source with a False Alarm Rate (FAR) probability of 5\%. In the following, these are referred to as the high-flux state (MJD~57591.76 -- MJD~57593.86) and the low-flux state (MJD~57594.76 -- MJD~57596.89).
Limiting the analysis to the high-flux state yields 3.6 live hours (3.1 live hours after correcting for acceptance) of data over the three nights. The analysis of the high-flux data subset results in an 8.8 $\sigma$ detection, using the same background estimation method in the ROI as for the whole dataset. We used a forward folding with maximum likelihood optimisation~\citep{2001A&A...374..895P}, assuming a power-law model. Results can be found in Table~\ref{table:VHEspectrum}. We used an unfolding technique to obtain the data points independently from the spectral fit~\citep{2007NIMPA.583..494A}.

The spectrum of the source during its high state is obtained for energies between 120\,GeV to 520\,GeV.
The minimum energy of the high-state spectrum is determined by selecting the energy at which the acceptance reaches 15\% of its maximum value. This minimum energy threshold arises due to the relatively large zenith angle under which the observations were conducted. The maximum energy is selected as the upper edge of the last significant 2 $\sigma$ bin. 
Systematic uncertainties for the spectral parameters $f_0$ and $\Gamma$, quoted in Table \ref{table:VHEspectrum}, are estimated following the approach described in~\cite{2017A&A...600A..89H} for a Reflected-Region Background method and a H.E.S.S. II mono analysis.
The low-flux state observations include a total of 8.1\,h of observation during the three last nights of the entire dataset. No detection is reported for this period, as significance only reached $\sim 2 \sigma$. For this reason, we follow ~\cite{1998PhRvD..57.3873F} to derive 99\% C.L. differential ULs  on the emission between 110\,GeV and 2\,TeV for the low-flux period. While the minimum value of this energy range is obtained following the approach used for the high-flux state, the maximum energy for the low-flux state is obtained as the maximum photon energy of the distribution during that period.

The H.E.S.S. $E>$~100\,GeV light curve for the six nights of data (see Fig.~\ref{fig:MWL_LC}) is obtained by fixing the spectral index found in the spectral analysis of the high state Bayesian Block (see Table \ref{table:VHEspectrum}), integrated from 100\,GeV up to 520\,GeV. The flux obtained for the night of MJD~57593 (24 July 2016), $(1.39\pm 0.23) \times 10^{-10} \,\textrm{ph}\,\textrm{cm}^{-2}\,\textrm{s}^{-1}$, represents the highest and best-constrained flux value for the flare in the H.E.S.S. dataset, having the smallest error bar due to the high detection significance that night.
Complementarily, as no significant signal is observed during the low-flux state, we compute a single UL using data from MJD~57594.76 -- MJD~57596.89 (25 July -- 27 July 2016), which we use to set constraints on the source emission. The spectral index resulting from the analysis of the high state, quoted in Table \ref{table:VHEspectrum}, is used to derive the flux UL, constraining a similar emission as that observed in the high flux state period.  The 99\% C.L. UL derived on the 100\,GeV flux of the source is \mbox{$8.76\times 10^{-11} \rm{ph}\,\textrm{cm}^{-2}\,\textrm{s}^{-1}$} (systematics not included).

\subsection{HE $\gamma$-rays}
\label{sec:HE_analyis}
The pair-conversion Large Area Telescope (LAT) on board the \textit{Fermi} satellite monitors the $\gamma$-ray sky in survey mode every three hours in the energy range from 20\,MeV to $>300$~GeV~\citep{2009ApJ...697.1071A}.
For this work, we used a ROI with a radius of 10$^\circ$ centred around OT~081~\citep[4FGL~J1751.5+0938, ][]{2020ApJS..247...33A}. The data sample was selected around the flare detected by LAT, MAGIC and H.E.S.S., from 6 to 31 July 2016 (MJD 57575-57600). We used the Fermitools software package \citep[v.2.0.0,][]{2019ascl.soft05011F} to analyse Pass8 \texttt{source} class (P8R3) events in the energy range 0.1-300\,GeV. 
To reduce Earth limb contamination, we applied a zenith angle cut of 90$^\circ$ to the data. The unbinned likelihood fit of the data was performed using the recommended Galactic diffuse emission model and isotropic component for the Pass 8 \texttt{source} event class.\footnote{\url{https://fermi.gsfc.nasa.gov/ssc/data/access/lat/BackgroundModels.html}}
The likelihood model included all the 4FGL \citep{2020ApJS..247...33A} sources located within 20$^\circ$ of the position of OT~081. 
For the likelihood minimisation, the spectral parameters of sources located within 10$^\circ$ of the ROI centre were left free to vary, as were the parameters for the Galactic diffuse and isotropic emission models. Parameters for sources outside of this region were fixed to their catalogue values. 
The unbinned likelihood fit was carried out in two steps. After a first fit, sources with $TS<$5 were removed from the model. After that cut, a final likelihood fit was carried out, assuming a power-law model for OT~081. The resulting model was used for the calculation of a daily and 3-day light curves. For these calculations, the spectral parameters of all sources were frozen to their overall fit values except for the diffuse components and the source of interest. OT~081 was modelled as a power-law. Moreover, the flux normalisation was also left free to vary for the only other variable source within the ROI, 4FGL~J1818.6+0903, (variability index of $\sim$52), which is located 6.7$^\circ$ from OT~081.

For the analysis of the data for the time period that we designate as P3, the data were integrated for a time period of one day centred around at MJD~57593.9 (24 Jul 2016). During this period in coincidence with the VHE detection, the \textit{Fermi}-LAT detected the source with a $TS=76.5$. Evidence for curvature was investigated using a likelihood ratio test to compare a power-law model with a log-parabola, resulting in only 1.5$\sigma$ C.L. preference for the latter model. Therefore, no significant curvature is found, and the spectrum can be well described by a power-law model with index $\Gamma = 1.98\pm0.16$. The measured integral flux was $(8.58\pm2.14)\times 10^{-7} \rm{ph}~\rm{cm}^{-2}\rm{s}^{-1}$ above 100\,MeV.

\subsection{X-rays}
\label{sec:Xray}
The {\it Neil Gehrels Swift Observatory} satellite \citep{2004ApJ...611.1005G} carried out 9 observations of OT~081 between 11 July (MJD~57580) and 28 July 2016 (MJD~57597). The observations were performed with all three instruments on board-the X-ray Telescope \citep[XRT;][0.2--10.0 keV]{2005SSRv..120..165B}, the Ultraviolet/Optical Telescope \citep[UVOT;][170--600 nm]{2005SSRv..120...95R} and the Burst Alert Telescope \citep[BAT;][15--150 keV]{2005SSRv..120..143B}. The hard X-ray flux of this source turned out to be below the sensitivity of the BAT instrument for such short exposures and therefore, the data from this instrument will not be used.

XRT observations were performed in photon counting (PC) mode. The XRT spectra were generated with the {\it Swift} XRT data products generator tool at the UK Swift Science Data Centre\footnote{http://www.swift.ac.uk/user\_objects} \citep[for details see][]{2009MNRAS.397.1177E}. Spectra having count rates higher than 0.5 counts s$^{-1}$ may be affected by pile-up. To correct for this effect the central region of the image was excluded, and the source image was extracted with an annular extraction region with an inner radius which depends on the level of pile-up \citep[see e.g.,][]{2005SPIE.5898..360M}. Ancillary response files were generated with \texttt{xrtmkarf} and accounted for different extraction regions as well as corrections for vignetting and the point spread function. We used the spectral redistribution matrices in the Calibration data base maintained by \texttt{HEASARC}. The spectra were rebinned such that there were at least 20 counts per bin, and we used the $\chi^2$ statistics. The X-ray spectral analysis was performed using the XSPEC 12.13.0c software package \citep{1996ASPC..101...17A}. We fit the spectrum with an absorbed power-law using the photoelectric absorption model \texttt{tbabs} \citep{2000ApJ...542..914W} with a neutral hydrogen column density fixed to the Galactic value \citep[9.99$\times$10$^{20}$ cm$^{-2}$;][]{2016A&A...594A.116H}. We also modelled the spectra with a log-parabola model with a pivot energy fixed at 1\,keV. We found that there is no significant statistical preference for a log-parabola model in any of the observations.

The 0.3-10 keV flux (corrected for Galactic absorption) varied by a factor of 2.5 in less than 3 weeks with values between 7.32 and 19.53 $\times$10$^{-12}$ erg cm$^{-2}$ s$^{-1}$. The peak flux was observed on 20 July 2016 (MJD~57590), four days before the highest VHE $\gamma$-ray flux observed. However, the X-ray flux was also high on 17 July 2016 (MJD~57587), the {\it Swift} observation nearest to the VHE peak. The X-ray photon index was hard ($<$ 1.6) for all periods. The hardest spectra ($\Gamma_{X}$ $\sim$ 1.2) were observed on MJD~57587 and MJD~57590.

\subsection{Optical band and polarimetry}
\label{sec:optical}
The UVOT (Ultraviolet/Optical Telescope) on board the \textit{Swift} satellite observes simultaneuosly with the XRT instrument. 
In this work, we considered the \textit{Swift} pointings mentioned  in Sec.~\ref{sec:Xray} for the standard processing of UVOT data \citep[see for instance][]{2015ApJ...812...65F} based on the photometry recipes reported in \cite{poole08}, and performed on the total exposures of each observation. We applied the aperture photometry analysis with the task included in the official software within the HEAsoft package (v6.23), selecting an aperture of radius 5$\arcsec$ to extract source counts and an annular aperture of internal radius 26$\arcsec$ and size 7$\arcsec$ to extract the background counts in all observations. Then, the count rate was converted to dereddened flux according to the official calibrations -- the standard Galactic E(B-V) value for the source \citep{2011ApJ...737..103S} and a mean interstellar extinction curve \citep{1999PASP..111...63F}.

The Automated Telescope for Optical Monitoring~\citep[ATOM][]{hauser2004atom}, is a 75-cm altazimuth telescope which is part of the H.E.S.S. project and serves as an automatic optical monitor of variable $\gamma$-ray sources and as a transmission monitor to help calibrate the Cherenkov shower image analysis. The data collected with the ATOM telescope from MJD~57591 (22 July 2016) to MJD~57595 (26 July 2016) are strictly simultaneous with the H.E.S.S. observations and were taken with $R$ and $B$ filters.

The Tuorla blazar monitoring program\footnote{\url{http://users.utu.fi/kani/1m/}} collects blazar optical light curves in the $R$ band from several observatories. The present work shows data from the 35-cm telescope at the Kungliga Vetenskapsakademien (KVA) observatory on La Palma, Canary Islands, Spain. The data are analysed using the semi-automatic pipeline described by  \citet{2018A&A...620A.185N}.\par
Optical images of OT~081 were also obtained with the Katzman Automatic Imaging Telescope \citep[KAIT;][]{2001ASPC..246..121F} at the Lick Observatory.  All images were reduced using a custom pipeline \citep{2010ApJS..190..418G} before carrying out the photometry. We applied a 9-pixel aperture (corresponding to $7.2''$) for photometry.
Several nearby stars were chosen from the Pan-STARRS1\footnote{\url{http://archive.stsci.edu/panstarrs/search.php}} catalogue for calibration with magnitudes transformed into Landolt magnitudes using the empirical prescription provided by Eq.~6 of ~\citet{2012ApJ...750...99T}. All of the KAIT images were taken in the $clear$ band (i.e., without using a filter), which is most similar to using an $R$ filter \citep[see][]{2012ApJ...750...99T}. We therefore calibrated all of the $clear$ band results to the Pan-STARRS1 $R$-band magnitude.

Optical photometric and polarimetric data provided by St.~Petersburg University are from the 70-cm AZT-8 telescope of the Crimean Astrophysical Observatory\footnote{In 1991, Ukraine, including the Crimean peninsula, became an independent state. While the Crimean Astrophysical Observatory became Ukrainian, the AZT-8 telescope located there continued to be operated jointly by the Crimean Observatory and by St. Petersburg group.}. 
Polarimetric observations were performed using two Savart plates rotated by 45$^\circ$ relative to each other \citep[see][]{2008A&A...492..389L}. Instrumental polarization was estimated using stars located near the object under the assumption that their radiation is unpolarized.

Additional optical polarization data were taken with the 2-m Liverpool Telescope \citep[LT;\,][]{2004SPIE.5489..679S} located on the Canary Island of La Palma. 
The RINGO3 polarimeter consists of a rotating polaroid (1 rotation every 4 seconds) which captures 8 images of the source at successive 45$^\circ$ rotations of the polaroid. These 8 exposures could be combined according to the equations in ~\citet{2002A&A...383..360C} to determine the degree and angle of polarization. The RINGO3 polarimeter acquires polarimetric measurements in three different passbands recorded in the so-called ``Red'',``Green'', and ``Blue'' cameras.\footnote{See the RINGO3 specifications at:\\ \url{http://telescope.livjm.ac.uk/TelInst/Inst/RINGO3/}}
The optical data were all corrected for Galactic extinction as in \cite{2011ApJ...737..103S}. The contribution of the host galaxy is negligible.

\subsection{Radio band}
The Atacama Large Millimeter/submillimeter Array (ALMA)\footnote{\url{https://www.eso.org/public/teles-instr/alma/}}, located on the Chajnantor plateau  of the Chilean Andes at 5000\,m altitude, observed OT~081 in the 250, 320, and 450\,GHz bands. ALMA consists of a giant array of 12-m antennas (the 12-m array with baselines up to 16\,km), and an additional compact array of 7-m and 12-m antennas to greatly enhance ALMA's ability to image extended targets.

The 15\,GHz data of OT~081 were obtained from the Owens Valley Radio Observatory (OVRO) 40-m Telescope blazar monitoring program \citep{2011ApJS..194...29R}. The OVRO 40-m radio telescope uses off-axis dual-beam optics and a cryogenic pseudo-correlation receiver with a 15~GHz centre frequency and 3\,GHz bandwidth. Calibration is achieved using a temperature-stable diode noise source to remove receiver gain drifts, and the flux density scale is derived from observations of 3C~286 assuming the \citet{1977A&A....61...99B} value of 3.44\,Jy at 15.0\,GHz. The systematic uncertainty of about 5\,\% in the flux density scale is not included in the error bars.  Complete details of the reduction and calibration procedure can be found in \citet{2011ApJS..194...29R}. 

Very Long Baseline Array (VLBA) observations of OT~081 were obtained at 43\,GHz for 23 epochs from  June 2015 to June 2017 as part of the VLBA-BU-Blazar program\footnote{\url{www.bu.edu/blazars/VLBAproject.html}} \citep{2022ApJS..260...12W}. The data reduction was performed using the Astronomical Image Processing System (AIPS) and {\it Difmap} software packages as described in \cite{2017ApJ...846...98J}. In order to study the jet kinematics, the source structure was modelled using a number of emission components consisting of Gaussian brightness distributions. This allowed us to determine the number of components needed to represent each image based on the $\chi^2$ statistic. Each component (knot) is characterized by the following parameters: flux density, $S$, distance from the core, $r$, position angle with respect to the core, $\Theta$, and size of the component, $a$ (FWHM of the Gaussian). The uncertainties of these parameters depend on the brightness temperature of the knot and are calculated using the relationships given in~\cite{2017ApJ...846...98J}. The detailed VLBA analysis is presented in Appendix~\ref{sec:appendix_vlba}.

\section{Multi-wavelength light curves}
\label{sec:LC}

\begin{figure*}
   \centering
      \includegraphics[height=22.5cm]{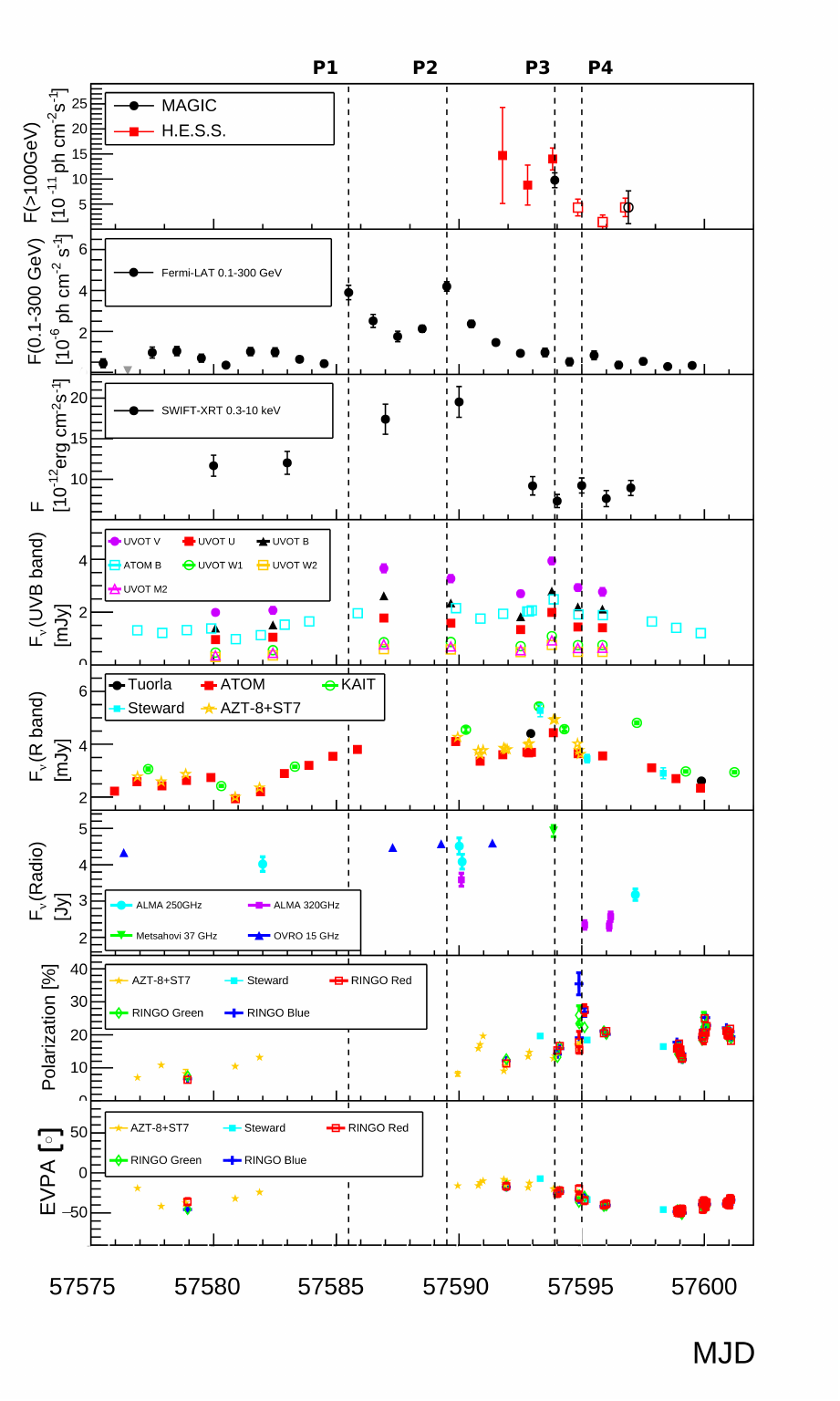}
      \caption{MWL light curves of OT~081 during the period from MJD~57575 to MJD~57602 (6 July to 02 August 2016). Vertical dashed lines P1 (MJD~57585.5 -- 16 July 2016), P2 (MJD 57589.5 -- 20 July 2016), P3 (MJD~57593.9 -- 24 July 2016) and P4 (MJD~57595 -- 26 July 2016) indicate the four states of the source which were identified.
      The empty circle and empty squares in the top panel correspond to  flux points derived for the low-significance dataset, compatible with the background-only hypothesis. In the second panel from the top, the downward-facing triangle corresponds to an UL.}
         \label{fig:MWL_LC}
\end{figure*}
In Fig.~\ref{fig:MWL_LC}, the collected MWL light curves are presented in order of decreasing energy starting from the top panel. The best time coverage comes from the \textit{Fermi}-LAT, and a double peak structure can be identified with compatible values of the flux at the peaks. This flare corresponds to the strongest activity detected from OT~081 during the entire \textit{Fermi} mission.\footnote{\url{https://fermi.gsfc.nasa.gov/ssc/data/access/lat/msl_lc/}} Unfortunately, due to moon and weather constraints, the VHE observations could only be performed once the flare was already in the decay phase of the LAT flux evolution. 

The four vertical dashed lines (P1, P2, P3, and P4) shown in Fig.~\ref{fig:MWL_LC} indicate four different states of activity.
P1, MJD~57585.5 (16 July 2016), corresponds to the first peak of the flare observed by \textit{Fermi}-LAT. No simultaneous MWL data are available, so we do not consider it for further modelling of the SED. 
P2 marks enhanced activity from \textit{Fermi}-LAT and \textit{Swift}-XRT on the same night, MJD~57589 (20 July 2016). P3, occurring on MJD~57593.9 (24 July 2016), is the period during which VHE $\gamma$-ray emission was detected by MAGIC and H.E.S.S.~\citep{2016ATel.9267....1M,2017ICRC...35..652S}. Finally, P4, (MJD~57595.5 -- 26 July 2016) is considered the post-VHE $\gamma$-ray detection state and is coincident with a peak in the optical polarization.
The broadband SEDs corresponding to P2, P3, and P4 are discussed within the framework of theoretical emission models in Sec.~\ref{sec:SED}.

In the top panel of Fig.~\ref{fig:MWL_LC}, we plot the VHE $\gamma$-ray light curves above 100\,GeV from MAGIC and H.E.S.S.
The six-night sampling of H.E.S.S. results in the detection of two different states of 3 days each, identified by a Bayesian Blocks algorithm. For detailed description of the analysis results, refer to Section~\ref{sec:HESS}. The H.E.S.S. data show that the source is no longer detected after the night of the 24th. The high-flux state correspond to the three first nights of the per-night binned lightcurve, while the low-flux state is represented by the last three points. The derived UL for the three following nights, quoted in Sec.~\ref{sec:HESS}, constrains the source flux during this period to at most two thirds of its flux during P3. 
The MAGIC observations where obtained during the high-flux period. The H.E.S.S. and MAGIC observations are complementary in terms of temporal coverage of the source during the night of 24 July 2016 (MJD~57593.9).  We note a slight decrease in flux, although compatible within systematic uncertainties. 
The detection of VHE $\gamma$-rays took place during the decay phase of the HE flare. The double-peaked HE flare reached flux peaks of $(3.90 \pm 0.35) \times 10^{-6}$ and  $(4.21 \pm 0.23)  \times 10^{-6} \,\textrm{ph}\,\textrm{cm}^{-2}\,\textrm{s}^{-1}$, respectively, for P1 and P2, around 30 times the average flux reported in the fourth \textit{Fermi}-LAT point source catalogue~\citep[4FGL,][]{2020arXiv200511208B}. 
The photon indices measured by the \textit{Fermi}-LAT were $\Gamma_{HE} = 2.02\pm 0.06$ for P1 and $\Gamma_{HE} = 1.89\pm 0.03$ for P2.
\cite{2018MNRAS.480.2324K} also reported the HE light curve in 3-day and weekly time bins with a peak flux of $2.9 \times 10^{-6}\,\textrm{ph}~\textrm{cm}^{-2} \mathrm{s}^{-1}$ centred at MJD~57588. Unfortunately, the sparse MWL data simultaneous to the most active period in the HE band do not allow a detailed study of the double structured flare. 
We have calculated the fractional variability ($F_{\textrm{var}}$) as in Eq. 3 of~\cite{2003MNRAS.345.1271V} for the light curves shown in Fig.~\ref{fig:MWL_LC}. The results are reported in Table~\ref{table:Fvar}. 
\begin{table}
\caption{$F_{\textrm{var}}$ measurements for the light curves in Fig.~\ref{fig:MWL_LC}.}
    \centering
   		\begin{tabular}{@{}l cc}
 		\hline
 		\hline
 			Instrument & $F_{\textrm{var}}$ & $ErF_{\textrm{var}}$  \\
		\hline
 		 \noalign{\smallskip}
        \textit{Fermi}-LAT & 0.86 & 0.03  \\
        \textit{Swift-XRT} & 0.52 & 0.01\\
        \textit{Swift-UVOT} & 1.12 & 0.01  \\
        KAIT & 0.565 & 0.003  \\
        KAIT & 0.670 & 0.003  \\
        ALMA (250\,GHz) & 0.28 & 0.01  \\
        ALMA (250\,GHz) & 0.336 & 0.004 \\
        OVRO & 0.150 & 0.002  \\
        \hline
         \end{tabular}
         \label{table:Fvar}
\end{table}

The \textit{Fermi}-LAT light curve has a $F_{\textrm{var}}$ of 0.86$\pm$0.03.
Over the 10 years of OT~081 data observed by \textit{Fermi} and reported in the 4FGL catalogue, the highest activity in HE $\gamma$-rays for this source was registered in the P1--P4 time range.
The highest value of the X-ray flux in the period of time considered in Fig.~\ref{fig:MWL_LC} is observed on MJD~57590 (20 July 2016), quasi-simultaneously to P2. The corresponding X-ray flux ($F_{0.3-10\,\text{keV}}$) was $(19.53 \pm 1.90)\times 10^{-12}\,\textrm{erg}~\textrm{cm}^{-2} \textrm{s}^{-1}$, and the photon index was $\Gamma_X = 1.23\pm 0.11$. The lowest X-ray flux is simultaneous to P3 and the VHE $\gamma$-ray detection. The $F_{\textrm{var}}$ of this light curve is 0.52$\pm$0.01.

The highest activity detected in the UV-optical band is reached during the VHE $\gamma$-ray flaring state centred on P3 and corresponds to 3.94$\pm$0.11\,mJy in the V band. The $F_{\textrm{var}}$ of the \textit{Swift}-UVOT light curve is the highest among the light curves presented in Fig.~\ref{fig:MWL_LC} with a value of 1.12$\pm$0.01.

The highest flux is reported by KAIT and corresponds to a value of 5.43$\pm$0.09\,mJy. 
The $F_{\textrm{var}}$ of the optical light curve of KAIT (which is the better sampled one among the optical light curves collected for this work) corresponds to a value of 0.565$\pm$0.003 for the time frame reported in Fig.~\ref{fig:MWL_LC}.

The radio light curve at high frequencies (ALMA, mm-radio band) typically shows strong variability for this source~\citep{1997A&AS..122..271R,1998A&AS..132..305T}. For the period of time considered in Fig.~\ref{fig:MWL_LC}, we calculated an $F_{\textrm{var}}$ of 0.28$\pm$0.01 for the ALMA light curve at 250~GHz. 
The optical polarization data show an increase of the percentage of polarization around the time of the detection in the VHE $\gamma$-ray band (between P3 and P4). The maximum value of the polarization percentage is 30\% in the RINGO3 red filter, which is consistent with the highest polarization percentage for this source reported in the literature~\citep{1986MNRAS.221..739B}. 
The highest polarization measurement is coincident with the radio flux decay. The $F_{\textrm{var}}$ of the polarization percentage has a value of 0.348$\pm$0.006.
The Electric Vector Position Angle (EVPA) is stable apart from a minor smooth decrease of a few degrees (from $\sim$-10$^\circ$to -47$^\circ$) after P3. This small rotation lasted about two days, starting from P3.

\section{Modelling the Broadband Spectral Energy Distribution}
\label{sec:SED}

\subsection{Leptonic emission model}
\label{sec:models_l}
The simplest emission model for blazars is the one-zone leptonic SSC model \citep[SSC,][]{1981ApJ...243..700K,1985ApJ...298..128B,1985ApJ...298..114M}.
In this framework, the emission in the radio to the UV or X-ray bands (depending on the type of blazar) is produced by a population of relativistic electrons via synchrotron radiation.
This low-energy photon field provides the seed photons for inverse-Compton (IC) scattering by the same population of leptons. The emission region is characterized as a homogeneous sphere with a radius $R$ and a bulk Lorentz factor $\Gamma$ that results in relativistic Doppler boosting by a Doppler factor $\delta = [\Gamma (1 - \beta_{\Gamma} \cos\theta_{\rm obs})]^{-1}$. In this region, the magnetic field $B$ is also uniform.

\begin{table*}
	\centering
	\caption{Log-parabola parameters for the combined HE-VHE SED (P3). The VHE spectra were corrected for EBL absorption. The significance C.L. reported in the last column is derived from a likelihood ratio test comparing a log-parabola fit to that of a power-law.}
	\label{tab:combined_parameters}
	\begin{tabular}{lccccr} 
		\hline
		  & $\alpha$ & $\beta$ & $E_{0}$  & $F_{0}$  & $\sigma$ \\
		  &  &  & [TeV]  & [$\rm erg\,cm^{-2}\,s^{-1}$] &  \\
		\hline
\textit{Fermi}-LAT--MAGIC  &  1.3 $\pm$ 0.2 &  
0.20 $\pm$ 0.06 &  1 & $(1.2 \pm 0.6) \times 10^{-12}$ & 3.2\\
\textit{Fermi}-LAT--H.E.S.S.  &  1.6 $\pm$ 0.3  &  0.28 $\pm$ 0.07 &  1  & $(1.5 \pm 1.2) \times 10^{-12}$  & 3.9 \\
		\hline
     \end{tabular}
\end{table*}

\citet{1998ApJ...509..608T} demonstrated that the SSC model parameters are constrained by the peaks of the synchroton ($\rm E_{\rm s,\,peak},\rm \nu L(\nu)_{\rm s,\,peak}$) and IC ($\rm E_{\rm IC,\,peak},\rm \nu L(\nu)_{\rm IC,\,peak}$) bumps. In our case, the synchrotron peak was estimated by fitting the data points with a second order polynomial function in log-log space as in \cite{2018RAA....18..120L}, yielding $\rm E_{\rm s,\,peak}=9.5\times 10^{-2}$ eV and $\rm \nu L(\nu)_{\rm s,\,peak}=9.2\times 10^{45}$ erg/s. 
In the $\gamma$-ray band, we fit the \textit{Fermi}-LAT and TeV data points by minimising $\chi^2$. We assumed the \textit{Fermi}-LAT data points to not be correlated, and the H.E.S.S. and MAGIC data points (and their associated covariance matrix) were computed using an unfolding algorithm. The correction for EBL absorption was applied using the \cite{2011MNRAS.410.2556D} model. The data points were fitted using a power-law and a log-parabola, and the minimisation was performed with the Markov Chain Monte Carlo method implemented in the emcee python package.\footnote{\url{https://emcee.readthedocs.io/en/stable/}} 
In both cases, the preferred fit was a log-parabola with significances of 3.2$\sigma$ for \textit{Fermi}-LAT--MAGIC and 3.9$\sigma$ for \textit{Fermi}-LAT--H.E.S.S. The respective parameters are listed in Table~\ref{tab:combined_parameters}. The significance C.L. is derived from a likelihood ratio test comparing the log-parabola fit to the power-law. 
Jointly fitting the \textit{Fermi}-LAT, MAGIC, and H.E.S.S. datasets (see Appendix \ref{sec:appendix:fitJ}) yields $\rm E_{\rm IC,\,peak}=2.5\times 10^{-2}$ GeV and $\rm \nu L(\nu)_{\rm IC,\,peak}=7.1\times 10^{46}$ erg/s.
Following~\citet{1998ApJ...509..608T}, using their Equations~4 and 11, the B-field and the Doppler factor $\delta$ are 
constrained to $B\delta \approx 10^{-3}\ {\rm G}$ and $B\delta^3 \approx 15{~\rm G}$.  This leads to values of $\delta$ $\approx$ 122 and B $\approx$ 8.2$\times 10^{-6}$~G, which are hard to reconcile with typical values found in the literature for blazars.
In fact, the value of B for BL Lacs is usually found to be in the range 0.1-1\,G~\citep[see e.g.][]{2010MNRAS.402..497G}, and values of $\delta$ are typically in the range of 5-25 \citep{2017ApJ...846...98J,2018ApJ...866..137L}.
Moreover, OT~081 is characterised by a high degree of Compton dominance (CD, defined as the ratio between the IC peak and the synchrotron peak fluxes) of the order of $\sim20$ during P3, which is more typical of FSRQs than of BL Lacs~\cite[see e.g.][]{2013ApJ...763..134F}. Therefore, a standard one-zone SSC model, which is the one typically used for BL Lacs, cannot properly account for the MWL emission of OT~081 during the P3 state. As with FSRQs, reproducing such high CD values requires an additional contribution from external photon fields.

\begin{figure}
\centering
\includegraphics[width=8.5 cm]{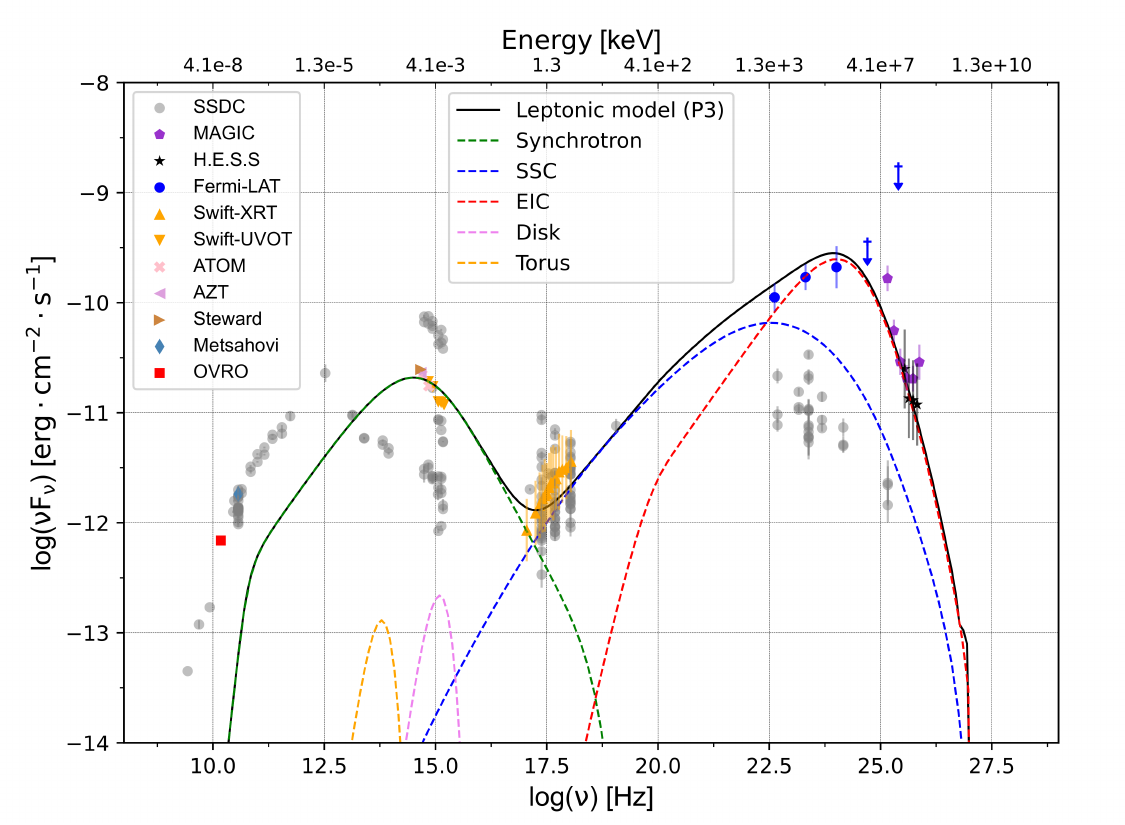} 
 \caption{Broadband SED of OT~081 (P3) together with the leptonic model. The filled markers are the simultaneous broadband data collected in P3 (MJD~57593.9 -- 24 July 2016). The solid black line represents the leptonic model. Grey circles are archival data from ASI/SSDC (\url{https://www.ssdc.asi.it/}).}
 \label{fig:Model_IR} 
\end{figure}

The detection of a relatively intense broad $H_{\beta}$ line in the optical spectrum \citep{1988A&A...191L..16S} suggests the presence of a well-developed BLR in the core of OT~081. If the active region of the jet lies within the BLR, the UV-optical photons from the clouds provide an intense field of target photons for the IC processes. However, in this case $\gamma$-ray photons with energies above $\sim$20 GeV will be efficiently absorbed by interacting with the same radiation field through the $\gamma\gamma\to e^{+}+e^{-} $ process. In light of the limits on the location of the emitting region from this dataset, described in Sec.~\ref{sec:location}, we assume for the modelling that the emitting region is either at the edge or beyond the BLR.

As for the few FSRQs detected in the VHE $\gamma$-ray band~\citep[3C279, PKS~1510-089, 4C+21.35, S3~0218+35, PKS~1441+25, PKS~0736+017, TON~0599, B2~1420+32, PKS~1413+135;][]{2008Sci...320.1752M,2013A&A...554A.107H,2011ApJ...730L...8A,2016A&A...595A..98A,2015ApJ...815L..22A, 2015ApJ...815L..23A,2020A&A...633A.162H,2017ATel11061....1M,2020ATel13412....1M,2022ATel15161....1B}, a plausible scenario  places the emission region outside of the BLR to a location at which the radiative environment is dominated by the thermal IR radiation field from the dusty torus. 
The same leptonic framework can be applied in the case of OT~081, including the contributions of both synchrotron and torus photons.

For the fully leptonic scenarios considered in this work, we use the model described in \cite{2003ApJ...593..667M}. The emission region is modelled as a sphere with comoving radius $R^\prime$,\footnote{We use the $\prime$ symbol to indicate quantities in the comoving frame.} that is moving with bulk Lorentz factor $\Gamma$ at a viewing angle $\theta_{\rm v}$ with respect to the observer's line of sight. The region carries a tangled magnetic field with field strength  $B^\prime$ and relativistic electrons following a smoothly broken power-law energy distribution with slopes $n^\prime_1$ and $n^\prime_2$ below and above the break at a Lorentz factor $\gamma^\prime_{\rm b}$. The energy density of the external field is modelled with a black body spectral shape with temperature $T=10^3$ K, total luminosity $L_{\rm IR}$, diluted within a radius $R_{\rm IR}$.\footnote{T, R and L are given in the galaxy frame.} The total luminosity can be fixed to the intercepted fraction $f_{\rm cov_{IR}}$ of the disk luminosity $L_{\rm disk}$, with a typical covering factor in the range $f_{\rm cov_{IR}}=0.4-0.6$. 

$L_{\rm disk}$ can be derived from the $H_{\beta}$ luminosity. In addition to the historical optical observations, more recent spectra are available in the Steward observatory database.\footnote{\url{http://james.as.arizona.edu/~psmith/Fermi/DATA/Objects/pks1749.html}} In particular, we use the spectrum taken on 30 August 2016 (MJD~57630) that confirms the detection of the emission lines with EW$_{\rm H_{\beta}}\sim3\,\textrm{\AA}$ and EW$_{[\rm O III]~5007}=5~\textrm{\AA}$. The H$_{\beta}$ line luminosity measured from this spectrum is $\sim$5$\times$10$^{41}$ erg/s. 
Using Eq. 6 of \cite{2006ApJ...641..689V}, we obtain a value of the mass of the BH $M_{\bullet}=10^{7.9} \times M_\odot $. 
The measured $H_{\beta}$ luminosity can be used to infer the total luminosity of the BLR using standard templates of BLR spectra \citep{1997MNRAS.286..415C}. Adopting the standard covering factor for the BLR, $f_{\rm cov_{BLR}}=0.1$, one then derives a disk luminosity $L_{\rm disk}\simeq 1.3\times 10^{44}$ erg s$^{-1}$. 
We follow the simplified prescription of \cite{2009MNRAS.397..985G} to estimate the torus size $R_{\rm IR}$ from the derived disk luminosity. We find $R_{\rm IR}\approx 2.5\times 10^{18} \,[L_{\rm disk}/(10^{45} \, {\rm erg} \, {\rm s}^{-1})]^{1/2} \,{\rm cm}\simeq 9\times 10^{17}$\,cm. The IR luminosity can be estimated as $L_{\rm IR}=f_{\rm cov_{IR}} L_{\rm disk}\simeq 6.5\times 10^{43} \, (f_{\rm cov_{IR}}/0.5)$ erg s$^{-1}$. These values are taken as indicative starting values which are then fine tuned to provide a good fit to the SED.
In Fig.~\ref{fig:Model_IR}, we present a possible model based on the scheme described above and the following parameters for the P3 source state: \mbox{$f_{\rm cov_{\rm IR}}=0.4$}, \mbox{$R_{\rm IR}=2.0\times 10^{18}$ cm}, $\Gamma=15$, \mbox{$\theta_{\rm v}=2.7^\circ$}, \mbox{$R^\prime=2.5\times 10^{16}$ cm}, $B^\prime=0.065$ G, $\gamma^\prime_{\rm e,min}=50$, $\gamma^\prime_{\rm e,break}=7\times10^{3}$, $\gamma^\prime_{\rm e,max}=8\times10^5$, $n^\prime_1=1.9$, $n^\prime_2=4.5$, $K^\prime=1.2\times 10^4$ cm$^{-3}$. The parameters of the model, including also P2 and P4 cases, are listed in Table~\ref{table:hybrid}. $K^\prime_{\rm}$ is a normalisation factor for the particle distributions (electrons or protons) as in \cite{2003ApJ...593..667M}, and $U^\prime_B$ is the energy density due to the magnetic field B$^\prime$. 
The parameters of the model, including those for the P2 and P4 cases, are listed in Table~\ref{table:leptonic}.

\begin{table}
\caption{Parameters of the leptonic model for periods P2 (MJD 57589.5 -- 20 July 2016) , P3 (MJD~57593.9 -- 24 July 2016), and P4 (MJD~57595 -- 26 July 2016). Parameters are described in the text. The quantities flagged with stars are derived quantities and not model parameters.}
    \centering
   		\begin{tabular}{@{}l ccc}
 		\hline
 		\hline
 				  & P2 & P3 & P4 \\
		\hline
 		 \noalign{\smallskip}
          $\delta$ & $20$ & $20$ & $20$ \\
          $R^\prime$ [10$^{16}$ cm] & $3.5$ & $2.5$ & $3.5$\\
 		 $B^\prime$ [10$^{-2}$\,G]  & $4.8$  & $6.5$  & $13$  \\
 		 $^\star U^\prime_B$ [10$^{-5}\,$erg$\,$cm$^{-3}$] &   $9.2$   & $17$  & $67$ \\
 		\hline
        $\gamma^\prime_{\rm e,min} $& $50$ & $50$ & $50$  \\
 		$\gamma^\prime_{\rm e,break}\ [10^3] $& $6$ & $7$& $3$ \\
        $\gamma^\prime_{\rm e,max}\ [10^5]$& $8$ & $8$ & $8$  \\
 		$n^\prime_{1}$ & $1.7 $ & $1.9 $ &  $1.9 $\\
 		$n^\prime_{2}$ & $4.7 $ & $4.5 $ &  $4.5 $\\
 		$K^\prime_{\rm e} $ [10$^{3}$\,cm$^{-3}$]  & $5$ & $12$ & $6$ \\
 		\hline
		\end{tabular}
		\label{table:leptonic}
\end{table}
\begin{table}
\caption{Parameters of lepto-hadronic model for periods P2 (MJD 57589.5 -- 20 July 2016) , P3 (MJD~57593.9 -- 24 July 2016), and P4 (MJD~57595 -- 26 July 2016). Parameters are described in the text. The quantities flagged with stars are derived quantities and not model parameters.}
    \centering
   		\begin{tabular}{@{}l ccc}
 		\hline
 		\hline
 				  & P2 & P3 & P4 \\
		\hline
 		 \noalign{\smallskip}
          $\delta$ & $20$ & $20$ & $20$ \\
          $R^\prime$ [10$^{16}$ cm] & $3.5$ & $2.5$ & $3.5$\\
 		 $B^\prime$ [10$^{-2}$\,G]  & $4.8$  & $6.5$  & $13$  \\
 		 $^\star U^\prime_B$ [10$^{-5}\,$erg$\,$cm$^{-3}$] &   $9.2$   & $17$  & $67$ \\
 		\hline
        $\gamma^\prime_{\rm e,min} $& $50$ & $50$ & $50$  \\
 		$\gamma^\prime_{\rm e,break}\ [10^3] $& $6$ & $7$& $3$ \\
        $\gamma^\prime_{\rm e,max}\ [10^5]$& $8$ & $8$ & $8$  \\
 		$n^\prime_{1}$ & $1.7 $ & $1.9 $ &  $1.9 $\\
 		$n^\prime_{2}$ & $4.7 $ & $4.5 $ &  $4.5 $\\
 		$K^\prime_{\rm e} $ [10$^{3}$\,cm$^{-3}$]  & $5$ & $12$ & $6$ \\
 		\hline
         $\gamma^\prime_{\rm p,min}$& $1$ & $1$ & $1$ \\
 		$\gamma^\prime_{\rm p,max} [10^7]$& $5.6$ & $5.6$ & $5.6$ \\
         $K_p^\prime$ & $3$ & $180$ & $40$ \\
 		$^\star U^\prime_{\rm p}$ [erg cm$^{-3}$] & $5.9$ & $22.6$ & $5.0$ \\
 		\hline
    	$^\star U^\prime_{\rm p}/U^\prime_{\rm B} $ [10$^{4}$] & $6.4$ & $13.5$ & $0.8$ \\
		$^\star L$ [10$^{47}$ erg s$^{-1}$] &  $1.7$ & $6.0$ & $1.4$ \\
        $^\star \nu_{\textrm{rate}} $ [s$^{-1}$] & $0.32$ & $0.22$ & $0.13$\\
 		\hline
		\end{tabular}
		\label{table:hybrid}
\end{table}

\subsection{Lepto-hadronic emission model}
\label{sec:models_lh}
\begin{figure*}
\centering
\begin{subfigure}[t]{0.5\textwidth}
\centering
\includegraphics[width=\linewidth]{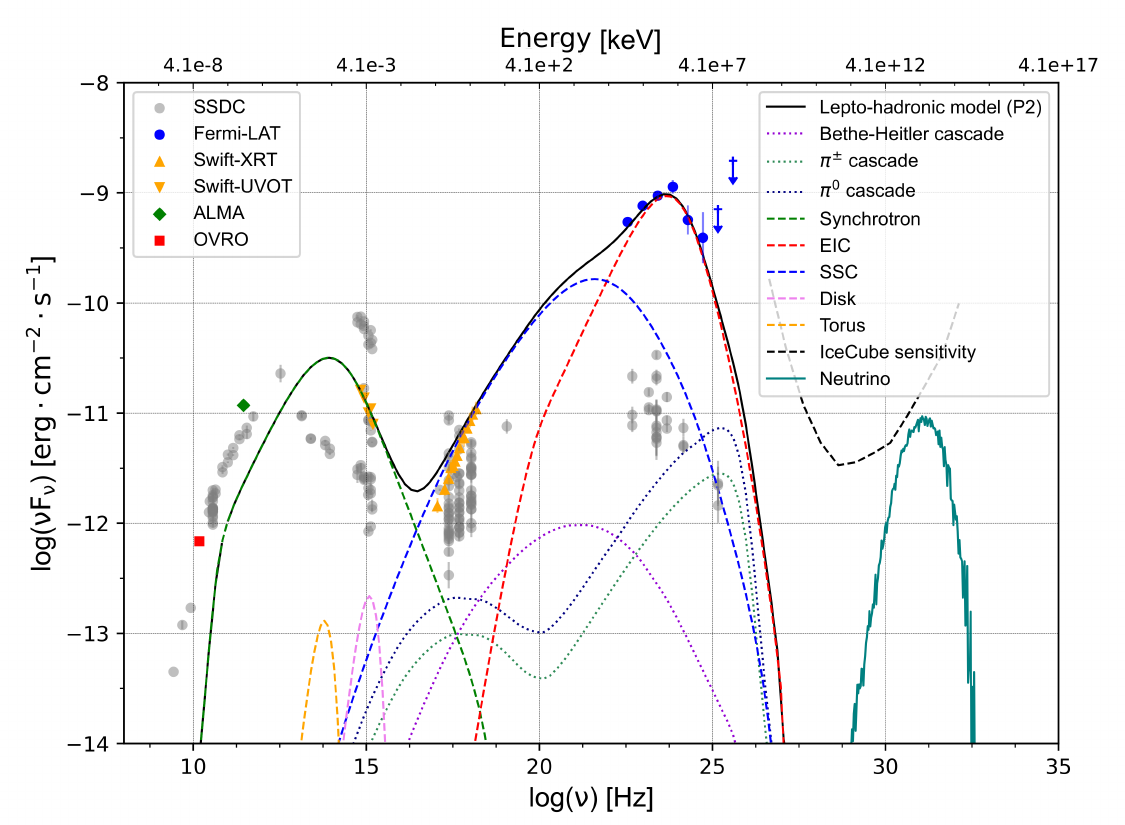}
\caption{P2: MJD~57589.5 (20 July 2016)}\label{fig:hybridP2}
\end{subfigure}%
\hfill
\begin{subfigure}[t]{0.5\textwidth}
\centering
\includegraphics[width=\linewidth]{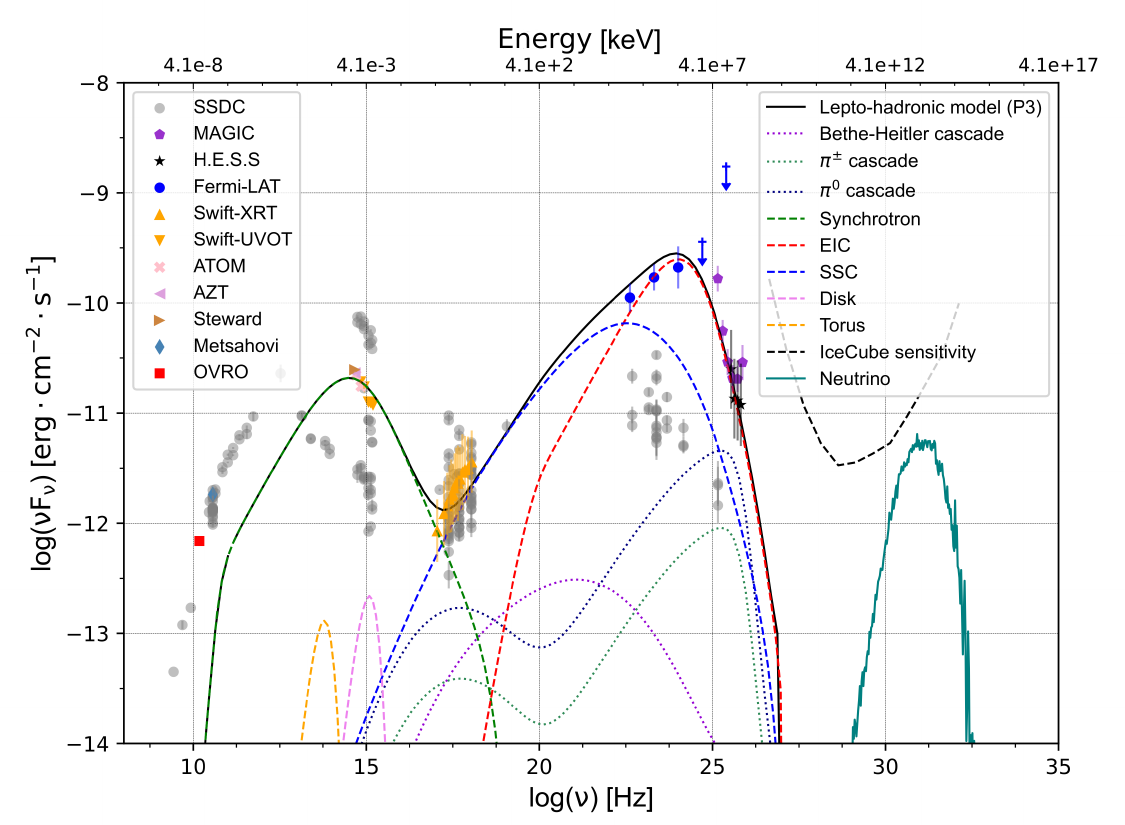}
\caption{P3: MJD~57593.9 (24 July 2016)} \label{fig:hybridP3}
\end{subfigure}%
\begin{subfigure}[t]{0.5\textwidth}
\centering
\includegraphics[width=\linewidth]{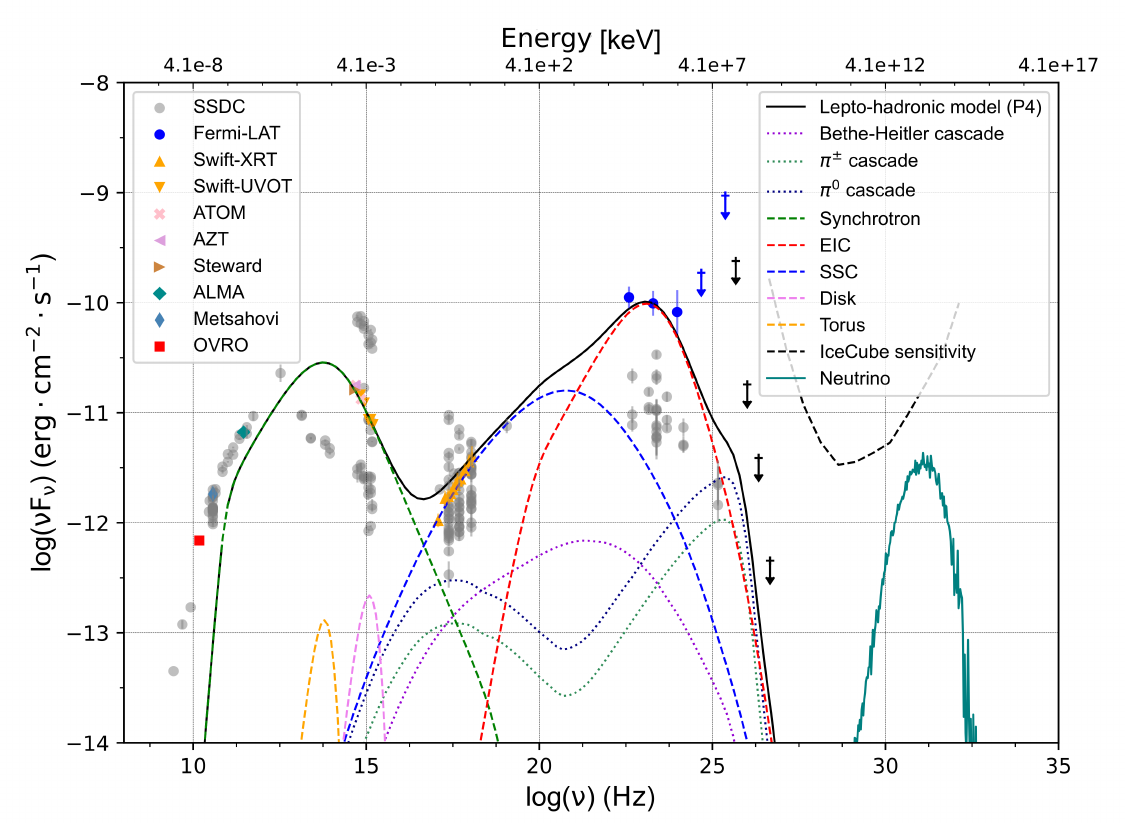}
\caption{P4: MJD~57595 (26 July 2016)} \label{fig:hybridP4}
\end{subfigure}%
 \caption{SEDs for the various states of OT~081 plotted with the best-fitting lepto-hadronic models (black solid line, parameters shown in  Table~\ref{table:hybrid}). The three states P2, P3, and P4 are shown in panels a), b), and c), respectively. The green dashed lines represent models of the synchrotron emission. The blue dashed lines are SSC components, the red dashed lines represent the EIC, the violet dotted curves are the Bethe-Heitler cascades, the blue and green dotted curves are respective cascades initiated by $\pi^0$-decay photons and $\pi^{\pm}$-decay, the teal third bumps to the right of the SEDs correspond to the calculated neutrino fluxes. Downward-facing arrows represent ULs. The IceCube sensitivity curve from~\citet{2019EPJC...79..234A} is shown as a black dashed line. Grey circles are archival data from ASI/SSDC.
 }
 \label{fig:Model_leha} 
\end{figure*}
Given the recent interest in multi-messenger photon and neutrino emission from blazars, we tested the effect of adding a hadronic component to the external inverse Compton (EIC) solution. This test was also driven by the intention of comparing such a model with the results obtained in \citet{2020PhRvL.124e1103A}, where the number of astrophysical neutrino events $\hat{n}_{s}$ are presented for many blazars in the 4FGL catalogue, including OT~081.
The $\gamma$ and neutrino emission from TXS~0506+056 during its 2017 flare was extensively studied as the first opportunity to constrain blazar hadronic models using multi-messenger information \citep{Ansoldi18, Keivani18, Cerruti19, Gao19}. The current consensus is that proton synchrotron emission is not compatible with the 2017 event, producing a neutrino flux that would be too low to be compatible with the IceCube detection. The most likely scenario is a Compton-dominated model (SSC or EIC) with a subdominant hadronic component that emerges in the X-ray and VHE band. The high proton power usually required by such a model can be alleviated if the target photons for the proton-photon interactions arise from an external photon field \citep{2019ApJ...881...46R}. With this result in mind, a potential hadronic contribution to the EIC emission model for OT~081 described above was estimated using the hadronic code described in~\citet{Cerruti15}. The numerical code simulates photon and neutrino emission from a spherical plasmoid (with radius $R^\prime$) in the jet with a homogeneous magnetic field $B^\prime$ that is moving towards the observer with Lorentz factor $\Gamma$. The plasmoid is filled with stationary populations of electrons and protons that are both parametrised with broken PWL functions (with indices $n^\prime_{e/p,1}$ and $n^\prime_{e/p,2}$) with exponential cutoffs. For P3, the same parameters as in the leptonic model were used for the electron distribution, the soft photon field, and the emitting region. For the proton distribution, we set the index equation to $n_1^\prime$, and the maximum proton Lorentz factor is $\gamma^\prime_{\rm p,max} = 5.6\times10^7$. The proton normalisation is fixed in order to not overshoot the VHE and X-ray fluxes via pair-cascade emission. The result is shown in Fig.~\ref{fig:Model_leha}, and the parameters are provided in Table~\ref{table:hybrid} for the periods P2, P3, and P4.

The equipartition factor $U^\prime_p/U^\prime_B$ reaches the value of $13.5\times 10^{4}$. 
The lepto-hadronic model describes a particle-dominated scenario with an energy density ratio far from equipartition. As discussed in~\cite{2016MNRAS.456.2374T}, the magnetic field and particle energy densities are expected to be more closely balanced in BL Lacs; however, it is common to find a particle energy density that dominates the magnetic field energy density by one or two orders of magnitude. On the other hand, the high Compton dominance of FSRQs could reflect a physical scenario that significantly departs from equipartition as in the lepto-hadronic model we tested. As discussed in \cite{2014ApJ...796L...5N}, such a situation can arise from either the geometries of the external radiation sources (broad-line region, hot-dust torus) being quasi-spherical rather than flat or most of the $\gamma$-ray radiation being produced in jet regions of low magnetisation.
The proton power for the lepto-hadronic scenario is $L_p = 6\times10^{47}$ erg s$^{-1}$ for P3, which is super-Eddington as long as $M_\bullet\simeq 10^8\ M_\odot$. If instead the BH mass is $M_\bullet = 10^9\ M_\odot$, the proton power becomes of the order of the Eddington luminosity.
The neutrino emission is peaking around 10\,PeV. 
For the lepto-hadronic model, the expected neutrino flux is much lower than the sensitivity of IceCube~\citep{2019EPJC...79..234A} or ANTARES~\citep{2020ApJ...892...92A}, consistent with the non-detection of any neutrinos from this source.
The best-fit number of astrophysical neutrino events reported in \cite{2020PhRvL.124e1103A} for OT~081 corresponds to a value of $\hat{n}_{s}$=12.2 (local pre-trial p-value
$-log10(p_{\rm{local}})=0.7$) assuming an astrophysical power-law spectral index of $\hat{\gamma}$=3.2. This value is compatible with the one we obtain from the lepto-hadronic model.
From the lepto-hadronic models of the three source states, shown in Fig.~\ref{fig:Model_leha}, we estimate CD values of 30, 21 and 3 for P2, P3 and P4, respectively. For comparison, we infer a CD of $\sim1$ for the low state of the source, based on a simple polynomial fit to the archival data by \cite{2018RAA....18..120L}. The position of both the low and HE SED peaks show a clear shift to higher energies with respect to the archival data. In particular, a major shift is found for P3 - the synchrotron peak is shifted by a factor $\sim$ 8 towards higher frequencies, the IC peak is shifted by more than 2 orders of magnitude with respect to the archival data. 

A proton-synchrotron dominated model has also been attempted. Such a model can almost reproduce the MWL SEDs for the different source states, but extreme physical parameters would be required. The proton luminosity would reach extreme values, $L_{p} = 5.6\times10^{48}$ erg s$^{-1}$, a factor $\sim$40 larger than the Eddington luminosity of a supermassive BH with mass $M_{\bullet} = 10^8\ M_{\odot}$ (see details in Appendix~\ref{sec:appendix}).

\subsection{Constraints on the location of the emitting region}
\label{sec:location}

The best joint fit to the HE and the intrinsic VHE spectra for P3 is described by a log-parabola function (see Sec.~\ref{sec:models_l}) for which the fit parameters are given in Table~\ref{tab:combined_parameters}. The intrinsic curvature implied by a log-parabola function can be attributed to different reasons, such as the energy-dependent Klein-Nishina cross section and/or intrinsic curvature in the particle distribution (e.g. acceleration/cooling processes) \citep[e.g.][]{2004A&A...422..103M,2006A&A...448..861M}.
Given the fact that OT~081 has a developed BLR, another cause of the curvature could be intrinsic absorption if the emitting region is located within the BLR photon field. We carried out a simple test to determine the maximum amount of absorption due to interactions with the BLR that is consistent with the P3 dataset. Note that for this test, we only considered absorption due to the BLR, so the result should be interpreted only as an upper limit on the opacity ($\tau_{\gamma \gamma}$) that is allowed by the observations.

The first step is to calculate the theoretical estimation of $\tau_{\gamma \gamma}$ for different energies and locations with respect to the BLR. These calculations are carried out using the agnpy code (revision 0113497b) \citep{2022A&A...660A..18N}. The implementation of the $\gamma \gamma$ absorption is based on \cite{2008ApJ...686..181F} and \cite{2009ApJ...692...32D}. For the calculations, we used the same parameters as in the leptonic modelling in Sec.~\ref{sec:models_l}. The BLR is assumed to be a shell located at a distance $R_{\rm BLR}\approx 10^{17} \,[L_{\rm disk}/(10^{45} \, {\rm erg} \, {\rm s}^{-1})]^{1/2} \,{\rm cm} \simeq 3.6 \times 10^{16} \,\rm{cm}$ \citep{2009MNRAS.397..985G}. With this configuration, we checked whether $\gamma \gamma$ absorption within the BLR affects the $\gamma$-ray emission for energies higher than $\sim20$\,GeV. 

\begin{figure}
\centering
\includegraphics[scale=0.53]{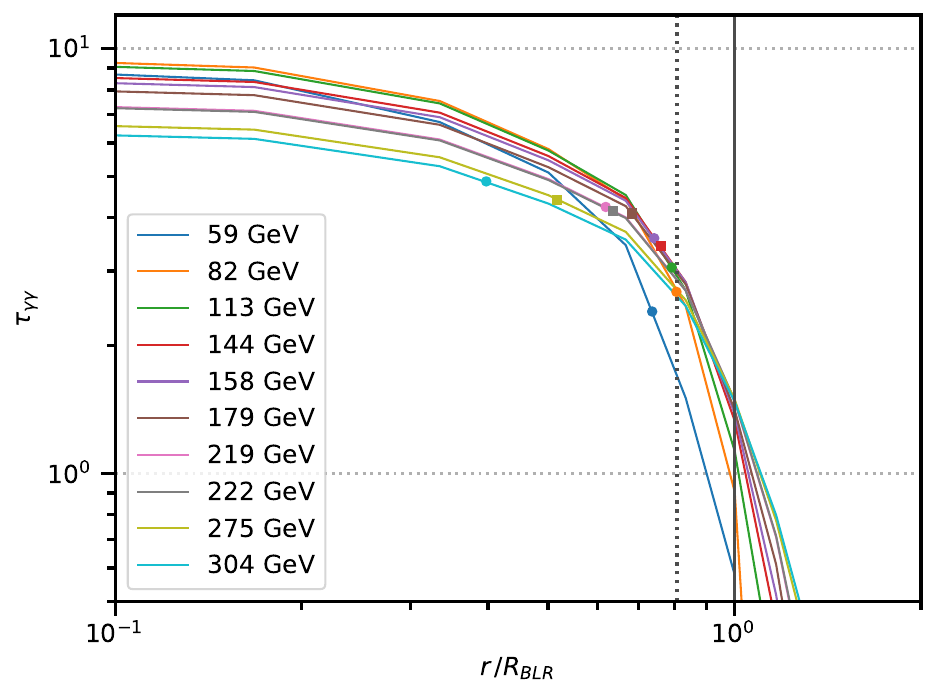}
 \caption{Optical depth due to $\gamma$-$\gamma$ absorption on the BLR radiation field as a function of distance from the central BH, normalised to the BLR radius, for each energy bin centre reported in the VHE $\gamma$-ray SEDs observed with MAGIC (dots) and H.E.S.S. (squares). For comparison, estimates of the absorption from theoretical models are plotted as colored lines. The vertical solid line denotes the BLR radius, while the dotted vertical line shows the lower limit on the distance of the emitting region allowed by the data.}
 \label{fig:tau_BLR} 
\end{figure}

In order to derive constraints from the observations, we consider the HE spectrum in the context of the VHE observations (after EBL correction). The maximum effective energy involved in the analysis of the \textit{Fermi}-LAT spectrum is a photon with an energy of 5.5\,GeV with a probability of 99.97\% of coming from OT~081. Therefore, the HE spectrum is not affected by the BLR absorption. Hence, we assume that the difference between the HE observed spectrum, which is compatible with a powerlaw function as described in Sec.~\ref{sec:HE_analyis}, extrapolated to the VHE band and the actual VHE measurements might be due to BLR absorption. We would also like to note that as the gamma-ray absorption within the BLR is starting to affect around 20\,GeV, this also does not affect the simple test on the BLR gamma-ray absorption. In order to perform the calculations, the HE spectrum is extrapolated to the VHE band, and the difference between the extrapolated spectrum ($F_{\rm{ext}}$) and the EBL-corrected measurement ($F_{\rm{obs}}$) is computed, obtaining an optical depth $\tau_{\gamma \gamma} (E)= \ln(F_{\rm{ext}}(E)/F_{\rm{obs}}(E))$. The HE spectral fit parameters are randomized assuming a normal distribution including the statistical uncertainty. For the VHE spectra observed with MAGIC, in addition to the statistical errors, 20\% and 15\% systematic uncertainties are included for the flux and energy scale, respectively. In the case of the spectrum derived from the H.E.S.S. observations, the statistical uncertainty together with 20\% and 19\% systematic uncertainties for the flux and energy scale are included. For each energy (associated with the centre of each spectral point), the limit on $\tau_{\gamma \gamma} (E)$ is estimated as the 95th percentile of the distribution for $10^{4}$ realizations.
By comparing our derived limits on $\tau_{\gamma \gamma} (E)$ at different energies to theoretical estimates of the absorption, we can place a limit on the distance to the BLR. Such limits depend on the energy, and the most restrictive value comes from the energy bin centred at $\sim82$\,GeV, as shown in Fig.~\ref{fig:tau_BLR}. 
Thus, we conclude that in the conservative scenario in which all the observed curvature in the intrinsic VHE spectra is caused by absorption within BLR, the location of the emitting region would be located at a distance of $>0.8\times R_{\rm{BLR}}$.
Therefore, in a more realistic scenario, the modelling assumption of the emitting region being located outside of the BLR is justified as well as consistent with similar scenarios for VHE-detected FSRQs.

\section{Discussion and Conclusions}
\label{sec:conclusions}
In this work, we present the first detection of OT~081 (\textit{z}=0.32) at VHE $\gamma$-rays with the MAGIC and H.E.S.S. Cherenkov telescopes. The VHE observations followed the flare in the HE band observed by \textit{Fermi}-LAT in July 2016. Due to moon and weather constraints, the VHE observations could be performed only a few days after the HE flare when the source was in a decay phase. A MWL study from radio to VHE $\gamma$-rays is carried out, and four different periods (named P1, P2, P3, and P4) were chosen for their MWL coverage. P1 and P2 correspond to the two peaks in the HE band, P3 represents the source state during the VHE detection, and P4 corresponds to the lower flux state after the flare, which is coincident with a high degree of polarization. 
Both the observed and EBL-corrected VHE spectra from OT~081 are well described by a power-law fit, and no sign of curvature was found. On the other hand, the joint HE-VHE spectra is best explained as a log-parabola with a peak located around 1\,GeV. 

In the VLBA study of the source, reported in Appendix ~\ref{sec:appendix_vlba}, events in the innermost region of the jet could explain the activity in HE $\gamma$-rays, which appears enhanced before (P1, MJD~57585.5/57856.9 -- 16/17 July 2016 and P2, MJD ~57589 -- 20  July  2016), rather than simultaneous with the VHE detection (P3, MJD~57593.9 -- 27 July 2016). However, because in this work we focus on the first detection of VHE $\gamma$ ray data from the source, we select simultaneous broadband data with a short time window (periods P2, P3 and P4) which do not allow us to establish a robust link to the VLBA analysis maps because of their larger time uncertainties with respect to the VHE $\gamma$-ray variability. In a separate work we will consider more complex alternative scenarios which could establish a link between VLBA and the broadband dataset here presented. Possible future flaring activity will certainly benefit from contemporaneous VLBA observation to shred light on the link between the observed features and VHE $\gamma$-ray flares.
The common categorization of blazars is based on the presence or absence of strong emission lines in their optical spectra. Sources presenting weak or absent emission lines are BL~Lacs, while strong lines are characteristic of FSRQs. The separation between the two categories is usually defined by an equivalent width of the emission line ($|EW_{\rm{rest}}| > 5~\textrm{\AA}$). The development of $\gamma$-ray telescopes and the many $\gamma$-ray detections of blazars brought attention to the broadband SEDs of such sources and to possible other categorizations of blazars not based solely on their optical spectra. The broadband SEDs of blazars, presenting a common doubled-bumped structure, has been used to further categorize this class of AGNs, depending on the positions of the synchrotron and IC peaks. The ratio  between the IC and synchrotron peak fluxes, the so-called Compton dominance, has been found to be higher for FSRQs than for BL~Lacs, making the modelling of the former sources more complex with respect to pure SSC leptonic models. 
Although OT~081 was classified as a BL Lac object and as LSP in the fourth catalogue of AGN detected by \textit{Fermi}-LAT \citep[4LAC][]{2020ApJ...892..105A}, it displays some characteristics more common of FSRQs. 
In particular, broad lines have been measured in its optical spectrum while in its low state \citep{1988A&A...191L..16S}. BL Lacertae, the prototypical source for the BL~Lac categorisation, also displays strong emission optical lines during low flux states ~\citep[see e.g.][]{1995ApJ...452L...5V}.

Using a spectrum of OT~081 taken on 30 August 2016 (MJD~57630), we estimated a  H$_{\beta}$ line luminosity  of $\sim$5$\times$10$^{41}$ erg/s, which (Eq. 6 of \cite{2006ApJ...641..689V}) provides a value of the mass of the BH of M$_{\bullet}=10^{7.9} \times M_\odot $.
In addition to the characteristics of the optical spectrum, the high CD of $\sim30$ (P2) reached by OT~081 during the flaring activity presented in this work is unusual for BL Lac objects. Moreover, the SED peaks are strongly shifted to higher energies during this period (the synchrotron peak by a factor of $\sim$~8, the IC peak by more than 2 orders of magnitude), especially during P3 where the VHE detection allows us to constrain the IC peak. 

The MWL data collected around the selected three source states are studied within the framework of state-of-the-art theoretical emission models.
As previously argued by \cite{2013MNRAS.436..304P}, in addition to the SSC emission, an external photon field can be considered as in the case of FSRQs. Guided by the presence of emission lines in the optical, we assume a well-developed BLR and dusty torus. The HE and VHE data are used to test the location of the emitting region for which a lower limit of $>0.8\,\rm R_{\rm BLR}$ is found. This indicates that the emitting region should be either at the BLR edge or outside to avoid $\gamma$-ray absorption. 
For the broadband SED modelling, the location of the emitting region is assumed to be outside of the BLR. A similar approach was taken in~\cite{2018A&A...617A..30M} where during a 2015 flare, the object S4~0954+65, which was classified in the literature as a BL~Lac object, exhibited characteristics typically associated with FSRQs or transitional blazars rather than BL~Lacs, including reaching a CD of $\simeq7$. In this case, the model that successfully describes the data is based on the assumption of a dusty torus as the source of external photons.

The emitting region in the model is assumed to be much closer to the torus than the location of interactions with K15 and K16, which is farther away (10 pc from the BH). If the scenario described by the VLBA analysis could be confirmed, the lepto-hadronic model would still work, though interactions with the knots would replace the torus as the source of the target photon field.

Given the recent interest in multi-messenger photon and neutrino emission from blazars, we tested the effect of adding a hadronic component to the leptonic EIC solution. This lepto-hadronic model successfully fits the data.   
We also tested a proton-synchrotron dominated solution based on the hadronic code of~\cite{Cerruti15} which is described in Appendix~\ref{sec:appendix}.
The proton-synchrotron dominated model we adopted fits the data well, but it requires a very high proton power of $L_{\rm{p}} = 5.6\times10^{48}$ erg s$^{-1}$, which is about a factor of $400$ times as large as the Eddington luminosity of a supermassive black hole with mass $M_{\bullet} = 10^8\ M_\odot$.

The first LBL detected at VHE was AP Librae \citep{2015A&A...573A..31H}. The broadband SED of AP Librae challenged single-zone leptonic models due to the extreme broadness of the HE component \citep{2015MNRAS.454.3229S,2015A&A...578A..69H}. A notable feature of both AP Librae and OT~081 is the hard X-ray spectrum, which has been interpreted as the onset of the high-energy component of the SED. In the case of AP~Librae, an extended X-ray jet was detected (\cite{2013ApJ...776...68K} and references mentioned above), which would explain the peculiar SED \citep{2015MNRAS.454.3229S,2016A&A...588A.110Z,2022ApJ...924...57R} by providing a source of external photons. A multi-zone SSC model \citep{2015A&A...578A..69H} as well as  hadronic models \citep{2017MNRAS.464.2213P} have been used to explain the electromagnetic emission of the source, providing in this case a satisfactory representation of the source's SED without the need for external photon fields.

In this work, we find that a single-zone SSC model is not sufficient to describe the broadband SEDs measured for OT~081 during the periods of activity we identified. Rather, an EIC contribution is required to describe the MWL dataset. In agreement with the presence of emission lines in the optical spectrum of the source, we model the source by assuming the presence of a well-developed BLR in the core as well as a dusty torus. We find that both a leptonic model and a lepto-hadronic model can successfully reproduce the dataset. 
Additionally, a pure hadronic model could be considered, but it would require relatively extreme values of the proton luminosity.
The VLBA analysis suggests another possible source of seed photons- the superluminal knot K15 that was ejected after a prolonged period of low-level non-thermal activity. Even if not the scope of this work, a future study, possibly including new data when available, could also explore two-zone modellings as another possible interpretation of the MWL emission.
This first detection of VHE $\gamma$-rays from OT~081 brings new insight into its MWL characterisation as well as widening our understanding of the emission mechanisms of high CD blazars.
The characteristics of the source's optical spectrum as well as the MWL observations and SED modelling suggest that this source is a transitional blazar, on the border between BL~Lacs and FSRQs. 

\section*{Authors' contributions}
In alphabetical order:\\
J. Becerra Gonzalez: paper editing and reviewing, \textit{Fermi}-LAT analysis, X-check of MAGIC analysis, investigation, visualization (Fig.~\ref{fig:tau_BLR}).\\
S. Jorstad: investigation (VLBA), visualization (Fig.~\ref{fig:1749_vlba2016}).\\
M. Manganaro: project leadership, paper drafting and editing, MAGIC data analysis, investigation, collection of the MWL data, correlations and variability study, visualization (Figs.~\ref{fig:VHE_spectrum},~\ref{fig:MWL_LC}).\\
S. Paiano: investigation (optical spectra measurements).\\
D. Sanchez: paper reviewing, investigation, Xcheck of H.E.S.S. analysis, visualization (Fig.~\ref{fig:joint}).\\
M. Seglar Arroyo: paper editing and reviewing, H.E.S.S. data analysis, investigation.\\
F. Tavecchio: theoretical modelling (leptonic models), theoretical interpretation\\
H. B. Xiao: testing of theoretical models (SSC and other leptonic models), visualization (Figs.~\ref{fig:Model_IR},~\ref{fig:Model_leha},~\ref{fig:psynch_all}).\\
The rest of the authors have contributed in one or several of the following ways: design, construction, maintenance and operation of the instrument(s) used to acquire the data; preparation and/or evaluation of the observation proposals; data acquisition, processing, calibration and/or reduction; production of analysis tools and/or related Monte Carlo simulations; discussion and approval of the contents of the draft, including internal reviewing.

\section*{Acknowledgements}
This work was supported in part by the Croatian Science Foundation under the project number IP-2022-10-4595.\\
M.~S-A. is supported by the grant FJC2020-044895-I funded by MCIN/AEI/10.13039/501100011033 and by the European Union NextGenerationEU/PRTR.\\
H.~B.~X. acknowledges the support from the National Natural Science Foundation of China (NSFC 12203034), the Shanghai Science and Technology Fund (22YF1431500), and the science research grants from the China Manned Space Project.
We would like to thank the Instituto de Astrof\'{\i}sica de Canarias for the excellent working conditions at the Observatorio del Roque de los Muchachos in La Palma. The financial support of the German BMBF, MPG and HGF; the Italian INFN and INAF; the Swiss National Fund SNF; the grants PID2019-104114RB-C31, PID2019-104114RB-C32, PID2019-104114RB-C33, PID2019-105510GB-C31, PID2019-107847RB-C41, PID2019-107847RB-C42, PID2019-107847RB-C44, PID2019-107988GB-C22 funded by MCIN/AEI/ 10.13039/501100011033; the Indian Department of Atomic Energy; the Japanese ICRR, the University of Tokyo, JSPS, and MEXT; the Bulgarian Ministry of Education and Science, National RI Roadmap Project DO1-400/18.12.2020 and the Academy of Finland grant nr. 320045 is gratefully acknowledged. This work was also been supported by Centros de Excelencia ``Severo Ochoa'' y Unidades ``Mar\'{\i}a de Maeztu'' program of the MCIN/AEI/ 10.13039/501100011033 (SEV-2016-0588, SEV-2017-0709, CEX2019-000920-S, CEX2019-000918-M, MDM-2015-0509-18-2) and by the CERCA institution of the Generalitat de Catalunya; by the Croatian Science Foundation (HrZZ) Project IP-2022-10-4595 and the University of Rijeka Project uniri-prirod-18-48; by the Deutsche Forschungsgemeinschaft (SFB1491 and SFB876); the Polish Ministry Of Education and Science grant No. 2021/WK/08; and by the Brazilian MCTIC, CNPq and FAPERJ.\\
The support of the Namibian authorities and of the University of Namibia in facilitating  the construction and operation of H.E.S.S. is gratefully acknowledged, as is the support by the German Ministry for Education and Research (BMBF), the Max Planck Society, the German Research Foundation (DFG), the Helmholtz Association, the Alexander von Humboldt Foundation, the French Ministry of Higher Education, Research and Innovation, the Centre National de la Recherche Scientifique (CNRS/IN2P3 and CNRS/INSU), the Commissariat \`a l'\'energie atomique et aux \'energies alternatives (CEA), the U.K. Science and Technology Facilities Council (STFC), the Knut and Alice Wallenberg Foundation, the National Science Centre, Poland grant no. 2016/22/M/ST9/00382, the South African Department of Science and Technology and National Research Foundation, the University of Namibia, the National Commission on Research, Science \& Technology of Namibia (NCRST), the Austrian Federal Ministry of Education, Science and Research and the Austrian Science Fund (FWF), the Australian Research Council (ARC), the Japan Society for the Promotion of Science and by the University of Amsterdam, the Science Committee of Armenia grant 21AG-1C085. We appreciate the excellent work of the technical support staff in Berlin, Zeuthen, Heidelberg, Palaiseau, Paris, Saclay, T\"ubingen and in Namibia in the construction and operation of the equipment. This work benefited from services provided by the H.E.S.S. Virtual Organisation, supported by the national resource providers of the EGI Federation. \\
The \textit{Fermi} LAT Collaboration acknowledges generous ongoing support from a number of agencies and institutes that have supported both the development and the operation of the LAT as well as scientific data analysis. These include the National Aeronautics and Space Administration and the Department of Energy in the United States, the Commissariat \`a l'Energie Atomique and the Centre National de la Recherche Scientifique / Institut National de Physique Nucl\'eaire et de Physique des Particules in France, the Agenzia Spaziale Italiana and the Istituto Nazionale di Fisica Nucleare in Italy, the Ministry of Education, Culture, Sports, Science and Technology (MEXT), High Energy Accelerator Research Organization (KEK) and Japan Aerospace Exploration Agency (JAXA) in Japan, and the K.~A.~Wallenberg Foundation, the Swedish Research Council and the Swedish National Space Board in Sweden.\\
Additional support for science analysis during the operations phase is gratefully acknowledged from the Istituto Nazionale di Astrofisica in Italy and the Centre National d'\'Etudes Spatiales in France. This work performed in part under DOE Contract DE-AC02-76SF00515.\\
Part of this work is based on archival data, software or online services provided by the Space Science Data Center ASI under contract ASI-INFN 2021-43-HH.0\\
This research has made use of data from the OVRO 40-m monitoring program \citep{2011ApJS..194...29R}, supported by private funding from the California Institute of Technology and the Max Planck Institute for Radio Astronomy, and by NASA grants NNX08AW31G, NNX11A043G, and NNX14AQ89G and NSF grants AST-0808050 and AST- 1109911. \\
S.K. acknowledges support from the European Research Council (ERC) under the European Unions Horizon 2020 research and innovation programme under grant agreement No.~771282.
W.M. gratefully acknowledges support by the ANID BASAL project FB210003 and FONDECYT 11190853.
R.R. is supported by ANID BASAL grant FB210003.
S.C. acknowledge support by the Italian Space Agency (Agenzia Spaziale Italiana, ASI) through contract ASI-OHBI-2017-12-
I.0, with agreement ASI-INFN 2021-43-HH.0, and its Space Science Data 
Center (SSDC).
This publication makes use of data obtained at the Mets\"ahovi Radio Observatory, operated by the Aalto University. 
This paper makes use of the following ALMA data: ADS/JAO.ALMA\#2015.1.00856.S. ALMA is a partnership of ESO (representing its member states), NSF (USA) and NINS (Japan), together with NRC (Canada), MOST and ASIAA (Taiwan), and KASI (Republic of Korea), in cooperation with the Republic of Chile. The Joint ALMA Observatory is operated by ESO, AUI/NRAO and NAOJ
This research has made use the TeVCat online source catalogue (http://tevcat.uchicago.edu). 
Part of this work is based on archival data, software or online services provided by the Space Science Data Center - ASI.
The research at Boston University was supported by NASA \textit{Fermi} Guest Investigator grants  80NSSC17K0649 and 80NSSC20K1567. The VLBA is an instrument of the National Radio Astronomy Observatory. The National Radio Astronomy Observatory is a facility of the National Science Foundation operated by Associated Universities, Inc.
Data from the Steward Observatory spectropolarimetric monitoring project were used. This program is supported by \textit{Fermi} Guest Investigator grants NNX08AW56G, NNX09AU10G, NNX12AO93G, and NNX15AU81G.
Part of this work is based on archival data, software or online services provided by the Space Science Data Center - ASI.
S.C. acknowledges support by the Italian Space Agency (Agenzia Spaziale Italiana, ASI) through contract ASI-OHBI-2017-12-
I.0, with agreement ASI-INFN 2021-43-HH.0, and its Space Science Data Center (SSDC).
A.V.F. and W.Z. received financial assistance from the Christopher R. Redlich Fund, as well as donations from Gary and Cynthia Bengier, Clark and Sharon Winslow, Alan Eustace (W.Z. is a Bengier-Winslow-Eustace Specialist in Astronomy), and numerous other donors. KAIT and its ongoing operation were made possible by donations from Sun Microsystems, Inc., the Hewlett-Packard Company, AutoScope Corporation, Lick Observatory, the U.S. National Science Foundation, the University of California, the Sylvia \& Jim Katzman Foundation, and the TABASGO Foundation. Research at Lick Observatory is partially supported by a generous gift from Google.
The authors wish to thank Matteo Cerruti (APC-Universit\'e Paris Cit\'e and ICC-Universitat de Barcelona) for providing the numerical results for the hadronic and lepto-hadronic models used in this work.

\section*{Affiliations}
\noindent
{\it
$^{1}$ {Japanese MAGIC Group: Institute for Cosmic Ray Research (ICRR), The University of Tokyo, Kashiwa, 277-8582 Chiba, Japan} \\
$^{2}$ {Instituto de Astrof\'isica de Canarias and Dpto. de  Astrof\'isica, Universidad de La Laguna, E-38200, La Laguna, Tenerife, Spain} \\
$^{3}$ {Instituto de Astrof\'isica de Andaluc\'ia-CSIC, Glorieta de la Astronom\'ia s/n, 18008, Granada, Spain} \\
$^{4}$ {National Institute for Astrophysics (INAF), I-00136 Rome, Italy} \\
$^{5}$ {Universit\`a di Udine and INFN Trieste, I-33100 Udine, Italy} \\
$^{6}$ {International Center for Relativistic Astrophysics (ICRA), Rome, Italy} \\
$^{7}$ {Max-Planck-Institut f\"ur Physik, D-80805 M\"unchen, Germany} \\
$^{8}$ {Universit\`a di Padova and INFN, I-35131 Padova, Italy} \\
$^{9}$ {Institut de F\'isica d'Altes Energies (IFAE), The Barcelona Institute of Science and Technology (BIST), E-08193 Bellaterra (Barcelona), Spain} \\
$^{10}$ {Technische Universit\"at Dortmund, D-44221 Dortmund, Germany} \\
$^{11}$ {Croatian MAGIC Group: University of Zagreb, Faculty of Electrical Engineering and Computing (FER), 10000 Zagreb, Croatia} \\
$^{12}$ {IPARCOS Institute and EMFTEL Department, Universidad Complutense de Madrid, E-28040 Madrid, Spain} \\
$^{13}$ {Centro Brasileiro de Pesquisas F\'isicas (CBPF), 22290-180 URCA, Rio de Janeiro (RJ), Brazil} \\
$^{14}$ {Centro de Investigaciones Energ\'eticas, Medioambientales y Tecnol\'ogicas, E-28040 Madrid, Spain} \\
$^{15}$ {ETH Z\"urich, CH-8093 Z\"urich, Switzerland} \\
$^{16}$ {Departament de F\'isica, and CERES-IEEC, Universitat Aut\`onoma de Barcelona, E-08193 Bellaterra, Spain} \\
$^{17}$ {Universit\`a di Pisa and INFN Pisa, I-56126 Pisa, Italy} \\
$^{18}$ {Universitat de Barcelona, ICCUB, IEEC-UB, E-08028 Barcelona, Spain} \\
$^{19}$ {Department for Physics and Technology, University of Bergen, All\'egaten 55, 5007 Bergen, Norway} \\
$^{20}$ {INFN MAGIC Group: INFN Sezione di Catania and Dipartimento di Fisica e Astronomia, University of Catania, I-95123 Catania, Italy} \\
$^{21}$ {Port d'Informaci\'o Cient\'ifica (PIC), E-08193 Bellaterra (Barcelona), Spain} \\
$^{22}$ {INFN MAGIC Group: INFN Sezione di Torino and Universit\`a degli Studi di Torino, I-10125 Torino, Italy} \\
$^{23}$ {INFN MAGIC Group: INFN Sezione di Bari and Dipartimento Interateneo di Fisica dell'Universit\`a e del Politecnico di Bari, I-70125 Bari, Italy} \\
$^{24}$ {Croatian MAGIC Group: University of Rijeka, Faculty of Physics, 51000 Rijeka, Croatia} \\
$^{25}$ {University of Geneva, Chemin d'Ecogia 16, CH-1290 Versoix, Switzerland} \\
$^{26}$ {Japanese MAGIC Group: Physics Program, Graduate School of Advanced Science and Engineering, Hiroshima University, 739-8526 Hiroshima, Japan} \\
$^{27}$ {Armenian MAGIC Group: ICRANet-Armenia, 0019 Yerevan, Armenia} \\
$^{28}$ {University of Lodz, Faculty of Physics and Applied Informatics, Department of Astrophysics, 90-236 Lodz, Poland} \\
$^{29}$ {Universit\"at Innsbruck, Institut f\"ur Astro- und Teilchenphysik, Technikerstraße 25, 6020 Innsbruck, Austria} \\  
$^{30}$ {Croatian MAGIC Group: Josip Juraj Strossmayer University of Osijek, Department of Physics, 31000 Osijek, Croatia} \\
$^{31}$ {Universit\"at W\"urzburg, D-97074 W\"urzburg, Germany} \\
$^{32}$ {Finnish MAGIC Group: Finnish Centre for Astronomy with ESO, University of Turku, FI-20014 Turku, Finland} \\
$^{33}$ {Department of Physics, University of Oslo, Sem S\ae landsvei 24 0371 Oslo, Norway} \\
$^{34}$ {Japanese MAGIC Group: Department of Physics, Tokai University, Hiratsuka, 259-1292 Kanagawa, Japan} \\
$^{35}$ {also at Dipartimento di Fisica, Universit\`a di Trieste, I-34127 Trieste, Italy} \\
$^{36}$ {Universit\`a di Siena and INFN Pisa, I-53100 Siena, Italy} \\
$^{37}$ {Saha Institute of Nuclear Physics, A CI of Homi Bhabha National Institute, Kolkata 700064, West Bengal, India} \\
$^{38}$ {Inst. for Nucl. Research and Nucl. Energy, Bulgarian Academy of Sciences, BG-1784 Sofia, Bulgaria} \\
$^{39}$ {Finnish MAGIC Group: Space Physics and Astronomy Research Unit, University of Oulu, FI-90014 Oulu, Finland} \\
$^{40}$ {Japanese MAGIC Group: Chiba University, ICEHAP, 263-8522 Chiba, Japan} \\
$^{41}$ {INAF Padova, Astronomical Observatory, vicolo dell’Osservatorio 5, I-35122 Padova, Italy} \\
$^{42}$ {Croatian MAGIC Group: Ru\dj{}er Bo\v{s}kovi\'c Institute, 10000 Zagreb, Croatia} \\
$^{43}$ {Japanese MAGIC Group: Institute for Space-Earth Environmental Research and Kobayashi-Maskawa Institute for the Origin of Particles and the Universe, Nagoya University, 464-6801 Nagoya, Japan} \\
$^{44}$ {Japanese MAGIC Group: Institute for Cosmic Ray Research (ICRR), The University of Tokyo, Kashiwa, 277-8582 Chiba, Japan} \\
$^{45}$ {INFN MAGIC Group: INFN Sezione di Perugia, I-06123 Perugia, Italy} \\
$^{46}$ {INFN Roma Tor Vergata, I-00133 Roma, Italy} \\ 
$^{47}$ {Department of Physics, Konan University, 8-9-1 Okamoto, Higashinada, Kobe, Hyogo 658-8501, Japan} \\  
$^{48}$ {Landessternwarte, Universit\"at Heidelberg, K\"onigstuhl, D 69117 Heidelberg, Germany} \\
$^{49}$ {Universit\'e de Paris, CNRS, Astroparticule et Cosmologie, F-75013 Paris, France} \\
$^{50}$ {Department of Physics and Electrical Engineering, Linnaeus University,  351 95 V\"axj\"o, Sweden} \\
$^{51}$ {Institut f\"ur Astronomie und Astrophysik, Universit\"at T\"ubingen, Sand 1, D 72076 T\"ubingen, Germany} \\
$^{52}$ {Centre for Space Research, North-West University, Potchefstroom 2520, South Africa} \\
$^{53}$ {Sorbonne Universit\'e, Universit\'e Paris Diderot, Sorbonne Paris Cit\'e, CNRS/IN2P3, Laboratoire de Physique Nucl\'eaire et de Hautes Energies, LPNHE, 4 Place Jussieu, F-75252 Paris, France} \\
$^{54}$ {University of Oxford, Department of Physics, Denys Wilkinson Building, Keble Road, Oxford OX1 3RH, UK}  \\
$^{55}$ {Astronomical Observatory, The University of Warsaw, Al. Ujazdowskie 4, 00-478 Warsaw, Poland}  \\
$^{56}$ {Instytut Fizyki J\c{a}drowej PAN, ul. Radzikowskiego 152, 31-342 Krak{\'o}w, Poland}  \\
$^{57}$ {Institut f\"ur Physik und Astronomie, Universit\"at Potsdam,  Karl-Liebknecht-Strasse 24/25, D 14476 Potsdam, Germany}  \\
$^{58}$ {School of Physical Sciences, University of Adelaide, Adelaide 5005, Australia}  \\
$^{59}$ {Aix Marseille Universit\'e, CNRS/IN2P3, CPPM, Marseille, France} \\
$^{60}$ {Laboratoire Leprince-Ringuet, École Polytechnique, CNRS, Institut Polytechnique de Paris, F-91128 Palaiseau, France}  \\
$^{61}$ {Obserwatorium Astronomiczne, Uniwersytet Jagiello{\'n}ski, ul. Orla 171, 30-244 Krak{\'o}w, Poland} \\
$^{62}$ {Friedrich-Alexander-Universit\"at Erlangen-N\"urnberg, Erlangen Centre for Astroparticle Physics, Nikolaus-Fiebiger-Str. 2, 91058 Erlangen, Germany} \\
$^{63}$ {University of Namibia, Department of Physics, Private Bag 13301, Windhoek 10005, Namibia}  \\
$^{64}$ {Institute of Astronomy, Faculty of Physics, Astronomy and Informatics, Nicolaus Copernicus University,  Grudziadzka 5, 87-100 Torun, Poland} \\
$^{65}$ {Nicolaus Copernicus Astronomical Center, Polish Academy of Sciences, ul. Bartycka 18, 00-716 Warsaw, Poland}  \\
$^{66}$ {IRFU, CEA, Universit\'e Paris-Saclay, F-91191 Gif-sur-Yvette, France}\\
$^{67}$ {Department of Physics and Astronomy, The University of Leicester, University Road, Leicester, LE1 7RH, United Kingdom}  \\
$^{68}$ {Yerevan Physics Institute, 2 Alikhanian Brothers St., 0036 Yerevan, Armenia} \\
$^{69}$ {Univ. Savoie Mont Blanc, CNRS, Laboratoire d' Annecy de Physique des Particules - IN2P3, 74000 Annecy, France}\\ 
$^{70}$ {Universit\"at Hamburg, Institut f\"ur Experimentalphysik, Luruper Chaussee 149, D 22761 Hamburg, Germany}  \\
$^{71}$ {Space Science Data Center, Agenzia Spaziale Italiana, Via del Politecnico snc, I-00133 Roma, Italy}\\ 
$^{72}$ {Astrophysics Science Division, NASA Goddard Space Flight Center, Greenbelt, MD, USA 20771}\\
$^{73}$ {INAF-IRA Bologna, Via P. Gobetti 101, I-40129, Bologna, Italy}\\ 
$^{74}$ {Universidad de Valpara\'iso. Instituto de F\'isica y Astronom\'ia de Valpara\'iso. Av. Gran Breta\~na 1111, Playa Ancha, Valpara\'iso, Chile.} \\
$^{75}$ {Finnish Center for Astronomy with ESO (FINCA), University of Turku, FI-20014, Turku, Finland}\\ 
$^{76}$ {Department of Astronomy, University of California, Berkeley, CA 94720-3411, USA}\\ 
$^{77}$ {Aalto University Mets\"ahovi Radio Observatory, Mets\"ahovintie 114, 02540 Kylm\"al\"a, Finland}\\ 
$^{78}$ {Astrophysics Research Institute, Liverpool John Moores University, Liverpool Science Park IC2, 146 Brownlow Hill}\\ 
$^{79}$ {Institute for Astrophysical Research, Boston University, 725 Commonwealth Avenue, Boston, MA 02215, USA}\\ 
$^{80}$ {Saint Petersburg State University, 7/9 Universitetskaya nab., St. Petersburg, 199034 Russia}\\ 
$^{81}$ {Institute of Astrophysics, Foundation for Research and Technology-Hellas, GR-71110 Heraklion, Greece}\\ 
$^{82}$ {Department of Physics, Univ. of Crete, GR-70013 Heraklion, Greece}\\ 
$^{83}$ {Aalto University Department of Electronics and Nanoengineering, P.O. BOX 15490, FI-00076 AALTO, Finland}\\ 
$^{84}$ {Departamento de Astronom\'ia, Universidad de Chile, Camino El Observatorio 1515, Las Condes, Santiago, Chile}\\ 
$^{85}$ {Owens Valley Radio Observatory, California Institute of Technology, Pasadena, CA 91125, USA}\\ 
$^{86}$ {CePIA, Astronomy Department, Universidad de Concepci\'on, Casilla 160-C, Concepci\'on, Chile}\\ 
$^{87}$ {INAF/OAR, via Frascati 33, I-00078 Monte Porzio Catone (RM), Italy}\\ 
$^{88}$ {Shanghai Key Lab for Astrophysics, Shanghai Normal University, Shanghai, 200234, People's Republic of China}\\ 
}


\section*{Data availability}
The FITS data from MAGIC in Fig.~\ref{fig:VHE_spectrum} will be available after publication at: \url{https://vobs.magic.pic.es/fits/}. 
The complete data set shown in Fig.~\ref{fig:MWL_LC}, the data points shown in Fig~\ref{fig:Model_IR}, Fig.~\ref{fig:Model_leha} and Fig.~\ref{fig:psynch_all}, and Table~\ref{table:VHEspectrum} will be available at the CDS (\url{http://cdsarc.u-strasbg.fr}) after publication. Other data underlying this article will be shared on reasonable request to the corresponding authors.



\bibliography{sample}

\begin{thebibliography}{}
\makeatletter
\relax
\def\mn@urlcharsother{\let\do\@makeother \do\$\do\&\do\#\do\^\do\_\do\%\do\~}
\def\mn@doi{\begingroup\mn@urlcharsother \@ifnextchar [ {\mn@doi@} {\mn@doi@[]}}
\def\mn@doi@[#1]#2{\def\@tempa{#1}\ifx\@tempa\@empty \href {http://dx.doi.org/#2} {doi:#2}\else \href {http://dx.doi.org/#2} {#1}\fi \endgroup}
\def\mn@eprint#1#2{\mn@eprint@#1:#2::\@nil}
\def\mn@eprint@arXiv#1{\href {http://arxiv.org/abs/#1} {{\tt arXiv:#1}}}
\def\mn@eprint@dblp#1{\href {http://dblp.uni-trier.de/rec/bibtex/#1.xml} {dblp:#1}}
\def\mn@eprint@#1:#2:#3:#4\@nil{\def\@tempa {#1}\def\@tempb {#2}\def\@tempc {#3}\ifx \@tempc \@empty \let \@tempc \@tempb \let \@tempb \@tempa \fi \ifx \@tempb \@empty \def\@tempb {arXiv}\fi \@ifundefined {mn@eprint@\@tempb}{\@tempb:\@tempc}{\expandafter \expandafter \csname mn@eprint@\@tempb\endcsname \expandafter{\@tempc}}}

\bibitem[\protect\citeauthoryear{{Aartsen} et~al.,}{{Aartsen} et~al.}{2019}]{2019EPJC...79..234A}
{Aartsen} M.~G.,  et~al., 2019, \mn@doi [European Physical Journal C] {10.1140/epjc/s10052-019-6680-0}, \href {https://ui.adsabs.harvard.edu/abs/2019EPJC...79..234A} {79, 234}

\bibitem[\protect\citeauthoryear{{Aartsen} et~al.,}{{Aartsen} et~al.}{2020}]{2020PhRvL.124e1103A}
{Aartsen} M.~G.,  et~al., 2020, \mn@doi [\prl] {10.1103/PhysRevLett.124.051103}, \href {https://ui.adsabs.harvard.edu/abs/2020PhRvL.124e1103A} {124, 051103}

\bibitem[\protect\citeauthoryear{{Abdo} et~al.,}{{Abdo} et~al.}{2009}]{2009ApJ...700..597A}
{Abdo} A.~A.,  et~al., 2009, \mn@doi [\apj] {10.1088/0004-637X/700/1/597}, \href {https://ui.adsabs.harvard.edu/\#abs/2009ApJ...700..597A} {700, 597}

\bibitem[\protect\citeauthoryear{{Abdo} et~al.,}{{Abdo} et~al.}{2010}]{2010ApJ...716...30A}
{Abdo} A.~A.,  et~al., 2010, \mn@doi [\apj] {10.1088/0004-637X/716/1/30}, \href {https://ui.adsabs.harvard.edu/abs/2010ApJ...716...30A} {716, 30}

\bibitem[\protect\citeauthoryear{{Abdollahi} et~al.,}{{Abdollahi} et~al.}{2020}]{2020ApJS..247...33A}
{Abdollahi} S.,  et~al., 2020, \mn@doi [\apjs] {10.3847/1538-4365/ab6bcb}, \href {https://ui.adsabs.harvard.edu/abs/2020ApJS..247...33A} {247, 33}

\bibitem[\protect\citeauthoryear{{Abeysekara} et~al.,}{{Abeysekara} et~al.}{2015}]{2015ApJ...815L..22A}
{Abeysekara} A.~U.,  et~al., 2015, \mn@doi [\apjl] {10.1088/2041-8205/815/2/L22}, \href {https://ui.adsabs.harvard.edu/abs/2015ApJ...815L..22A} {815, L22}

\bibitem[\protect\citeauthoryear{{Ahnen} et~al.,}{{Ahnen} et~al.}{2015}]{2015ApJ...815L..23A}
{Ahnen} M.~L.,  et~al., 2015, \mn@doi [\apjl] {10.1088/2041-8205/815/2/L23}, \href {https://ui.adsabs.harvard.edu/abs/2015ApJ...815L..23A} {815, L23}

\bibitem[\protect\citeauthoryear{{Ahnen} et~al.,}{{Ahnen} et~al.}{2016}]{2016A&A...595A..98A}
{Ahnen} M.~L.,  et~al., 2016, \mn@doi [\aap] {10.1051/0004-6361/201629461}, \href {https://ui.adsabs.harvard.edu/abs/2016A&A...595A..98A} {595, A98}

\bibitem[\protect\citeauthoryear{{Ajello} et~al.,}{{Ajello} et~al.}{2017}]{2017ApJS..232...18A}
{Ajello} M.,  et~al., 2017, \mn@doi [\apjs] {10.3847/1538-4365/aa8221}, \href {https://ui.adsabs.harvard.edu/\#abs/2017ApJS..232...18A} {232, 18}

\bibitem[\protect\citeauthoryear{{Ajello} et~al.,}{{Ajello} et~al.}{2020}]{2020ApJ...892..105A}
{Ajello} M.,  et~al., 2020, \mn@doi [\apj] {10.3847/1538-4357/ab791e}, \href {https://ui.adsabs.harvard.edu/abs/2020ApJ...892..105A} {892, 105}

\bibitem[\protect\citeauthoryear{{Albert} et~al.,}{{Albert} et~al.}{2007}]{2007NIMPA.583..494A}
{Albert} J.,  et~al., 2007, \mn@doi [Nuclear Instruments and Methods in Physics Research A] {10.1016/j.nima.2007.09.048}, \href {https://ui.adsabs.harvard.edu/abs/2007NIMPA.583..494A} {583, 494}

\bibitem[\protect\citeauthoryear{{Albert} et~al.,}{{Albert} et~al.}{2020}]{2020ApJ...892...92A}
{Albert} A.,  et~al., 2020, \mn@doi [\apj] {10.3847/1538-4357/ab7afb}, \href {https://ui.adsabs.harvard.edu/abs/2020ApJ...892...92A} {892, 92}

\bibitem[\protect\citeauthoryear{{Aleksi{\'c}} et~al.,}{{Aleksi{\'c}} et~al.}{2011}]{2011ApJ...730L...8A}
{Aleksi{\'c}} J.,  et~al., 2011, \mn@doi [\apjl] {10.1088/2041-8205/730/1/L8}, \href {https://ui.adsabs.harvard.edu/abs/2011ApJ...730L...8A} {730, L8}

\bibitem[\protect\citeauthoryear{{Aleksi{\'c}} et~al.,}{{Aleksi{\'c}} et~al.}{2016}]{2016APh....72...76A}
{Aleksi{\'c}} J.,  et~al., 2016, \mn@doi [Astroparticle Physics] {10.1016/j.astropartphys.2015.02.005}, \href {http://adsabs.harvard.edu/abs/} {72, 76}

\bibitem[\protect\citeauthoryear{{Aller}, {Aller}  \& {Hodge}}{{Aller} et~al.}{1981}]{1981AJ.....86..325A}
{Aller} H.~D.,  {Aller} M.~F.,   {Hodge} P.~E.,  1981, \mn@doi [\aj] {10.1086/112892}, \href {https://ui.adsabs.harvard.edu/abs/1981AJ.....86..325A} {86, 325}

\bibitem[\protect\citeauthoryear{{Ansoldi} et~al.,}{{Ansoldi} et~al.}{2018}]{Ansoldi18}
{Ansoldi} S.,  et~al., 2018, \mn@doi [\apjl] {10.3847/2041-8213/aad083}, \href {https://ui.adsabs.harvard.edu/abs/2018ApJ...863L..10A} {863, L10}

\bibitem[\protect\citeauthoryear{{Arnaud}}{{Arnaud}}{1996}]{1996ASPC..101...17A}
{Arnaud} K.~A.,  1996, in {Jacoby} G.~H.,  {Barnes} J.,  eds,  Astronomical Society of the Pacific Conference Series Vol. 101, Astronomical Data Analysis Software and Systems V. p.~17

\bibitem[\protect\citeauthoryear{{Atwood} et~al.,}{{Atwood} et~al.}{2009}]{2009ApJ...697.1071A}
{Atwood} W.~B.,  et~al., 2009, \mn@doi [\apj] {10.1088/0004-637X/697/2/1071}, \href {https://ui.adsabs.harvard.edu/abs/2009ApJ...697.1071A} {697, 1071}

\bibitem[\protect\citeauthoryear{{Baars}, {Genzel}, {Pauliny-Toth}  \& {Witzel}}{{Baars} et~al.}{1977}]{1977A&A....61...99B}
{Baars} J.~W.~M.,  {Genzel} R.,  {Pauliny-Toth} I.~I.~K.,   {Witzel} A.,  1977, \aap, \href {https://ui.adsabs.harvard.edu/abs/1977A&A....61...99B} {500, 135}

\bibitem[\protect\citeauthoryear{{Ballet}, {Burnett}, {Digel}  \& {Lott}}{{Ballet} et~al.}{2020}]{2020arXiv200511208B}
{Ballet} J.,  {Burnett} T.~H.,  {Digel} S.~W.,   {Lott} B.,  2020, arXiv e-prints, \href {https://ui.adsabs.harvard.edu/abs/2020arXiv200511208B} {p. arXiv:2005.11208}

\bibitem[\protect\citeauthoryear{{Balonek} et~al.,}{{Balonek} et~al.}{2016}]{2016ATel.9259....1B}
{Balonek} T.~J.,  et~al., 2016, The Astronomer's Telegram, \href {https://ui.adsabs.harvard.edu/abs/2016ATel.9259....1B} {9259, 1}

\bibitem[\protect\citeauthoryear{{Band} \& {Grindlay}}{{Band} \& {Grindlay}}{1985}]{1985ApJ...298..128B}
{Band} D.~L.,  {Grindlay} J.~E.,  1985, \mn@doi [\apj] {10.1086/163593}, \href {https://ui.adsabs.harvard.edu/abs/1985ApJ...298..128B} {298, 128}

\bibitem[\protect\citeauthoryear{{Baring}, {B{\"o}ttcher}  \& {Summerlin}}{{Baring} et~al.}{2017}]{2017MNRAS.464.4875B}
{Baring} M.~G.,  {B{\"o}ttcher} M.,   {Summerlin} E.~J.,  2017, \mn@doi [\mnras] {10.1093/mnras/stw2344}, \href {https://ui.adsabs.harvard.edu/abs/2017MNRAS.464.4875B} {464, 4875}

\bibitem[\protect\citeauthoryear{{Barthelmy} et~al.,}{{Barthelmy} et~al.}{2005}]{2005SSRv..120..143B}
{Barthelmy} S.~D.,  et~al., 2005, \mn@doi [\ssr] {10.1007/s11214-005-5096-3}, \href {https://ui.adsabs.harvard.edu/abs/2005SSRv..120..143B} {120, 143}

\bibitem[\protect\citeauthoryear{{Becerra Gonzalez}, {Thompson}  \& {Fermi-LAT Collaboration}}{{Becerra Gonzalez} et~al.}{2016}]{2016ATel.9231....1B}
{Becerra Gonzalez} J.,  {Thompson} D.,   {Fermi-LAT Collaboration} 2016, The Astronomer's Telegram, \href {http://adsabs.harvard.edu/abs/2016ATel.9231....1B} {9231}

\bibitem[\protect\citeauthoryear{{Berge}, {Funk}  \& {Hinton}}{{Berge} et~al.}{2007}]{2007A&A...466.1219B}
{Berge} D.,  {Funk} S.,   {Hinton} J.,  2007, \mn@doi [\aap] {10.1051/0004-6361:20066674}, \href {https://ui.adsabs.harvard.edu/abs/2007A&A...466.1219B} {466, 1219}

\bibitem[\protect\citeauthoryear{{Blanch}, {Sitarek}  \& {Striskovic}}{{Blanch} et~al.}{2022}]{2022ATel15161....1B}
{Blanch} O.,  {Sitarek} J.,   {Striskovic} J.,  2022, The Astronomer's Telegram, \href {https://ui.adsabs.harvard.edu/abs/2022ATel15161....1B} {15161, 1}

\bibitem[\protect\citeauthoryear{{B{\"o}ttcher}}{{B{\"o}ttcher}}{2007}]{2007Ap&SS.309...95B}
{B{\"o}ttcher} M.,  2007, \mn@doi [\apss] {10.1007/s10509-007-9404-0}, \href {http://cdsads.u-strasbg.fr/abs/2007Ap%26SS.309...95B} {309, 95}

\bibitem[\protect\citeauthoryear{{Brindle}, {Hough}, {Bailey}, {Axon}  \& {Hyland}}{{Brindle} et~al.}{1986}]{1986MNRAS.221..739B}
{Brindle} C.,  {Hough} J.~H.,  {Bailey} J.~A.,  {Axon} D.~J.,   {Hyland} A.~R.,  1986, \mn@doi [\mnras] {10.1093/mnras/221.3.739}, \href {https://ui.adsabs.harvard.edu/\#abs/1986MNRAS.221..739B} {221, 739}

\bibitem[\protect\citeauthoryear{{Burbidge} \& {Hewitt}}{{Burbidge} \& {Hewitt}}{1987}]{1987AJ.....93....1B}
{Burbidge} G.,  {Hewitt} A.,  1987, \mn@doi [\aj] {10.1086/114283}, \href {https://ui.adsabs.harvard.edu/abs/1987AJ.....93....1B} {93, 1}

\bibitem[\protect\citeauthoryear{{Burrows} et~al.,}{{Burrows} et~al.}{2005}]{2005SSRv..120..165B}
{Burrows} D.~N.,  et~al., 2005, \mn@doi [\ssr] {10.1007/s11214-005-5097-2}, \href {https://ui.adsabs.harvard.edu/abs/2005SSRv..120..165B} {120, 165}

\bibitem[\protect\citeauthoryear{{Celotti}, {Padovani}  \& {Ghisellini}}{{Celotti} et~al.}{1997}]{1997MNRAS.286..415C}
{Celotti} A.,  {Padovani} P.,   {Ghisellini} G.,  1997, \mn@doi [\mnras] {10.1093/mnras/286.2.415}, \href {https://ui.adsabs.harvard.edu/abs/1997MNRAS.286..415C} {286, 415}

\bibitem[\protect\citeauthoryear{{Cerruti}, {Zech}, {Boisson}  \& {Inoue}}{{Cerruti} et~al.}{2015}]{Cerruti15}
{Cerruti} M.,  {Zech} A.,  {Boisson} C.,   {Inoue} S.,  2015, \mn@doi [\mnras] {10.1093/mnras/stu2691}, \href {https://ui.adsabs.harvard.edu/abs/2015MNRAS.448..910C} {448, 910}

\bibitem[\protect\citeauthoryear{{Cerruti}, {Zech}, {Boisson}, {Emery}, {Inoue}  \& {Lenain}}{{Cerruti} et~al.}{2019}]{Cerruti19}
{Cerruti} M.,  {Zech} A.,  {Boisson} C.,  {Emery} G.,  {Inoue} S.,   {Lenain} J.~P.,  2019, \mn@doi [\mnras] {10.1093/mnrasl/sly210}, \href {https://ui.adsabs.harvard.edu/abs/2019MNRAS.483L..12C} {483, L12}

\bibitem[\protect\citeauthoryear{{Ciprini}, {Becerra Gonz{\'a}lez}, {Pivato}  \& {Thompson}}{{Ciprini} et~al.}{2016}]{2016ATel.9260....1C}
{Ciprini} S.,  {Becerra Gonz{\'a}lez} J.,  {Pivato} G.,   {Thompson} D.~J.,  2016, The Astronomer's Telegram, \href {http://adsabs.harvard.edu/abs/2016ATel.9260....1C} {9260}

\bibitem[\protect\citeauthoryear{{Clarke} \& {Neumayer}}{{Clarke} \& {Neumayer}}{2002}]{2002A&A...383..360C}
{Clarke} D.,  {Neumayer} D.,  2002, \mn@doi [\aap] {10.1051/0004-6361:20011717}, \href {https://ui.adsabs.harvard.edu/abs/2002A&A...383..360C} {383, 360}

\bibitem[\protect\citeauthoryear{{Clements}, {Smith}, {Aller}  \& {Aller}}{{Clements} et~al.}{1995}]{1995AJ....110..529C}
{Clements} S.~D.,  {Smith} A.~G.,  {Aller} H.~D.,   {Aller} M.~F.,  1995, \mn@doi [\aj] {10.1086/117540}, \href {https://ui.adsabs.harvard.edu/\#abs/1995AJ....110..529C} {110, 529}

\bibitem[\protect\citeauthoryear{{Dallacasa}, {Stanghellini}, {Centonza}  \& {Fanti}}{{Dallacasa} et~al.}{2000}]{2000A&A...363..887D}
{Dallacasa} D.,  {Stanghellini} C.,  {Centonza} M.,   {Fanti} R.,  2000, \aap, \href {https://ui.adsabs.harvard.edu/\#abs/2000A&A...363..887D} {363, 887}

\bibitem[\protect\citeauthoryear{{Dermer}, {Finke}, {Krug}  \& {B{\"o}ttcher}}{{Dermer} et~al.}{2009}]{2009ApJ...692...32D}
{Dermer} C.~D.,  {Finke} J.~D.,  {Krug} H.,   {B{\"o}ttcher} M.,  2009, \mn@doi [\apj] {10.1088/0004-637X/692/1/32}, \href {https://ui.adsabs.harvard.edu/abs/2009ApJ...692...32D} {692, 32}

\bibitem[\protect\citeauthoryear{{Dom{\'\i}nguez} et~al.,}{{Dom{\'\i}nguez} et~al.}{2011}]{2011MNRAS.410.2556D}
{Dom{\'\i}nguez} A.,  et~al., 2011, \mn@doi [\mnras] {10.1111/j.1365-2966.2010.17631.x}, \href {https://ui.adsabs.harvard.edu/abs/2011MNRAS.410.2556D} {410, 2556}

\bibitem[\protect\citeauthoryear{{Evans} et~al.,}{{Evans} et~al.}{2009}]{2009MNRAS.397.1177E}
{Evans} P.~A.,  et~al., 2009, \mn@doi [\mnras] {10.1111/j.1365-2966.2009.14913.x}, \href {https://ui.adsabs.harvard.edu/#abs/2009MNRAS.397.1177E} {397, 1177}

\bibitem[\protect\citeauthoryear{{Falomo}, {Pian}  \& {Treves}}{{Falomo} et~al.}{2014}]{2014A&ARv..22...73F}
{Falomo} R.,  {Pian} E.,   {Treves} A.,  2014, \mn@doi [\aapr] {10.1007/s00159-014-0073-z}, \href {https://ui.adsabs.harvard.edu/abs/2014A&ARv..22...73F} {22, 73}

\bibitem[\protect\citeauthoryear{{Feldman} \& {Cousins}}{{Feldman} \& {Cousins}}{1998}]{1998PhRvD..57.3873F}
{Feldman} G.~J.,  {Cousins} R.~D.,  1998, \mn@doi [\prd] {10.1103/PhysRevD.57.3873}, \href {https://ui.adsabs.harvard.edu/abs/1998PhRvD..57.3873F} {57, 3873}

\bibitem[\protect\citeauthoryear{{Fermi Science Support Development Team}}{{Fermi Science Support Development Team}}{2019}]{2019ascl.soft05011F}
{Fermi Science Support Development Team} 2019, {Fermitools: Fermi Science Tools}, Astrophysics Source Code Library, record ascl:1905.011 (\mn@eprint {ascl} {1905.011})

\bibitem[\protect\citeauthoryear{{Filippenko}, {Li}, {Treffers}  \& {Modjaz}}{{Filippenko} et~al.}{2001}]{2001ASPC..246..121F}
{Filippenko} A.~V.,  {Li} W.~D.,  {Treffers} R.~R.,   {Modjaz} M.,  2001, in {Paczynski} B.,  {Chen} W.-P.,   {Lemme} C.,  eds,  Astronomical Society of the Pacific Conference Series Vol. 246, IAU Colloq. 183: Small Telescope Astronomy on Global Scales. p.~121

\bibitem[\protect\citeauthoryear{{Finke}}{{Finke}}{2013}]{2013ApJ...763..134F}
{Finke} J.~D.,  2013, \mn@doi [\apj] {10.1088/0004-637X/763/2/134}, \href {https://ui.adsabs.harvard.edu/abs/2013ApJ...763..134F} {763, 134}

\bibitem[\protect\citeauthoryear{{Finke}, {Dermer}  \& {B{\"o}ttcher}}{{Finke} et~al.}{2008}]{2008ApJ...686..181F}
{Finke} J.~D.,  {Dermer} C.~D.,   {B{\"o}ttcher} M.,  2008, \mn@doi [\apj] {10.1086/590900}, \href {https://ui.adsabs.harvard.edu/abs/2008ApJ...686..181F} {686, 181}

\bibitem[\protect\citeauthoryear{{Fitzpatrick}}{{Fitzpatrick}}{1999}]{1999PASP..111...63F}
{Fitzpatrick} E.~L.,  1999, \mn@doi [\pasp] {10.1086/316293}, \href {https://ui.adsabs.harvard.edu/abs/1999PASP..111...63F} {111, 63}

\bibitem[\protect\citeauthoryear{{Fomin}, {Stepanian}, {Lamb}, {Lewis}, {Punch}  \& {Weekes}}{{Fomin} et~al.}{1994}]{1994APh.....2..137F}
{Fomin} V.~P.,  {Stepanian} A.~A.,  {Lamb} R.~C.,  {Lewis} D.~A.,  {Punch} M.,   {Weekes} T.~C.,  1994, \mn@doi [Astroparticle Physics] {10.1016/0927-6505(94)90036-1}, \href {https://ui.adsabs.harvard.edu/abs/1994APh.....2..137F} {2, 137}

\bibitem[\protect\citeauthoryear{{Furniss} et~al.,}{{Furniss} et~al.}{2015}]{2015ApJ...812...65F}
{Furniss} A.,  et~al., 2015, \mn@doi [\apj] {10.1088/0004-637X/812/1/65}, \href {https://ui.adsabs.harvard.edu/abs/2015ApJ...812...65F} {812, 65}

\bibitem[\protect\citeauthoryear{{Ganeshalingam} et~al.,}{{Ganeshalingam} et~al.}{2010}]{2010ApJS..190..418G}
{Ganeshalingam} M.,  et~al., 2010, \mn@doi [\apjs] {10.1088/0067-0049/190/2/418}, \href {https://ui.adsabs.harvard.edu/abs/2010ApJS..190..418G} {190, 418}

\bibitem[\protect\citeauthoryear{{Gao}, {Fedynitch}, {Winter}  \& {Pohl}}{{Gao} et~al.}{2019}]{Gao19}
{Gao} S.,  {Fedynitch} A.,  {Winter} W.,   {Pohl} M.,  2019, \mn@doi [Nature Astronomy] {10.1038/s41550-018-0610-1}, \href {https://ui.adsabs.harvard.edu/abs/2019NatAs...3...88G} {3, 88}

\bibitem[\protect\citeauthoryear{{Gehrels} et~al.,}{{Gehrels} et~al.}{2004}]{2004ApJ...611.1005G}
{Gehrels} N.,  et~al., 2004, \mn@doi [\apj] {10.1086/422091}, \href {https://ui.adsabs.harvard.edu/abs/2004ApJ...611.1005G} {611, 1005}

\bibitem[\protect\citeauthoryear{{Ghisellini} \& {Tavecchio}}{{Ghisellini} \& {Tavecchio}}{2009}]{2009MNRAS.397..985G}
{Ghisellini} G.,  {Tavecchio} F.,  2009, \mn@doi [\mnras] {10.1111/j.1365-2966.2009.15007.x}, \href {https://ui.adsabs.harvard.edu/abs/2009MNRAS.397..985G} {397, 985}

\bibitem[\protect\citeauthoryear{{Ghisellini}, {Tavecchio}, {Foschini}, {Ghirlanda}, {Maraschi}  \& {Celotti}}{{Ghisellini} et~al.}{2010}]{2010MNRAS.402..497G}
{Ghisellini} G.,  {Tavecchio} F.,  {Foschini} L.,  {Ghirlanda} G.,  {Maraschi} L.,   {Celotti} A.,  2010, \mn@doi [\mnras] {10.1111/j.1365-2966.2009.15898.x}, \href {https://ui.adsabs.harvard.edu/abs/2010MNRAS.402..497G} {402, 497}

\bibitem[\protect\citeauthoryear{{Ghisellini}, {Righi}, {Costamante}  \& {Tavecchio}}{{Ghisellini} et~al.}{2017}]{2017MNRAS.469..255G}
{Ghisellini} G.,  {Righi} C.,  {Costamante} L.,   {Tavecchio} F.,  2017, \mn@doi [\mnras] {10.1093/mnras/stx806}, \href {https://ui.adsabs.harvard.edu/abs/2017MNRAS.469..255G} {469, 255}

\bibitem[\protect\citeauthoryear{{Giannios}, {Uzdensky}  \& {Begelman}}{{Giannios} et~al.}{2009}]{2009MNRAS.395L..29G}
{Giannios} D.,  {Uzdensky} D.~A.,   {Begelman} M.~C.,  2009, \mn@doi [\mnras] {10.1111/j.1745-3933.2009.00635.x}, \href {http://adsabs.harvard.edu/abs/2009MNRAS.395L..29G} {395, L29}

\bibitem[\protect\citeauthoryear{{Gregorini}, {Mantovani}, {Eckart}, {Biermann}, {Witzel}  \& {Kuhr}}{{Gregorini} et~al.}{1984}]{1984AJ.....89..323G}
{Gregorini} L.,  {Mantovani} F.,  {Eckart} A.,  {Biermann} P.,  {Witzel} A.,   {Kuhr} H.,  1984, \mn@doi [\aj] {10.1086/113516}, \href {https://ui.adsabs.harvard.edu/abs/1984AJ.....89..323G} {89, 323}

\bibitem[\protect\citeauthoryear{{H.~E.~S.~S. Collaboration} et~al.,}{{H.~E.~S.~S. Collaboration} et~al.}{2013}]{2013A&A...554A.107H}
{H.~E.~S.~S. Collaboration} et~al., 2013, \mn@doi [\aap] {10.1051/0004-6361/201321135}, \href {https://ui.adsabs.harvard.edu/abs/2013A&A...554A.107H} {554, A107}

\bibitem[\protect\citeauthoryear{{H.~E.~S.~S. Collaboration} et~al.,}{{H.~E.~S.~S. Collaboration} et~al.}{2015}]{2015A&A...573A..31H}
{H.~E.~S.~S. Collaboration} et~al., 2015, \mn@doi [\aap] {10.1051/0004-6361/201321436}, \href {https://ui.adsabs.harvard.edu/abs/2015A&A...573A..31H} {573, A31}

\bibitem[\protect\citeauthoryear{{H.~E.~S.~S. Collaboration} et~al.,}{{H.~E.~S.~S. Collaboration} et~al.}{2017}]{2017A&A...600A..89H}
{H.~E.~S.~S. Collaboration} et~al., 2017, \mn@doi [\aap] {10.1051/0004-6361/201629427}, \href {https://ui.adsabs.harvard.edu/abs/2017A&A...600A..89H} {600, A89}

\bibitem[\protect\citeauthoryear{{H.~E.~S.~S. Collaboration} et~al.,}{{H.~E.~S.~S. Collaboration} et~al.}{2020}]{2020A&A...633A.162H}
{H.~E.~S.~S. Collaboration} et~al., 2020, \mn@doi [\aap] {10.1051/0004-6361/201935906}, \href {https://ui.adsabs.harvard.edu/abs/2020A&A...633A.162H} {633, A162}

\bibitem[\protect\citeauthoryear{{HI4PI Collaboration} et~al.,}{{HI4PI Collaboration} et~al.}{2016}]{2016A&A...594A.116H}
{HI4PI Collaboration} et~al., 2016, \mn@doi [\aap] {10.1051/0004-6361/201629178}, \href {https://ui.adsabs.harvard.edu/abs/2016A&A...594A.116H} {594, A116}

\bibitem[\protect\citeauthoryear{Hauser, M{\"o}llenhoff, P{\"u}hlhofer, Wagner, Hagen  \& Knoll}{Hauser et~al.}{2004}]{hauser2004atom}
Hauser M.,  M{\"o}llenhoff C.,  P{\"u}hlhofer G.,  Wagner S.,  Hagen H.-J.,   Knoll M.,  2004, Astronomische Nachrichten, 325, 659

\bibitem[\protect\citeauthoryear{{Healey}, {Romani}, {Taylor}, {Sadler}, {Ricci}, {Murphy}, {Ulvestad}  \& {Winn}}{{Healey} et~al.}{2007}]{2007ApJS..171...61H}
{Healey} S.~E.,  {Romani} R.~W.,  {Taylor} G.~B.,  {Sadler} E.~M.,  {Ricci} R.,  {Murphy} T.,  {Ulvestad} J.~S.,   {Winn} J.~N.,  2007, \mn@doi [\apjs] {10.1086/513742}, \href {https://ui.adsabs.harvard.edu/\#abs/2007ApJS..171...61H} {171, 61}

\bibitem[\protect\citeauthoryear{{Hervet}, {Boisson}  \& {Sol}}{{Hervet} et~al.}{2015}]{2015A&A...578A..69H}
{Hervet} O.,  {Boisson} C.,   {Sol} H.,  2015, \mn@doi [\aap] {10.1051/0004-6361/201425330}, \href {https://ui.adsabs.harvard.edu/abs/2015A&A...578A..69H} {578, A69}

\bibitem[\protect\citeauthoryear{{Holler} et~al.,}{{Holler} et~al.}{2015}]{2015arXiv150902902H}
{Holler} M.,  et~al., 2015, \mn@doi [arXiv e-prints] {10.48550/arXiv.1509.02902}, \href {https://ui.adsabs.harvard.edu/abs/2015arXiv150902902H} {p. arXiv:1509.02902}

\bibitem[\protect\citeauthoryear{{Iguchi}, {Fujisawa}, {Kameno}, {Inoue}, {Shen}, {Hirotani}  \& {Miyoshi}}{{Iguchi} et~al.}{2000}]{2000PASJ...52.1037I}
{Iguchi} S.,  {Fujisawa} K.,  {Kameno} S.,  {Inoue} M.,  {Shen} Z.-Q.,  {Hirotani} K.,   {Miyoshi} M.,  2000, \mn@doi [\pasj] {10.1093/pasj/52.6.1037}, \href {https://ui.adsabs.harvard.edu/abs/2000PASJ...52.1037I} {52, 1037}

\bibitem[\protect\citeauthoryear{{Jorstad} et~al.,}{{Jorstad} et~al.}{2017}]{2017ApJ...846...98J}
{Jorstad} S.~G.,  et~al., 2017, \mn@doi [\apj] {10.3847/1538-4357/aa8407}, \href {http://adsabs.harvard.edu/abs/2017ApJ...846...98J} {846, 98}

\bibitem[\protect\citeauthoryear{{Kaufmann}, {Wagner}  \& {Tibolla}}{{Kaufmann} et~al.}{2013}]{2013ApJ...776...68K}
{Kaufmann} S.,  {Wagner} S.~J.,   {Tibolla} O.,  2013, \mn@doi [\apj] {10.1088/0004-637X/776/2/68}, \href {https://ui.adsabs.harvard.edu/abs/2013ApJ...776...68K} {776, 68}

\bibitem[\protect\citeauthoryear{{Keivani} et~al.,}{{Keivani} et~al.}{2018}]{Keivani18}
{Keivani} A.,  et~al., 2018, \mn@doi [\apj] {10.3847/1538-4357/aad59a}, \href {https://ui.adsabs.harvard.edu/abs/2018ApJ...864...84K} {864, 84}

\bibitem[\protect\citeauthoryear{{Kim} et~al.,}{{Kim} et~al.}{2018}]{2018MNRAS.480.2324K}
{Kim} D.-W.,  et~al., 2018, \mn@doi [\mnras] {10.1093/mnras/sty1993}, \href {https://ui.adsabs.harvard.edu/abs/2018MNRAS.480.2324K} {480, 2324}

\bibitem[\protect\citeauthoryear{{Konigl}}{{Konigl}}{1981}]{1981ApJ...243..700K}
{Konigl} A.,  1981, \mn@doi [\apj] {10.1086/158638}, \href {https://ui.adsabs.harvard.edu/abs/1981ApJ...243..700K} {243, 700}

\bibitem[\protect\citeauthoryear{{Kovalev}, {Nizhelsky}, {Kovalev}, {Berlin}, {Zhekanis}, {Mingaliev}  \& {Bogdantsov}}{{Kovalev} et~al.}{1999}]{1999A&AS..139..545K}
{Kovalev} Y.~Y.,  {Nizhelsky} N.~A.,  {Kovalev} Y.~A.,  {Berlin} A.~B.,  {Zhekanis} G.~V.,  {Mingaliev} M.~G.,   {Bogdantsov} A.~V.,  1999, \mn@doi [\aaps] {10.1051/aas:1999406}, \href {https://ui.adsabs.harvard.edu/\#abs/1999A&AS..139..545K} {139, 545}

\bibitem[\protect\citeauthoryear{{Larionov} et~al.,}{{Larionov} et~al.}{2008}]{2008A&A...492..389L}
{Larionov} V.~M.,  et~al., 2008, \mn@doi [\aap] {10.1051/0004-6361:200810937}, \href {https://ui.adsabs.harvard.edu/abs/2008A&A...492..389L} {492, 389}

\bibitem[\protect\citeauthoryear{{Ledden} \& {O'Dell}}{{Ledden} \& {O'Dell}}{1985}]{1985ApJ...298..630L}
{Ledden} J.~E.,  {O'Dell} S.~L.,  1985, \mn@doi [\apj] {10.1086/163647}, \href {https://ui.adsabs.harvard.edu/abs/1985ApJ...298..630L} {298, 630}

\bibitem[\protect\citeauthoryear{Li \& Ma}{Li \& Ma}{1983}]{li1983analysis}
Li T.-P.,  Ma Y.-Q.,  1983, \apj, 272, 317

\bibitem[\protect\citeauthoryear{{Li}, {Cai}, {Yang}, {Luo}, {Yan}, {He}  \& {Wang}}{{Li} et~al.}{2021}]{2021MNRAS.506.1540L}
{Li} X.-P.,  {Cai} Y.,  {Yang} H.-T.,  {Luo} Y.-H.,  {Yan} Y.,  {He} J.-Y.,   {Wang} L.-S.,  2021, \mn@doi [\mnras] {10.1093/mnras/stab1834}, \href {https://ui.adsabs.harvard.edu/abs/2021MNRAS.506.1540L} {506, 1540}

\bibitem[\protect\citeauthoryear{{Lin} \& {Fan}}{{Lin} \& {Fan}}{2018}]{2018RAA....18..120L}
{Lin} C.,  {Fan} J.-H.,  2018, \mn@doi [Research in Astronomy and Astrophysics] {10.1088/1674-4527/18/10/120}, \href {https://ui.adsabs.harvard.edu/abs/2018RAA....18..120L} {18, 120}

\bibitem[\protect\citeauthoryear{{Liodakis}, {Hovatta}, {Huppenkothen}, {Kiehlmann}, {Max-Moerbeck}  \& {Readhead}}{{Liodakis} et~al.}{2018}]{2018ApJ...866..137L}
{Liodakis} I.,  {Hovatta} T.,  {Huppenkothen} D.,  {Kiehlmann} S.,  {Max-Moerbeck} W.,   {Readhead} A. C.~S.,  2018, \mn@doi [\apj] {10.3847/1538-4357/aae2b7}, \href {https://ui.adsabs.harvard.edu/abs/2018ApJ...866..137L} {866, 137}

\bibitem[\protect\citeauthoryear{{MAGIC Collaboration} et~al.,}{{MAGIC Collaboration} et~al.}{2008}]{2008Sci...320.1752M}
{MAGIC Collaboration} et~al., 2008, \mn@doi [Science] {10.1126/science.1157087}, \href {https://ui.adsabs.harvard.edu/abs/2008Sci...320.1752M} {320, 1752}

\bibitem[\protect\citeauthoryear{{MAGIC Collaboration} et~al.,}{{MAGIC Collaboration} et~al.}{2018}]{2018A&A...617A..30M}
{MAGIC Collaboration} et~al., 2018, \mn@doi [\aap] {10.1051/0004-6361/201832624}, \href {https://ui.adsabs.harvard.edu/abs/2018A&A...617A..30M} {617, A30}

\bibitem[\protect\citeauthoryear{{Maraschi} \& {Tavecchio}}{{Maraschi} \& {Tavecchio}}{2003}]{2003ApJ...593..667M}
{Maraschi} L.,  {Tavecchio} F.,  2003, \mn@doi [\apj] {10.1086/342118}, \href {https://ui.adsabs.harvard.edu/abs/2003ApJ...593..667M} {593, 667}

\bibitem[\protect\citeauthoryear{{Marscher}}{{Marscher}}{2014}]{2014ApJ...780...87M}
{Marscher} A.~P.,  2014, \mn@doi [\apj] {10.1088/0004-637X/780/1/87}, \href {https://ui.adsabs.harvard.edu/abs/2014ApJ...780...87M} {780, 87}

\bibitem[\protect\citeauthoryear{{Marscher} \& {Gear}}{{Marscher} \& {Gear}}{1985}]{1985ApJ...298..114M}
{Marscher} A.~P.,  {Gear} W.~K.,  1985, \mn@doi [\apj] {10.1086/163592}, \href {https://ui.adsabs.harvard.edu/abs/1985ApJ...298..114M} {298, 114}

\bibitem[\protect\citeauthoryear{{Massaro}, {Perri}, {Giommi}, {Nesci}  \& {Verrecchia}}{{Massaro} et~al.}{2004}]{2004A&A...422..103M}
{Massaro} E.,  {Perri} M.,  {Giommi} P.,  {Nesci} R.,   {Verrecchia} F.,  2004, \mn@doi [\aap] {10.1051/0004-6361:20047148}, \href {https://ui.adsabs.harvard.edu/abs/2004A&A...422..103M} {422, 103}

\bibitem[\protect\citeauthoryear{{Massaro}, {Tramacere}, {Perri}, {Giommi}  \& {Tosti}}{{Massaro} et~al.}{2006}]{2006A&A...448..861M}
{Massaro} E.,  {Tramacere} A.,  {Perri} M.,  {Giommi} P.,   {Tosti} G.,  2006, \mn@doi [\aap] {10.1051/0004-6361:20053644}, \href {https://ui.adsabs.harvard.edu/abs/2006A&A...448..861M} {448, 861}

\bibitem[\protect\citeauthoryear{{Miller}, {French}  \& {Hawley}}{{Miller} et~al.}{1978}]{1978ApJ...219L..85M}
{Miller} J.~S.,  {French} H.~B.,   {Hawley} S.~A.,  1978, \mn@doi [\apj] {10.1086/182612}, \href {https://ui.adsabs.harvard.edu/\#abs/1978ApJ...219L..85M} {219, L85}

\bibitem[\protect\citeauthoryear{{Mirzoyan}}{{Mirzoyan}}{2016}]{2016ATel.9267....1M}
{Mirzoyan} R.,  2016, The Astronomer's Telegram, \href {http://adsabs.harvard.edu/abs/2016ATel.9267....1M} {9267}

\bibitem[\protect\citeauthoryear{{Mirzoyan}}{{Mirzoyan}}{2017}]{2017ATel11061....1M}
{Mirzoyan} R.,  2017, The Astronomer's Telegram, \href {https://ui.adsabs.harvard.edu/abs/2017ATel11061....1M} {11061, 1}

\bibitem[\protect\citeauthoryear{{Mirzoyan}}{{Mirzoyan}}{2020}]{2020ATel13412....1M}
{Mirzoyan} R.,  2020, The Astronomer's Telegram, \href {https://ui.adsabs.harvard.edu/abs/2020ATel13412....1M} {13412, 1}

\bibitem[\protect\citeauthoryear{{Moretti} et~al.,}{{Moretti} et~al.}{2005}]{2005SPIE.5898..360M}
{Moretti} A.,  et~al., 2005, in {Siegmund} O. H.~W.,  ed.,  Society of Photo-Optical Instrumentation Engineers (SPIE) Conference Series Vol. 5898, UV, X-Ray, and Gamma-Ray Space Instrumentation for Astronomy XIV. pp 360--368, \mn@doi{10.1117/12.617164}

\bibitem[\protect\citeauthoryear{{Nalewajko}, {Sikora}  \& {Begelman}}{{Nalewajko} et~al.}{2014}]{2014ApJ...796L...5N}
{Nalewajko} K.,  {Sikora} M.,   {Begelman} M.~C.,  2014, \mn@doi [\apjl] {10.1088/2041-8205/796/1/L5}, \href {https://ui.adsabs.harvard.edu/abs/2014ApJ...796L...5N} {796, L5}

\bibitem[\protect\citeauthoryear{{Nigro}, {Sitarek}, {Gliwny}, {Sanchez}, {Tramacere}  \& {Craig}}{{Nigro} et~al.}{2022}]{2022A&A...660A..18N}
{Nigro} C.,  {Sitarek} J.,  {Gliwny} P.,  {Sanchez} D.,  {Tramacere} A.,   {Craig} M.,  2022, \mn@doi [\aap] {10.1051/0004-6361/202142000}, \href {https://ui.adsabs.harvard.edu/abs/2022A&A...660A..18N} {660, A18}

\bibitem[\protect\citeauthoryear{{Nilsson} et~al.,}{{Nilsson} et~al.}{2018}]{2018A&A...620A.185N}
{Nilsson} K.,  et~al., 2018, \mn@doi [\aap] {10.1051/0004-6361/201833621}, \href {https://ui.adsabs.harvard.edu/abs/2018A&A...620A.185N} {620, A185}

\bibitem[\protect\citeauthoryear{{Padovani} \& {Giommi}}{{Padovani} \& {Giommi}}{1995}]{1995ApJ...444..567P}
{Padovani} P.,  {Giommi} P.,  1995, \mn@doi [\apj] {10.1086/175631}, \href {https://ui.adsabs.harvard.edu/abs/1995ApJ...444..567P} {444, 567}

\bibitem[\protect\citeauthoryear{{Parsons} \& {Hinton}}{{Parsons} \& {Hinton}}{2014}]{2014APh....56...26P}
{Parsons} R.~D.,  {Hinton} J.~A.,  2014, \mn@doi [Astroparticle Physics] {10.1016/j.astropartphys.2014.03.002}, \href {https://ui.adsabs.harvard.edu/abs/2014APh....56...26P} {56, 26}

\bibitem[\protect\citeauthoryear{{Petropoulou}, {Vasilopoulos}  \& {Giannios}}{{Petropoulou} et~al.}{2017}]{2017MNRAS.464.2213P}
{Petropoulou} M.,  {Vasilopoulos} G.,   {Giannios} D.,  2017, \mn@doi [\mnras] {10.1093/mnras/stw2453}, \href {https://ui.adsabs.harvard.edu/abs/2017MNRAS.464.2213P} {464, 2213}

\bibitem[\protect\citeauthoryear{{Pica}, {Smith}, {Webb}, {Leacock}, {Clements}  \& {Gombola}}{{Pica} et~al.}{1988}]{1988AJ.....96.1215P}
{Pica} A.~J.,  {Smith} A.~G.,  {Webb} J.~R.,  {Leacock} R.~J.,  {Clements} S.,   {Gombola} P.~P.,  1988, \mn@doi [\aj] {10.1086/114875}, \href {https://ui.adsabs.harvard.edu/abs/1988AJ.....96.1215P} {96, 1215}

\bibitem[\protect\citeauthoryear{{Piron} et~al.,}{{Piron} et~al.}{2001}]{2001A&A...374..895P}
{Piron} F.,  et~al., 2001, \mn@doi [\aap] {10.1051/0004-6361:20010798}, \href {https://ui.adsabs.harvard.edu/abs/2001A&A...374..895P} {374, 895}

\bibitem[\protect\citeauthoryear{{Poole}, {Breeveld}, {Page}  \& {et al.}}{{Poole} et~al.}{2008}]{poole08}
{Poole} T.~S.,  {Breeveld} A.~A.,  {Page} M.~J.,   {et al.} 2008, \mnras, 383, 627

\bibitem[\protect\citeauthoryear{{Potter} \& {Cotter}}{{Potter} \& {Cotter}}{2013}]{2013MNRAS.436..304P}
{Potter} W.~J.,  {Cotter} G.,  2013, \mn@doi [\mnras] {10.1093/mnras/stt1569}, \href {https://ui.adsabs.harvard.edu/abs/2013MNRAS.436..304P} {436, 304}

\bibitem[\protect\citeauthoryear{{Pursimo}, {Nilsson}, {Takalo}, {Sillanp{\"a}{\"a}}, {Heidt}  \& {Pietil{\"a}}}{{Pursimo} et~al.}{2002}]{2002A&A...381..810P}
{Pursimo} T.,  {Nilsson} K.,  {Takalo} L.~O.,  {Sillanp{\"a}{\"a}} A.,  {Heidt} J.,   {Pietil{\"a}} H.,  2002, \mn@doi [\aap] {10.1051/0004-6361:20011564}, \href {https://ui.adsabs.harvard.edu/abs/2002A&A...381..810P} {381, 810}

\bibitem[\protect\citeauthoryear{{Reimer}, {B{\"o}ttcher}  \& {Buson}}{{Reimer} et~al.}{2019}]{2019ApJ...881...46R}
{Reimer} A.,  {B{\"o}ttcher} M.,   {Buson} S.,  2019, \mn@doi [\apj] {10.3847/1538-4357/ab2bff}, \href {https://ui.adsabs.harvard.edu/abs/2019ApJ...881...46R} {881, 46}

\bibitem[\protect\citeauthoryear{{Reuter} et~al.,}{{Reuter} et~al.}{1997}]{1997A&AS..122..271R}
{Reuter} H.~P.,  et~al., 1997, \mn@doi [\aaps] {10.1051/aas:1997333}, \href {https://ui.adsabs.harvard.edu/\#abs/1997A&AS..122..271R} {122, 271}

\bibitem[\protect\citeauthoryear{{Richards} et~al.,}{{Richards} et~al.}{2011}]{2011ApJS..194...29R}
{Richards} J.~L.,  et~al., 2011, \mn@doi [\apjs] {10.1088/0067-0049/194/2/29}, \href {http://cdsads.u-strasbg.fr/abs/2011ApJS..194...29R} {194, 29}

\bibitem[\protect\citeauthoryear{{Rieger}, {Bosch-Ramon}  \& {Duffy}}{{Rieger} et~al.}{2007}]{Rieger07}
{Rieger} F.~M.,  {Bosch-Ramon} V.,   {Duffy} P.,  2007, \mn@doi [\apss] {10.1007/s10509-007-9466-z}, \href {https://ui.adsabs.harvard.edu/abs/2007Ap&SS.309..119R} {309, 119}

\bibitem[\protect\citeauthoryear{{Roming} et~al.,}{{Roming} et~al.}{2005}]{2005SSRv..120...95R}
{Roming} P. W.~A.,  et~al., 2005, \mn@doi [\ssr] {10.1007/s11214-005-5095-4}, \href {https://ui.adsabs.harvard.edu/abs/2005SSRv..120...95R} {120, 95}

\bibitem[\protect\citeauthoryear{{Roychowdhury}, {Meyer}, {Georganopoulos}, {Breiding}  \& {Petropoulou}}{{Roychowdhury} et~al.}{2022}]{2022ApJ...924...57R}
{Roychowdhury} A.,  {Meyer} E.~T.,  {Georganopoulos} M.,  {Breiding} P.,   {Petropoulou} M.,  2022, \mn@doi [\apj] {10.3847/1538-4357/ac34f1}, \href {https://ui.adsabs.harvard.edu/abs/2022ApJ...924...57R} {924, 57}

\bibitem[\protect\citeauthoryear{{Sambruna}, {Ghisellini}, {Hooper}, {Kollgaard}, {Pesce}  \& {Urry}}{{Sambruna} et~al.}{1999}]{1999ApJ...515..140S}
{Sambruna} R.~M.,  {Ghisellini} G.,  {Hooper} E.,  {Kollgaard} R.~I.,  {Pesce} J.~E.,   {Urry} C.~M.,  1999, \mn@doi [\apj] {10.1086/307005}, \href {https://ui.adsabs.harvard.edu/abs/1999ApJ...515..140S} {515, 140}

\bibitem[\protect\citeauthoryear{{Sanchez} et~al.,}{{Sanchez} et~al.}{2015}]{2015MNRAS.454.3229S}
{Sanchez} D.~A.,  et~al., 2015, \mn@doi [\mnras] {10.1093/mnras/stv2151}, \href {https://ui.adsabs.harvard.edu/abs/2015MNRAS.454.3229S} {454, 3229}

\bibitem[\protect\citeauthoryear{{Scargle}, {Norris}, {Jackson}  \& {Chiang}}{{Scargle} et~al.}{2013}]{2013ApJ...764..167S}
{Scargle} J.~D.,  {Norris} J.~P.,  {Jackson} B.,   {Chiang} J.,  2013, \mn@doi [\apj] {10.1088/0004-637X/764/2/167}, \href {https://ui.adsabs.harvard.edu/abs/2013ApJ...764..167S} {764, 167}

\bibitem[\protect\citeauthoryear{{Scarpa} \& {Falomo}}{{Scarpa} \& {Falomo}}{1997}]{1997A&A...325..109S}
{Scarpa} R.,  {Falomo} R.,  1997, \aap, \href {https://ui.adsabs.harvard.edu/abs/1997A&A...325..109S} {325, 109}

\bibitem[\protect\citeauthoryear{{Schlafly} \& {Finkbeiner}}{{Schlafly} \& {Finkbeiner}}{2011}]{2011ApJ...737..103S}
{Schlafly} E.~F.,  {Finkbeiner} D.~P.,  2011, \mn@doi [\apj] {10.1088/0004-637X/737/2/103}, \href {https://ui.adsabs.harvard.edu/abs/2011ApJ...737..103S} {737, 103}

\bibitem[\protect\citeauthoryear{{Sch{\"u}ssler} et~al.,}{{Sch{\"u}ssler} et~al.}{2017}]{2017ICRC...35..652S}
{Sch{\"u}ssler} F.,  et~al., 2017, in 35th International Cosmic Ray Conference (ICRC2017). p.~652 (\mn@eprint {arXiv} {1708.01083})

\bibitem[\protect\citeauthoryear{{Sikora}, {Stawarz}, {Moderski}, {Nalewajko}  \& {Madejski}}{{Sikora} et~al.}{2009}]{Sikora09}
{Sikora} M.,  {Stawarz} {\L}.,  {Moderski} R.,  {Nalewajko} K.,   {Madejski} G.~M.,  2009, \mn@doi [\apj] {10.1088/0004-637X/704/1/38}, \href {https://ui.adsabs.harvard.edu/abs/2009ApJ...704...38S} {704, 38}

\bibitem[\protect\citeauthoryear{{Steele} et~al.,}{{Steele} et~al.}{2004}]{2004SPIE.5489..679S}
{Steele} I.~A.,  et~al., 2004, in {Oschmann} Jacobus~M. J.,  ed.,  Society of Photo-Optical Instrumentation Engineers (SPIE) Conference Series Vol. 5489, \procspie. pp 679--692, \mn@doi{10.1117/12.551456}

\bibitem[\protect\citeauthoryear{{Stickel}, {Fried}  \& {Kuehr}}{{Stickel} et~al.}{1988}]{1988A&A...191L..16S}
{Stickel} M.,  {Fried} J.~W.,   {Kuehr} H.,  1988, \aap, \href {https://ui.adsabs.harvard.edu/abs/1988A&A...191L..16S} {191, L16}

\bibitem[\protect\citeauthoryear{{Stickel}, {Padovani}, {Urry}, {Fried}  \& {Kuehr}}{{Stickel} et~al.}{1991}]{1991ApJ...374..431S}
{Stickel} M.,  {Padovani} P.,  {Urry} C.~M.,  {Fried} J.~W.,   {Kuehr} H.,  1991, \mn@doi [\apj] {10.1086/170133}, \href {https://ui.adsabs.harvard.edu/abs/1991ApJ...374..431S} {374, 431}

\bibitem[\protect\citeauthoryear{{Stocke}, {Morris}, {Gioia}, {Maccacaro}, {Schild}, {Wolter}, {Fleming}  \& {Henry}}{{Stocke} et~al.}{1991}]{1991ApJS...76..813S}
{Stocke} J.~T.,  {Morris} S.~L.,  {Gioia} I.~M.,  {Maccacaro} T.,  {Schild} R.,  {Wolter} A.,  {Fleming} T.~A.,   {Henry} J.~P.,  1991, \mn@doi [\apjs] {10.1086/191582}, \href {https://ui.adsabs.harvard.edu/abs/1991ApJS...76..813S} {76, 813}

\bibitem[\protect\citeauthoryear{{Tavecchio} \& {Ghisellini}}{{Tavecchio} \& {Ghisellini}}{2016}]{2016MNRAS.456.2374T}
{Tavecchio} F.,  {Ghisellini} G.,  2016, \mn@doi [\mnras] {10.1093/mnras/stv2790}, \href {https://ui.adsabs.harvard.edu/abs/2016MNRAS.456.2374T} {456, 2374}

\bibitem[\protect\citeauthoryear{{Tavecchio}, {Maraschi}  \& {Ghisellini}}{{Tavecchio} et~al.}{1998}]{1998ApJ...509..608T}
{Tavecchio} F.,  {Maraschi} L.,   {Ghisellini} G.,  1998, \mn@doi [\apj] {10.1086/306526}, \href {http://adsabs.harvard.edu/abs/1998ApJ...509..608T} {509, 608}

\bibitem[\protect\citeauthoryear{{Teraesranta} et~al.,}{{Teraesranta} et~al.}{1998}]{1998A&AS..132..305T}
{Teraesranta} H.,  et~al., 1998, \mn@doi [\aaps] {10.1051/aas:1998297}, \href {https://ui.adsabs.harvard.edu/\#abs/1998A&AS..132..305T} {132, 305}

\bibitem[\protect\citeauthoryear{{Tonry} et~al.,}{{Tonry} et~al.}{2012}]{2012ApJ...750...99T}
{Tonry} J.~L.,  et~al., 2012, \mn@doi [\apj] {10.1088/0004-637X/750/2/99}, \href {https://ui.adsabs.harvard.edu/abs/2012ApJ...750...99T} {750, 99}

\bibitem[\protect\citeauthoryear{{Torniainen}, {Tornikoski}, {Ter{\"a}sranta}, {Aller}  \& {Aller}}{{Torniainen} et~al.}{2005}]{2005A&A...435..839T}
{Torniainen} I.,  {Tornikoski} M.,  {Ter{\"a}sranta} H.,  {Aller} M.~F.,   {Aller} H.~D.,  2005, \mn@doi [\aap] {10.1051/0004-6361:20041886}, \href {https://ui.adsabs.harvard.edu/\#abs/2005A&A...435..839T} {435, 839}

\bibitem[\protect\citeauthoryear{{Urry} \& {Padovani}}{{Urry} \& {Padovani}}{1995}]{1995PASP..107..803U}
{Urry} C.~M.,  {Padovani} P.,  1995, \mn@doi [\pasp] {10.1086/133630}, \href {https://ui.adsabs.harvard.edu/abs/1995PASP..107..803U} {107, 803}

\bibitem[\protect\citeauthoryear{{Urry}, {Sambruna}, {Worrall}, {Kollgaard}, {Feigelson}, {Perlman}  \& {Stocke}}{{Urry} et~al.}{1996}]{1996ApJ...463..424U}
{Urry} C.~M.,  {Sambruna} R.~M.,  {Worrall} D.~M.,  {Kollgaard} R.~I.,  {Feigelson} E.~D.,  {Perlman} E.~S.,   {Stocke} J.~T.,  1996, \mn@doi [\apj] {10.1086/177259}, \href {https://ui.adsabs.harvard.edu/\#abs/1996ApJ...463..424U} {463, 424}

\bibitem[\protect\citeauthoryear{{Vaughan}, {Edelson}, {Warwick}  \& {Uttley}}{{Vaughan} et~al.}{2003}]{2003MNRAS.345.1271V}
{Vaughan} S.,  {Edelson} R.,  {Warwick} R.~S.,   {Uttley} P.,  2003, \mn@doi [\mnras] {10.1046/j.1365-2966.2003.07042.x}, \href {https://ui.adsabs.harvard.edu/abs/2003MNRAS.345.1271V} {345, 1271}

\bibitem[\protect\citeauthoryear{{Vermeulen}, {Ogle}, {Tran}, {Browne}, {Cohen}, {Readhead}, {Taylor}  \& {Goodrich}}{{Vermeulen} et~al.}{1995}]{1995ApJ...452L...5V}
{Vermeulen} R.~C.,  {Ogle} P.~M.,  {Tran} H.~D.,  {Browne} I.~W.~A.,  {Cohen} M.~H.,  {Readhead} A.~C.~S.,  {Taylor} G.~B.,   {Goodrich} R.~W.,  1995, \mn@doi [\apjl] {10.1086/309716}, \href {https://ui.adsabs.harvard.edu/abs/1995ApJ...452L...5V} {452, L5}

\bibitem[\protect\citeauthoryear{{Vestergaard} \& {Peterson}}{{Vestergaard} \& {Peterson}}{2006}]{2006ApJ...641..689V}
{Vestergaard} M.,  {Peterson} B.~M.,  2006, \mn@doi [\apj] {10.1086/500572}, \href {https://ui.adsabs.harvard.edu/abs/2006ApJ...641..689V} {641, 689}

\bibitem[\protect\citeauthoryear{{Wakely} \& {Horan}}{{Wakely} \& {Horan}}{2008}]{2008ICRC....3.1341W}
{Wakely} S.~P.,  {Horan} D.,  2008, International Cosmic Ray Conference, \href {http://adsabs.harvard.edu/abs/2008ICRC....3.1341W} {3, 1341}

\bibitem[\protect\citeauthoryear{{Weaver} et~al.,}{{Weaver} et~al.}{2022}]{2022ApJS..260...12W}
{Weaver} Z.~R.,  et~al., 2022, \mn@doi [\apjs] {10.3847/1538-4365/ac589c}, \href {https://ui.adsabs.harvard.edu/abs/2022ApJS..260...12W} {260, 12}

\bibitem[\protect\citeauthoryear{{Wilms}, {Allen}  \& {McCray}}{{Wilms} et~al.}{2000}]{2000ApJ...542..914W}
{Wilms} J.,  {Allen} A.,   {McCray} R.,  2000, \mn@doi [\apj] {10.1086/317016}, \href {https://ui.adsabs.harvard.edu/abs/2000ApJ...542..914W} {542, 914}

\bibitem[\protect\citeauthoryear{{Zacharias} \& {Wagner}}{{Zacharias} \& {Wagner}}{2016}]{2016A&A...588A.110Z}
{Zacharias} M.,  {Wagner} S.~J.,  2016, \mn@doi [\aap] {10.1051/0004-6361/201526698}, \href {https://ui.adsabs.harvard.edu/abs/2016A&A...588A.110Z} {588, A110}

\bibitem[\protect\citeauthoryear{Zanin, Carmona, Sitarek  et~al.}{Zanin et~al.}{2013}]{zanin2013}
Zanin R.,  Carmona E.,  Sitarek J.,   et~al., 2013, {\it Proc. of the 33th International Cosmic Ray Conference (ICRC)} 2-9 July 2013, Rio de Janeiro, Brazil, 773

\bibitem[\protect\citeauthoryear{{Zdziarski} \& {Boettcher}}{{Zdziarski} \& {Boettcher}}{2015}]{Zdziarski15}
{Zdziarski} A.~A.,  {Boettcher} M.,  2015, \mn@doi [\mnras] {10.1093/mnrasl/slv039}, \href {https://ui.adsabs.harvard.edu/abs/2015MNRAS.450L..21Z} {450, L21}

\bibitem[\protect\citeauthoryear{{de Naurois} \& {Rolland}}{{de Naurois} \& {Rolland}}{2009}]{2009APh....32..231D}
{de Naurois} M.,  {Rolland} L.,  2009, \mn@doi [Astroparticle Physics] {10.1016/j.astropartphys.2009.09.001}, \href {https://ui.adsabs.harvard.edu/abs/2009APh....32..231D} {32, 231}

\makeatother
\end{thebibliography}



\appendix
\section{VLBA study: jet evolution}
\label{sec:appendix_vlba}
\begin{figure*}
   \centering
   \begin{subfigure}[t]{19cm}
          \centering
          \includegraphics[width=19cm]{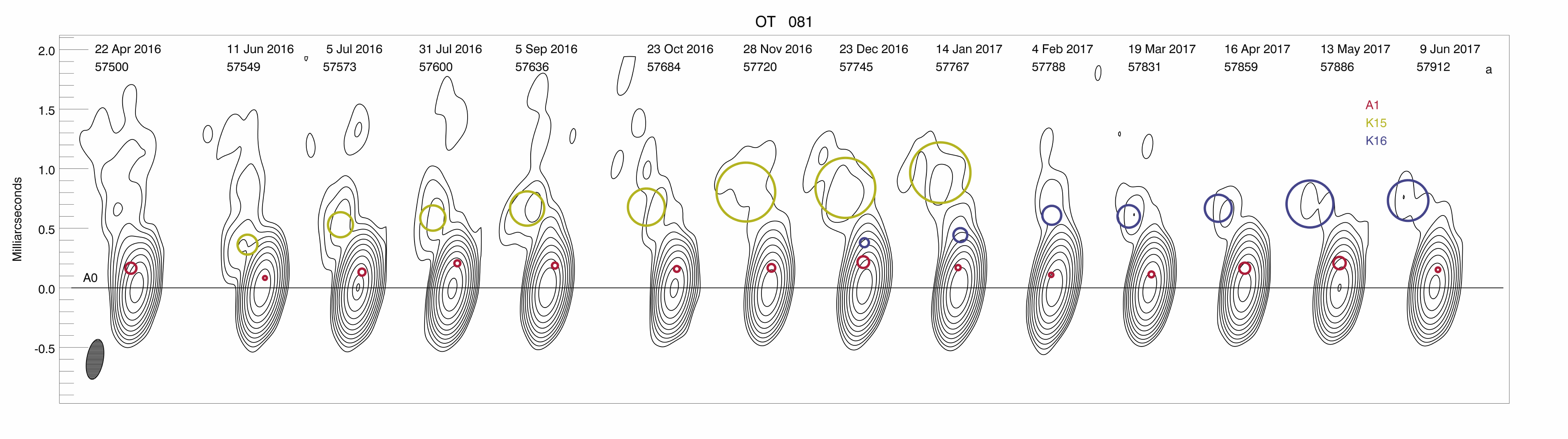}
          \caption{Total intensity images shown by contours, with the global total intensity peak 4389~mJy/beam and levels of 0.15, 0.3, 0.6,..,76.8, and 98\% of the peak; the black horizontal line indicates the position of the core, A0, while the red, green, and blue circles mark the positions of knots A1, K15, and K16, respectively, according to model fits.
          \label{fig:1749_vlba2016a}}
    \end{subfigure}
    \begin{subfigure}[t]{19cm}
           \centering
           \includegraphics[width=19 cm]{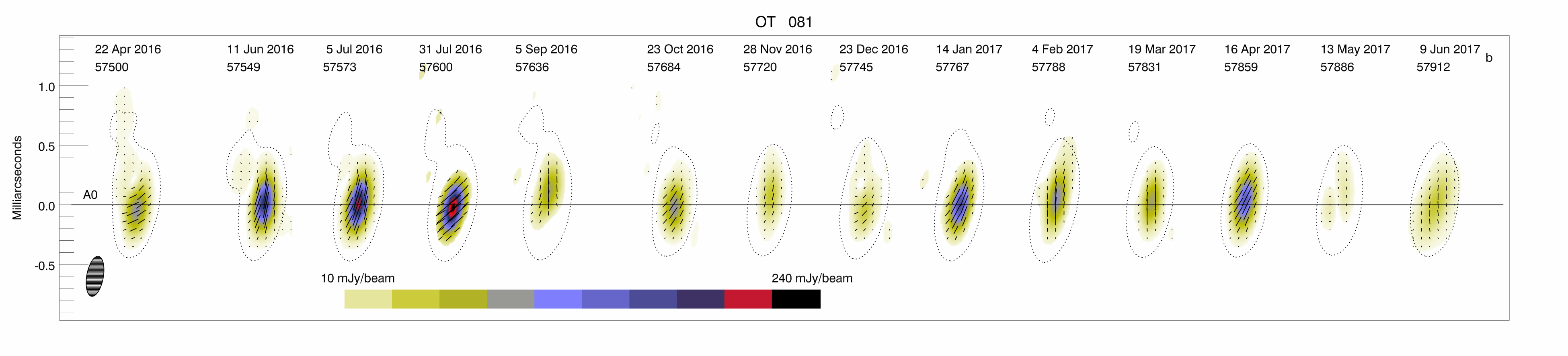} 
           \caption{Polarized intensity images plotted by color scale, with a 0.5\% contour of the total intensity peak shown by black dotted curves and black line segments within each image indicating the direction of the polarization electric vector. }
           \label{fig:1749_vlba2016b}
    \end{subfigure}
 \caption{Sequence of OT~081 VLBA images at 43\,GHz, convolved with a beam of FWHM dimensions 0.34$\times$0.14 mas$^{2}$ along the polarization angle PA=-10$^{\circ}$. Separated images/maps can be visible at \url{https://www.bu.edu/blazars/VLBA_GLAST/1749.html}}
 \label{fig:1749_vlba2016} 
\end{figure*}
\begin{table*}
  \begin{threeparttable}
	\centering
	\caption{Kinematic properties of knots K15 and K16}
	\label{tab:Speed}
	\begin{tabular}{lcc} 
		\hline
		Kinematic parameters  & K15 & K16 \\
		\hline
Proper motion (mas/yr)     &   0.826 $\pm$ 0.068                   &  0.918 $\pm$ 0.079               \\
Apparent speed (c)       &   16.83 $\pm$ 1.41                    &  18.59 $\pm$ 1.59                \\
Ejection time (MJD)      &   57346 $\pm$ 55 (20 November, 2015)  &  57565 $\pm$ 58 (26 June, 2016)  \\
T$_{\textrm{A1}}$ (MJD)    &     57408$\pm$21(20 January, 2016)       & 57620$\pm$19 (19 August 2016)  \\
		\hline
    \end{tabular}
    \begin{tablenotes}
      \item T$_{\textrm{A1}}$ is the time of the passage of moving knots through the stationary feature A1. The uncertainty in T$_{\textrm{A1}}$ is calculated using the uncertainties in the position of A1 and the proper motion of a component. The uncertainty in the ejection time is not included in the calculation of the uncertainty in T$_{\textrm{A1}}$.
    \end{tablenotes}
  \end{threeparttable} 
\end{table*}

\begin{table*}
	\centering
	\caption{Average parameters of the main features shown in Fig.~\ref{fig:1749_vlba2016}}
	\label{tab:AvePar}
\begin{tabular}{lcccc} 
		\hline
		Parameter& A0 & A1 & K15 & K16\\
		\hline
Number of epochs                &   23                          &    23                            &      9                                &  7                             \\
Average flux (Jy)       &   2.62 $\pm$ 0.87           &    0.48 $\pm$ 0.45             &   0.087 $\pm$ 0.014                   &  0.043 $\pm$ 0.013              \\
Maximum flux (Jy)       &   4.32 $\pm$ 0.06           &    1.85 $\pm$ 0.06             &   0.10  $\pm$  0.01                 &  0.06  $\pm$  0.01            \\
Average distance (mas)  &                               &    0.14 $\pm$ 0.04             &   0.70 $\pm$ 0.18                   &  0.60 $\pm$ 0.14              \\
Average PA (deg)        &                               &    6 $\pm$ 10                &   13 $\pm$ 4                      &  6 $\pm$ 6                 \\
Average Size (mas)      &   0.03 $\pm$ 0.01           &    0.06  $\pm$  0.03           &   0.33  $\pm$  0.13                 &  0.21  $\pm$  0.12            \\
		\hline
	\end{tabular}
\end{table*}   

The parsec-scale jet of OT~081 is strongly core-dominated at 43\,GHz. Fig.~\ref{fig:1749_vlba2016} shows the total and polarized intensity VLBA images of the blazar from April 2016 to June 2017. The very compact VLBI (Very Large Baseline Interferometry) core, A0, is located at the southern end of the jet, and it likely is a stationary physical structure in the jet. A quasi-stationary feature located 0.14$\pm$0.04~mas\footnote{milli-arc seconds: at the redshift $z=0.322$, 1~mas corresponds to 4.675~pc with the cosmological parameters adopted in this work (see Sec.~\ref{sec:introduction}).}
downstream of the core is detected at all 23 epochs. This knot is identified as feature A1 reported by \cite{2017ApJ...846...98J} that is located 0.11$\pm$0.04~mas from the core, although its position angle varies significantly throughout the epochs.
During the period analysed here, two superluminal knots, K15 and K16, were detected. Their positions according to the modelling are marked on the images presented in Fig.~\ref{fig:1749_vlba2016}.  
Table~\ref{tab:Speed} gives the kinematic properties of K15 and K16 while Table~\ref{tab:AvePar} lists the average parameters of the main features shown in Fig.~\ref{fig:1749_vlba2016}. Extrapolating the motions of the knots suggests that K15 and K16 coincided with the VLBI core on MJD~57346$\pm$55 ($\sim$20 November 2015) and MJD~57565$\pm$58 ($\sim$26 June 2016), respectively. 

According to Tables~\ref{tab:Speed} and \ref{tab:AvePar}, knots K15 and K16 have very similar properties. The derived ejection times of both knots are associated with periods of significant brightening of the core and an increase in its degree of polarization. 
In the Turbulent Extreme Multi-Zone (TEMZ) model proposed by \cite{2014ApJ...780...87M}, rapid VHE $\gamma$-ray flares result from a temporary alignment of the turbulent magnetic field with a direction relative to a shock front that is most favourable to extremely efficient particle acceleration \citep[e.g.][]{2017MNRAS.464.4875B}. 
In the magnetic reconnection model of \cite{2009MNRAS.395L..29G}, plasma zones with oppositely directed magnetic fields come into contact (perhaps also driven by turbulence), creating ``mini-jets'' of VHE particles that stream at velocities near $c$ relative to the ambient jet plasma.
Although the above interpretations appear to be a viable explanation for the $\gamma$-ray activity of OT~081 during the 2016 outburst, an unanswered question is why knot K15, with properties similar to those of K16, did not trigger $\gamma$-ray activity. One explanation is that K15 was ejected after a prolonged period with a low level of nonthermal activity, providing few potential target photons for Compton scattering. Since K16 was ejected only $\sim$200~days after K15, it is possible that K15 generated optical-infrared seed photons (synchrotron or emission lines or dust from nearby clouds heated by a UV flare from K15) that the electrons in K16 later scattered to $\gamma$-ray energies.

\newpage

\section{Proton-synchrotron dominated model}
\label{sec:appendix}
\begin{figure*}
\centering
\begin{subfigure}[t]{0.5\textwidth}
\centering
\includegraphics[width=\linewidth]{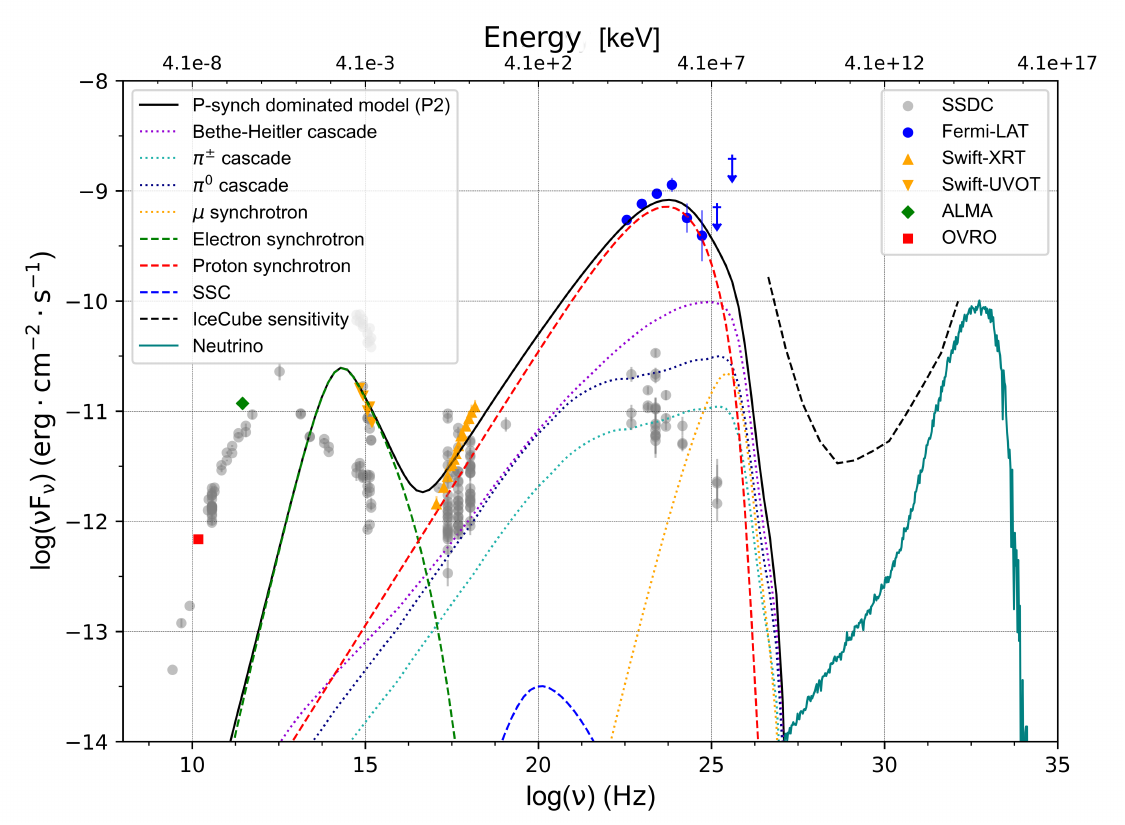}
\caption{P2: MJD~57589.5 (20 July 2016)}\label{fig:psynch_P2}
\end{subfigure}%
\hfill
\begin{subfigure}[t]{0.5\textwidth}
\centering
\includegraphics[width=\linewidth]{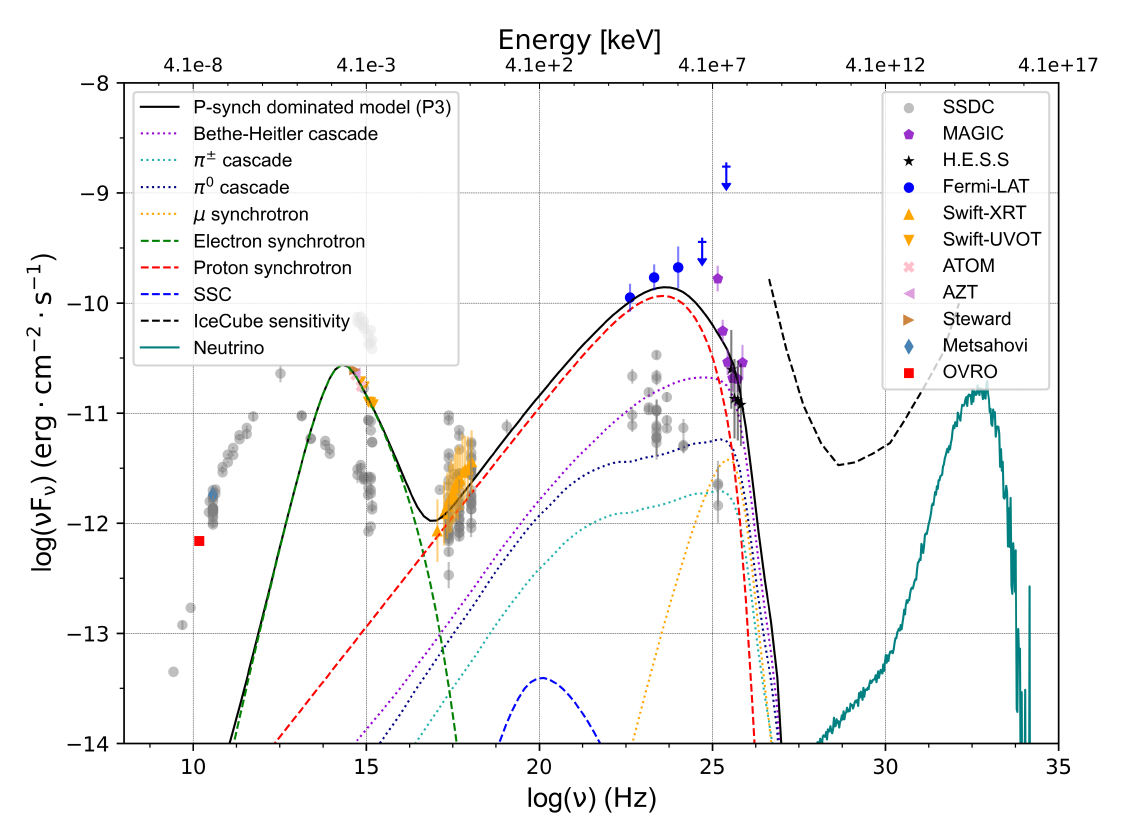}
\caption{P3: MJD~57593.9 (24 July 2016)} \label{fig:psynch_P3}
\end{subfigure}%
\begin{subfigure}[t]{0.5\textwidth}
\centering
\includegraphics[width=\linewidth]{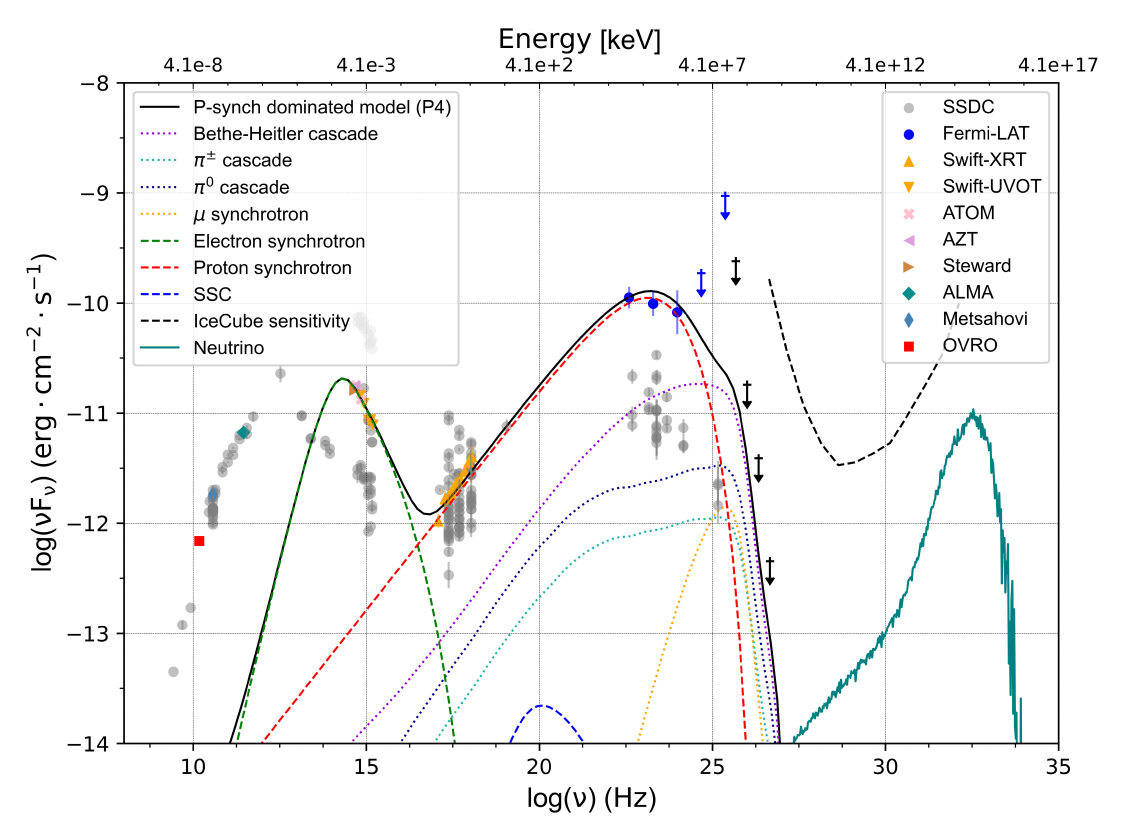}
\caption{P4: MJD~57595 (26 July 2016)} \label{fig:psynch_P4}
\end{subfigure}%

 \caption{SEDs for the various states of OT~081 plotted with the best-fitting proton-synchrotron models (black solid line). The states P2, P3, and P4 are shown. The green dashed lines represent the synchrotron components. The blue dashed lines are SSC components, the purple dotted lines are the Bethe-Heitler cascades, the blue and cyan dotted lines are the respective cascades from $\pi^0$ and $\pi^{\pm}$, the red dashed lines are the proton-synchrotron models, and the solid teal lines represent the calculated neutrino fluxes. The parameters used for this model are listed in Table~\ref{table:psyncmodel}. The IceCube sensitivity curve from ~\citet{2019EPJC...79..234A} is shown as a black dashed line. Grey circles are archival data from ASI/SSDC.}
 \label{fig:psynch_all} 
\end{figure*}

In Fig.~\ref{fig:psynch_all}, we show proton-synchrotron dominated models of the SED for the various states of activity for OT~081. These models were obtained using the hadronic code described in~\citet{Cerruti15}. In contrast with leptonic models in which the two SED components are related, proton-synchrotron scenarios result in SED components that are independent, resulting in a degenerate set of solutions. To reduce the parameter space, we make two physical assumptions: (1) electrons and protons share the same acceleration mechanism, resulting in the same injection index $\alpha^\prime_{e/p,1}$ for their distributions; (2) the maximum proton Lorentz factor $\gamma^\prime_{\rm p,max}$ is determined by the balance between the acceleration time scale $\tau_{acc} \simeq 10\ (m_p c / eB)\ \gamma^\prime_p$ \citep[see e.g.][]{Rieger07} and the adiabatic time scale $\tau_{ad} \simeq R^\prime/c$, which is the shortest cooling time scale for protons. Even with these assumptions, we can only provide a sample solution that describes the SED of OT~081 that uses typical values for the Doppler factor $\delta = 30$ and the magnetic field $B^\prime=10$ G. The particle distribution spectra follow a power-law with an exponential cutoff. The complete list of parameter values is provided in Table~\ref{table:psyncmodel}. 

In this hadronic scenario, the $\gamma$-ray emission of OT~081 is dominated by proton-synchrotron radiation in the GeV band while the VHE emission is produced by pair cascades (from Bethe-Heitler interactions and pion decays) and muon-synchrotron radiation. Arguably, the most important drawback of hadronic blazar models is the amount of power in protons needed to fit the data \citep{Sikora09, Zdziarski15}.  The model shown here is indeed characterised by a large value of $L_p = 5.6\times10^{48}$ erg s$^{-1}$, which is about a factor of $400$ times as large as the Eddington luminosity of a super massive black hole with mass $M_\bullet = 10^8\ M_\odot$. The corresponding calculated neutrino flux peaks in the EeV band with a flux of $10^{-11}$ erg cm$^{-2}$ s$^{-1}$, at the level of the secondary photons from pion decay. \\

\begin{table}
\caption{Parameters of the proton-synchrotron models for periods P2 (MJD 57589.5 -- 20 July 2016), P3 (MJD~57593.9 -- 24 July 2016), and P4 (MJD~57595 -- 26 July 2016). Parameters are described in the text. The quantities flagged with stars are derived quantities and not model parameters.}
    \centering
   		\begin{tabular}{@{}l ccc}
 		\hline
 		\hline
		 		 & P2 & P3 & P4 \\
 		\hline
 		 \noalign{\smallskip}
          $\delta$ & $30$  & $30$  & $30$ \\
      $R^\prime$ [10$^{15}$ cm] & $5$ & $5$ & $5$  \\
 $^\star \tau_{\rm obs}$ [hours] & $2$ & $2$  & $2$ \\ 
 		\hline
         $B^\prime$ [G] & $10$ & $10$ & $10$\\
   $^\star U^\prime_B$ [erg cm$^{-3}$] &  $4.0$  &  $4.0$  &  $4.0$ \\
 		\hline
     $\gamma^\prime_{\rm e,min} $& $300$ & $300$ & $300$ \\
   $\gamma^\prime_{\rm e,break} $&  $=\gamma_{\rm e,min}$ &  $=\gamma_{\rm e,min}$ &  $=\gamma_{\rm e,min}$ \\
        $\gamma^\prime_{\rm e,max}$&  $3000$ &  $3000$ &  $3000$ \\
 		$\alpha^\prime_{\rm e,1}=\alpha^\prime_{p,1}$ & $2.0$ & $2.2$  & $2.2$ \\
   $\alpha^\prime_{\rm e,2}$ & $3.9$ & $3.9$ & $3.9$ \\
 		$K^\prime_{\rm e}$ [cm$^{-3}$]   & $2300$ & $8000$ & $6000$ \\
 		$^\star u^\prime_{\rm e}$ [10$^{-4}\,$erg$\,$cm$^{-3}$] &  $8.2$ & $9.1$& $6.8$ \\
 		\hline
         $\gamma^\prime_{\rm p,min}$& $1$ & $1$ &  $1$\\
 		$\gamma^\prime_{\rm p,max} [10^9]$ & $1.2$ & $1.3$  & $0.8$ \\
        $K^\prime_{\rm p}$ [10$^{3}$ cm$^{-3}$]   & $759$ & $720$ & $1020$ \\
 		$^\star U^\prime_{\rm p}$ [10$^{3}\,$erg cm$^{-3}$] & $2.3$ & $5.3$ & $7.5$ \\
 		\hline
    	$^\star U^\prime_{\rm p}/U_{\rm B} [10^{3}]$ & $0.6$ & $1.3$ & $1.9$\\
 		$^\star L$ [10$^{48}$ erg s$^{-1}$] & $2.5$ & $5.6$   & $8.0$\\
        $^\star \nu_{\textrm{rate}} $ [s$^{-1}$] & $0.78$ & $0.055$ & $0.34$ \\
 		\hline
 		\hline
 		
		\end{tabular}
 	 \newline
 		      	 \label{table:psyncmodel}
 		\end{table}


\newpage
\section{Joint fit of the HE and VHE data}\label{sec:appendix:fitJ}
\begin{figure*}
\centering
\begin{subfigure}[t]{0.5\textwidth}
\centering
\includegraphics[width=\linewidth]{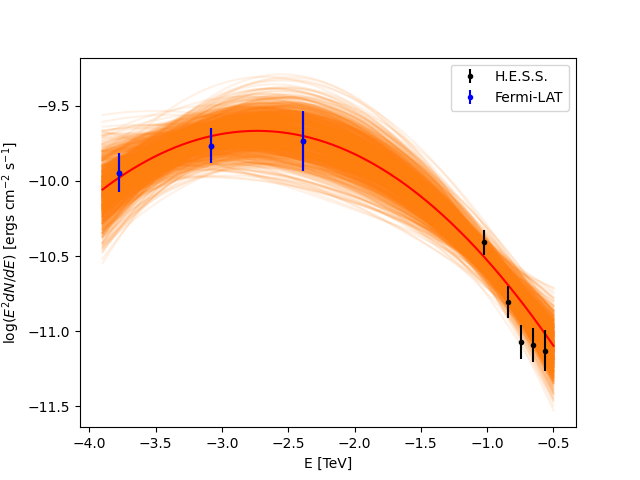}
\end{subfigure}%
\hfill
\begin{subfigure}[t]{0.5\textwidth}
\centering
\includegraphics[width=\linewidth]{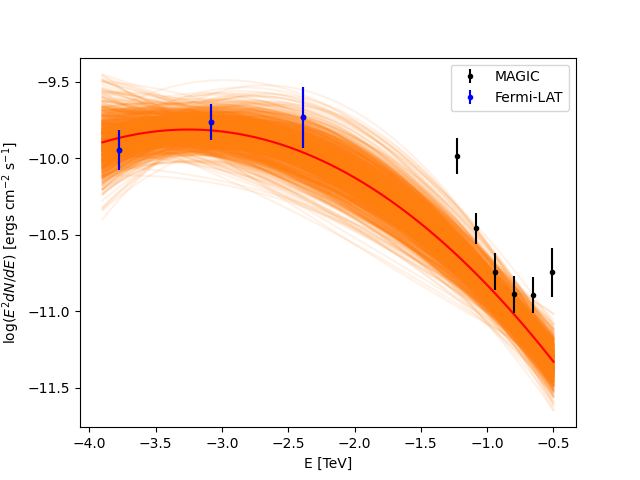}

\end{subfigure}%
 \caption{The \textit{Fermi}-LAT and VHE SEDs of OT~081. Plotted at left is the SED using H.E.S.S. data, and plotted at right is the SED using MAGIC data. The red lines are the best-fit models, and the orange areas are the corresponding 1$\sigma$ error contours.}\label{fig:joint}
 
\end{figure*}

We have performed joint fits of \textit{Fermi}-LAT data points with those from either H.E.S.S. or MAGIC (see Fig.~\ref{fig:joint}). To perform the fit we use $\chi^2$ statistic, and the energy correlation between the VHE points is taken into account by using a covariance matrix. The \textit{Fermi}-LAT points are considered to not be correlated. For the minimisation, we used an MCMC method that was implemented in the emcee python package.

The VHE data points are corrected for EBL absorption, and the fit is then performed in log-log space. We used three models in fitting the data - a power-law, a log-parabola and a power-law with an exponential cutoff. The best-fit results are shown in Fig.~\ref{fig:joint}.

\bsp	
\label{lastpage}
\end{document}